\documentstyle[11pt,aaspp4]{article}
\def\kms{km s$^{-1}$}
\def\dla{Ly$\alpha$}
\def\nav{$N_{\rm a}(v)$}
\def\lya{Ly$\alpha$\ }
\def\zabs{$z_{\rm abs}$}
\begin{document}

\title{High Signal-to-Noise Echelle Spectroscopy \\ of QSO 
Absorption Line Systems with Metals \\ in the Direction of HS
1700+6416\footnote{Based on observations obtained with the
Kitt Peak National Observatory 4 m telescope, which is operated
by the Association of Universities for Research in Astronomy
(AURA) under contract with the National Science Foundation.}}

\author{Todd M. Tripp\altaffilmark{2,}\altaffilmark{3}, Limin 
Lu\altaffilmark{4,}\altaffilmark{5}, and Blair D. 
Savage\altaffilmark{2}}
\altaffiltext{2}{Department of Astronomy, University of Wisconsin 
- Madison, 475 N. Charter St., Madison, WI 53706 - 1582, 
Electronic mail: tripp@astro.princeton.edu, 
savage@madraf.astro.wisc.edu}

\altaffiltext{3}{Current address: Princeton University Observatory, 
Peyton Hall, Princeton, NJ 08544}

\altaffiltext{4}{Astronomy Department, 105-24, California Institute 
of Technology, Pasadena, CA 91125, Electronic mail: 
ll@astro.caltech.edu}

\altaffiltext{5}{Hubble Fellow}

\begin{abstract}
We have obtained a high signal-to-noise (30 $\leq$ S/N $\leq$ 70) 
high resolution (FWHM = 20 \kms ) spectrum of the radio-quiet 
QSO HS 1700+6416 ($z_{\rm em}$ = 2.72) with the echelle 
spectrograph on the KPNO 4m telescope. We detect 13 metal 
systems in the optical spectrum of this QSO, including six systems 
with associated optically thin Lyman limit absorption in the {\it 
HST} spectrum obtained by Reimers et al. We use the apparent 
column density technique and profile fitting to measure the heavy 
element column densities and to evaluate the impact of unresolved 
absorption saturation. Profile fitting indicates that four of the 
\ion{C}{4} systems are narrow with $b \ <$ 8 \kms , which implies 
that these absorbers are relatively cool and probably photoionized.

The dense cluster of \ion{C}{4} doublets at 2.432 $<$ \zabs\ $<$ 
2.441 shows the weak line of one \ion{C}{4} absorber apparently 
aligned with the strong line of a different \ion{C}{4} doublet, i.e., 
line locked, for two pairs of \ion{C}{4} absorbers. Line locking has 
been detected previously in \zabs\ $\approx$ $z_{\rm em}$ 
absorbers where radiation pressure is likely to play a role, but it is 
surprising in this case since this \ion{C}{4} complex is displaced by 
$\sim$24000 \kms\ from the QSO emission redshift. This may be 
the remnant (or precursor) of a broad absorption line (BAL) outflow. 
However, it is possible that these alignments are chance alignments 
rather than true line locking.

The high ion column density ratios in the multicomponent Lyman 
limit absorber at \zabs\ = 2.3150 suggest that the ionization 
conditions in this absorber differ significantly from the conditions in 
the gaseous halo of the Milky Way. From photoionization models 
we derive [Si/H] $\geq$ --0.95 and [Al/H] $\geq$ --0.96 for the 
stongest component of this absorber. These are conservative lower 
limits derived from lower ionization stages only; photoionization 
models in agreement with the observed low and high ionization 
stages require [M/H] $\approx$ --0.45. In contrast, Vogel \& 
Reimers derive [N/H] $<$ --1.04 and [O/H] = --1.52 for this 
absorber. We suggest that the discrepancy is due to the low 
resolution of the Vogel \& Reimers data (FWHM $\approx$ 300 
\kms ) which introduces serious blending and saturation problems. 
The photoionized model with [M/H] = --0.45 has a particle density 
$n_{\rm H} \ \approx$ 0.02 cm$^{-3}$, a size of a few hundred pc, 
and a mass of roughly 1 $\times \ 10^{5} M_{\odot}$, assuming the 
absorber is spherical.

We detect unsaturated \ion{C}{4} and rather strong \ion{N}{5} 
``associated'' absorption at \zabs\ = 2.7125. The apparent column 
density of the weak \ion{N}{5} 1242.8 \AA\ line is greater than the 
apparent column density of the stronger \ion{N}{5} 1238.8 \AA\ 
line in this absorber, which indicates that the \ion{N}{5} profile is 
affected by unresolved saturation or that the \ion{N}{5} absorbing 
gas does not completely cover the QSO emission source. If the latter 
interpretation is correct, then the associated absorbing gas must be 
close to the QSO. We have used the observed spectral energy 
distribution of the QSO, corrected for intervening Lyman limit 
absorption, for photoionization modeling of the associated absorber, 
and we derive [N/H] $\geq$ --0.65 and [C/H] $\geq$ --0.82.
 Other corrections (e.g., to account for dust 
in intervening absorbers or `Lyman valley' attenuation) will make 
the spectral energy distribution harder and increase the metallicity 
estimates. However, the absorption profiles suggest that the constant 
density single slab model is too simplistic: we obtain 
$b$(\ion{N}{5}) = 25.2$\pm$1.3 \kms\ and $b$(\ion{C}{4}) = 
11.4$\pm$1.1 \kms\ from profile fitting, and therefore the gas which 
produces the \ion{N}{5} absorption does not have the same 
temperature and non-thermal motions as the \ion{C}{4} gas.

Finally, we briefly examine the number of \ion{Mg}{2} systems 
detected per unit redshift, and we tentatively conclude that $dN/dz$ 
is dominated by weak \ion{Mg}{2} lines with $W_{\rm r} \ <$ 0.3 
\AA .
\end{abstract}


\section{Introduction}

The successful deployment of the {\it Hubble Space Telescope 
(HST)} has opened many windows for the study of the high redshift 
universe. In the spectroscopic arena, the {\it HST} Faint Object 
Spectrograph (FOS) observation of the luminous radio-quiet QSO 
HS 1700+6416 ($z_{\rm em}$ = 2.72) by Reimers et 
al.\markcite{r5} (1992) has provided an extraordinary spectrum 
which demonstrates the potential of the {\it HST} for the study of 
high $z$ QSO absorption lines. This FOS spectrum has yielded 
results on topics about which little or nothing was known before the 
launch of the {\it HST}, including (1) the helium content of QSO 
absorbers at \zabs\ $>$ 1.84, and (2) an ultra-highly ionized phase of 
QSO absorbers traced by species such as \ion{Ne}{8} (see also 
Hamann et al.\markcite{h1} 1995), which may reveal the presence 
of gas with $T \ \gg \ 10^{5}$ K (Verner, Tytler, \& 
Barthel\markcite{v1} 1994). These new topics are observationally 
challenging because the resonance lines of \ion{He}{1}, 
\ion{He}{2}, and the ultra-highly ionized species (e.g., 
\ion{Ne}{8}, \ion{Mg}{10}, and \ion{Si}{12}) all occur at rest 
wavelengths $\lambda _{\rm r} \ <$ 912 \AA , and only a handful of 
QSOs are currently known with continuum flux detected at 
$\lambda _{\rm r} \ll$ 912 \AA\ (c.f. Beaver et al.\markcite{b2} 
1991; Jakobsen et al.\markcite{j1} 1994; \S 6.1 in Tripp, Bechtold, 
\& Green\markcite{t1} 1994; Reimers et al.\markcite{r4} 1995b; 
Lyons et al.\markcite{l4} 1995; Tytler et al.\markcite{t4} 1995). 
The continua of background QSOs with sufficient redshift to show 
these absorption lines are usually undetectable due to attenuation by 
intervening absorption, either optically thick Lyman limit (LL) 
absorption or the cumulative effect of many absorbers along the line 
of sight (c.f., 
M\o ller \& Jakobsen\markcite{m6} 1990; Picard \& 
Jakobsen\markcite{p7} 1993). The HS 1700+6416 sight line has its 
share of LL absorbers --- seven LL systems are well-detected in the 
Reimers et al. FOS spectrum --- but remarkably all of the LLs are 
optically thin ($\tau _{\rm LL} \lesssim$ 1) and the quasar 
continuum is detected from $\sim$1200 to 3250 \AA , which 
corresponds to $\sim$320 to 870 \AA\ in the QSO restframe.

In addition to the ground-breaking {\it HST} spectrum, this sight 
line is intriguing because a deep ROSAT PSPC image obtained by 
Reimers et al.\markcite{r1} (1995a) shows two extended X-ray 
sources within a few arcminutes of the QSO. Both of these X-ray 
objects are flagged as greater than 3$\sigma$ extended sources by 
the SHARC survey (see Romer et al.\markcite{r8} 1996a, 
1996b\markcite{ro}). One of these X-ray sources is Abell 2246, a 
galaxy cluster at $z$ = 0.225. The other extended source is less 
easily identified and may be a higher redshift galaxy cluster. The 
close proximity of these X-ray objects to the sight line offers the 
tantalizing possibility of studying the detailed properties of galaxy 
intracluster media detected in absorption in the spectrum of HS 
1700+6416.

HS 1700+6416 is also an excellent target for ground-based studies 
of LL absorbers. Since the LL systems in the {\it HST} spectrum are 
all optically thin, the Lyman continuum optical depth can be 
measured, and this provides an accurate estimate of the \ion{H}{1} 
column density. Previous studies of LL absorbers (e.g., 
Steidel\markcite{s11} 1990) have mostly used samples of optically 
thick LL systems and have been forced to rely on elaborate methods 
to estimate $N$(\ion{H}{1}). To study the abundances and physical 
conditions in LL systems, and to complement space-based programs, 
we have obtained a high resolution (FWHM = 20 \kms ) high 
signal-to-noise optical spectrum of HS 1700+6416, and in this paper 
we present the data and analyses of the absorption systems. The 
paper is organized as follows. We briefly review the observations 
and data reduction in \S 2 and the line selection and column density 
measurements in \S 3. In \S 4 we provide some comments on the 
individual absorption systems including assessments of absorption 
line saturation. In this section we also comment on a dense cluster of 
\ion{C}{4} absorbers at 2.432 $<$ \zabs\ $<$ 2.441 in which the 
weak lines of some \ion{C}{4} absorbers are well-aligned with the 
strong lines of different \ion{C}{4} doublets. In \S 5 we discuss the 
physical conditions and abundances in the absorption systems. In \S 
6 we examine the number of \ion{Mg}{2} absorbers detected per 
unit redshift including weak absorbers. After some comments on the 
abundances derived by Vogel \& Reimers\markcite{v3} (1995) from 
the low resolution FOS spectrum in \S 7, we summarize the paper in 
\S 8.

\section{Observations and Data Reduction}

The observations and data reduction techniques are fully described 
in Tripp et al.\markcite{t2} (1996); we summarize a few of the 
pertinent details here. HS 1700+6416 was observed during darktime 
(18--20 May 1993) with the echelle spectrograph and the T2KB 
2048$\times$2048 CCD on the 4m telescope at Kitt Peak National 
Observatory (KPNO). With a 1$\farcs 33 \times 9\farcs7$ entrance 
aperture, the instrument setup provided a resolution of 20 \kms\ with 
$\sim$2 pixels/FWHM and spectral coverage from 4300 to 8350 
\AA\ with no gaps between echelle orders. Five one-hour 
observations were obtained over the course of three nights. 

The preliminary data reduction (overscan and bias subtraction, 
flatfielding, etc.) was done with IRAF following the standard 
procedure. The ``optimal'' spectrum extraction routine described by 
Horne\markcite{h5} (1986; see also Robertson\markcite{r6} 1986) 
was used to extract the final spectra from the individual 
observations, and then the individual spectra were rebinned to the 
same wavelength scale and coadded. In addition to the flux 
spectrum, the optimal extraction program yields a 1$\sigma$ error 
spectrum which includes the contribution from read noise as well as 
statistical noise, and this error spectrum was propagated throughtout 
the data reduction and analysis procedures. The coadded spectrum 
was normalized by fitting cubic splines to regions free of absorption 
lines. Overlapping regions of adjacent orders were weighted by their 
signal-to-noise ratios and coadded. The final normalized spectrum of 
HS 1700+6416 is shown in Figure 1. The signal-to-noise ratio of the 
final spectrum ranges from $\sim$30 to 70 per resolution element at 
$\lambda >$ 4550 \AA\ (the S/N is somewhat lower within the \lya 
forest).

\begin{figure}
\caption{(following pages) Normalized spectrum of HS 1700+6416 obtained 
with the echelle spectrograph on the KPNO 4m telescope, plotted as a 
function of vacuum heliocentric wavelength. Absorption lines due to 
extragalactic metals are marked with a number that identifies the 
line in column 1 of Table 1. At $\lambda <$ 4560 \AA , most of the 
absorption lines are due to the \dla\ forest. The resolution is 20 \kms\ 
with 30 $\lesssim S/N \lesssim$ 70 per resolution element (the S/N 
decreases somewhat below 4500 \AA ). The zero level is indicated 
with a dashed line, and the solid line near zero is the 1$\sigma$ 
uncertainty due to photon statistics and read noise. The 
discontinuous increase in the flux uncertainty at $\lambda \approx$ 
4520 \AA\ is due to a bad column in the CCD which prevented 
coaddition of overlapping orders. Telluric absorption lines have not 
been removed from this spectrum.}
\end{figure}
\begin{figure}
\figurenum{1}
\plotone{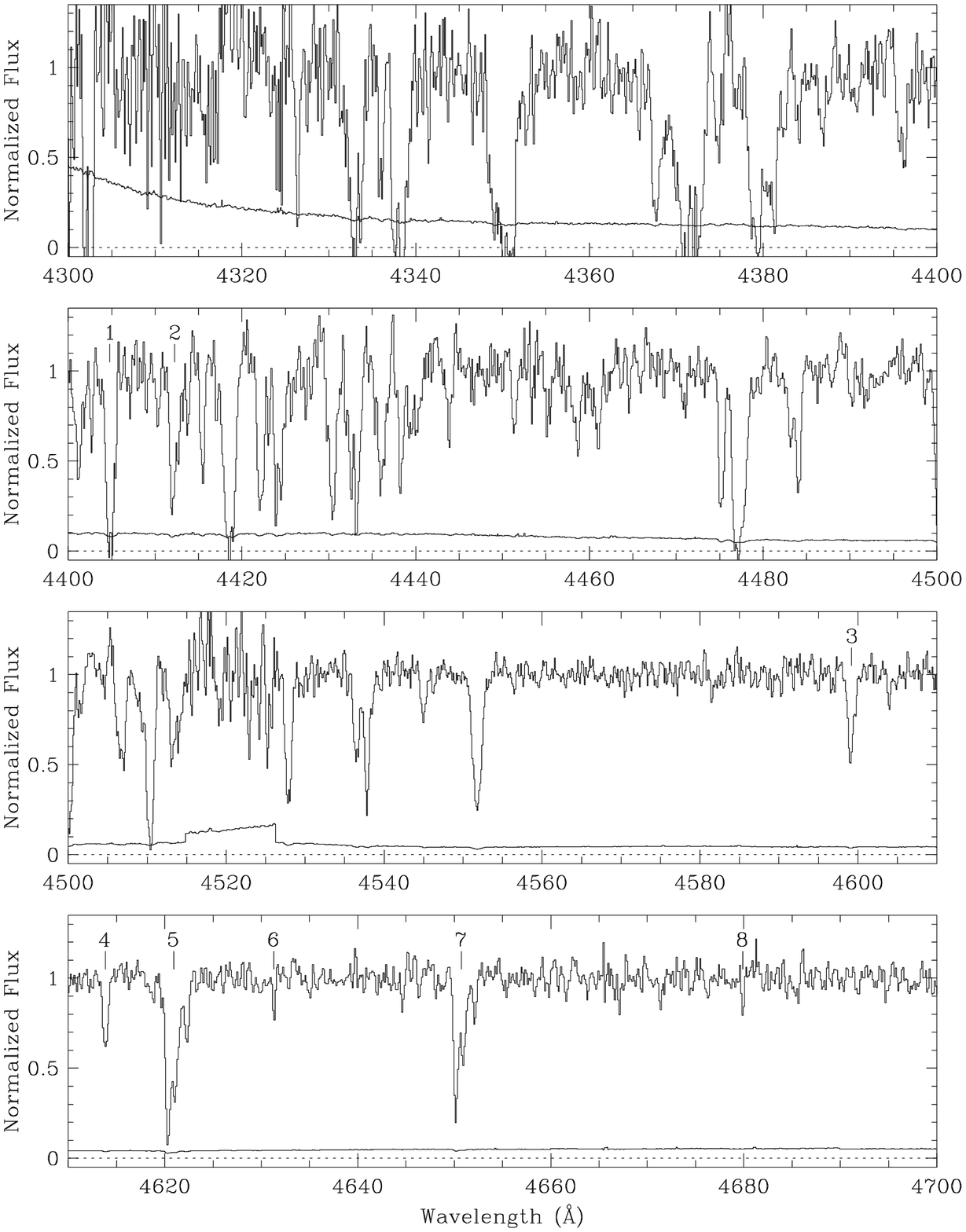}
\caption{continued}
\end{figure}
\begin{figure}
\figurenum{1}
\plotone{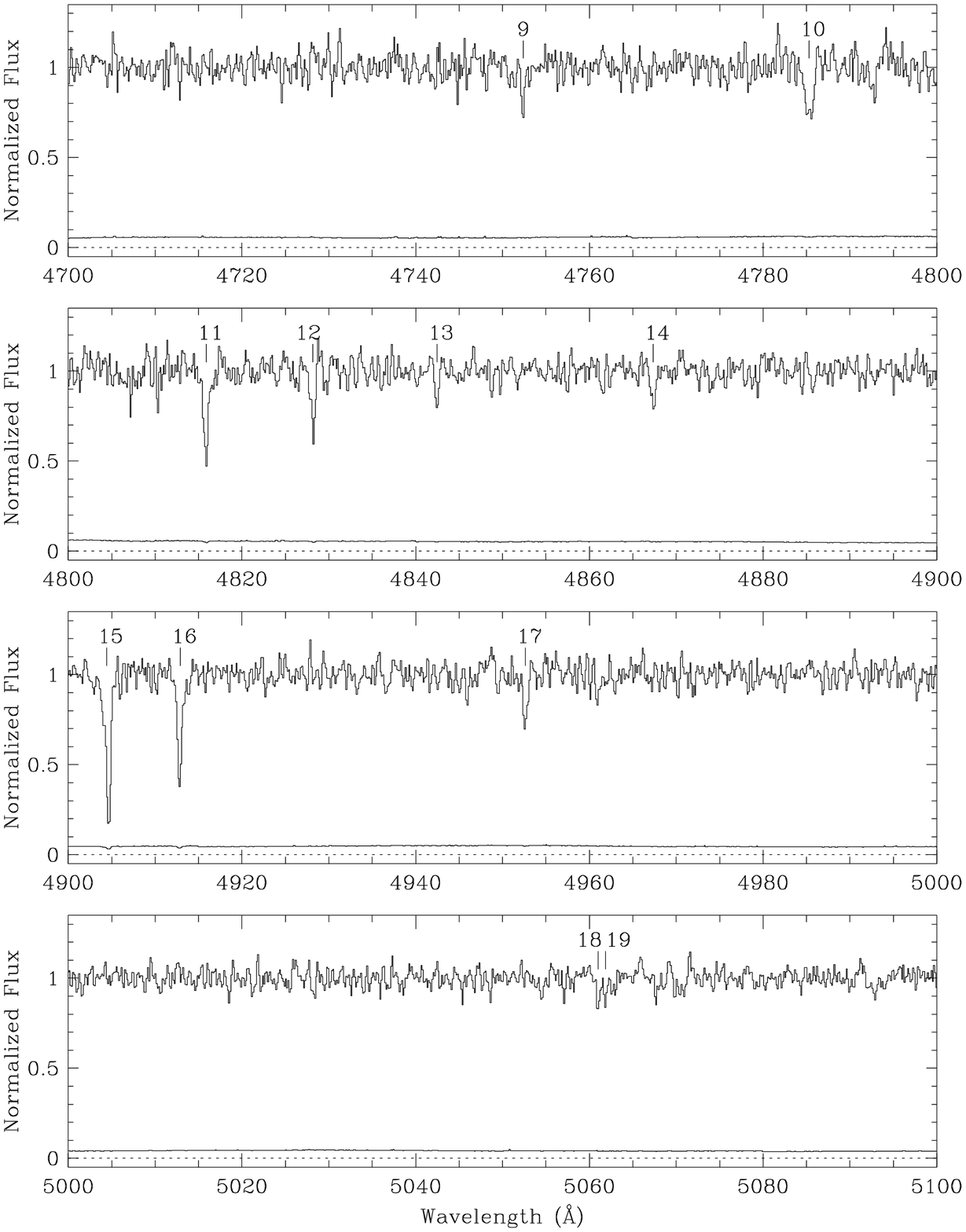}
\caption{continued}
\end{figure}
\begin{figure}
\figurenum{1}
\plotone{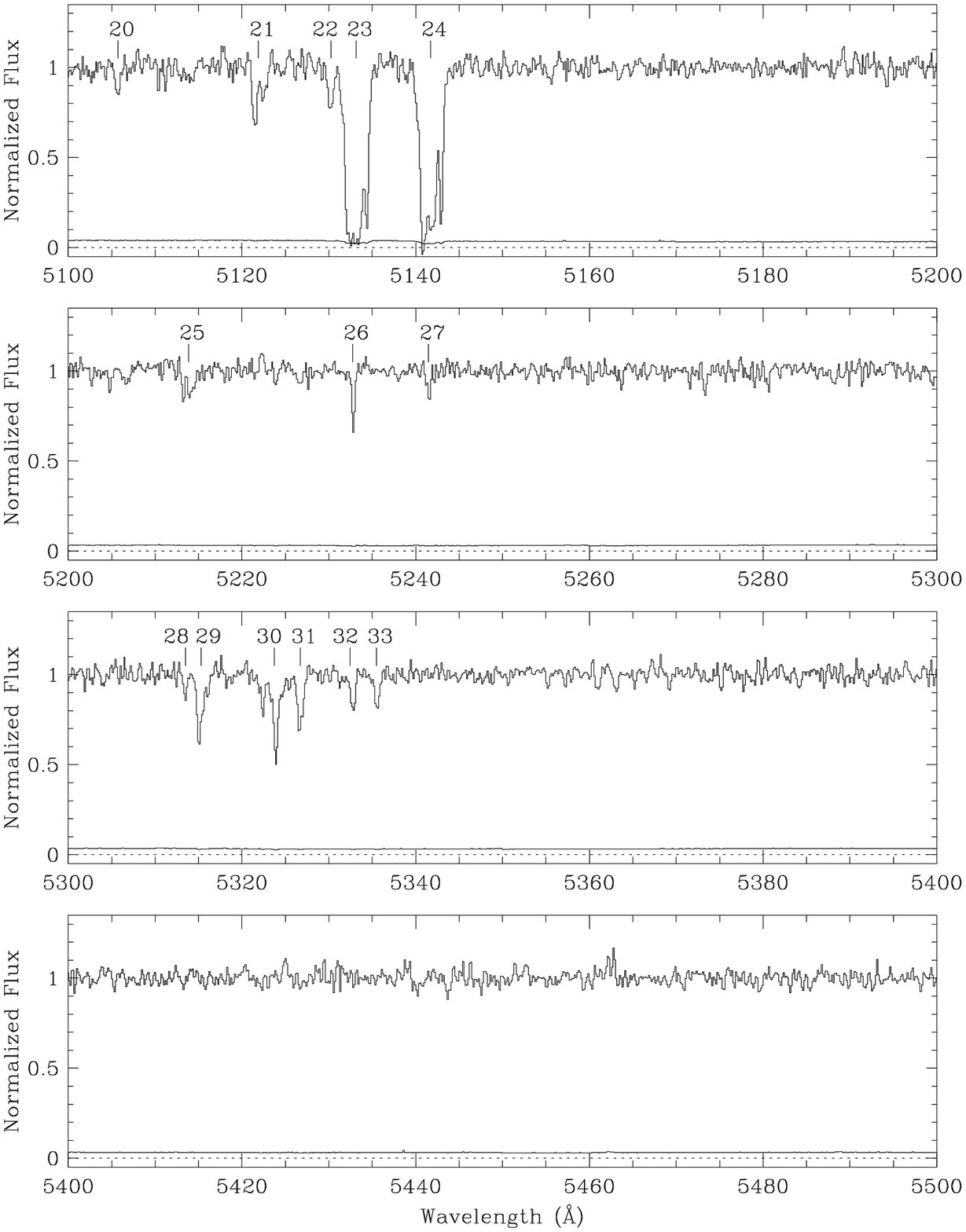}
\caption{continued}
\end{figure}
\begin{figure}
\figurenum{1}
\plotone{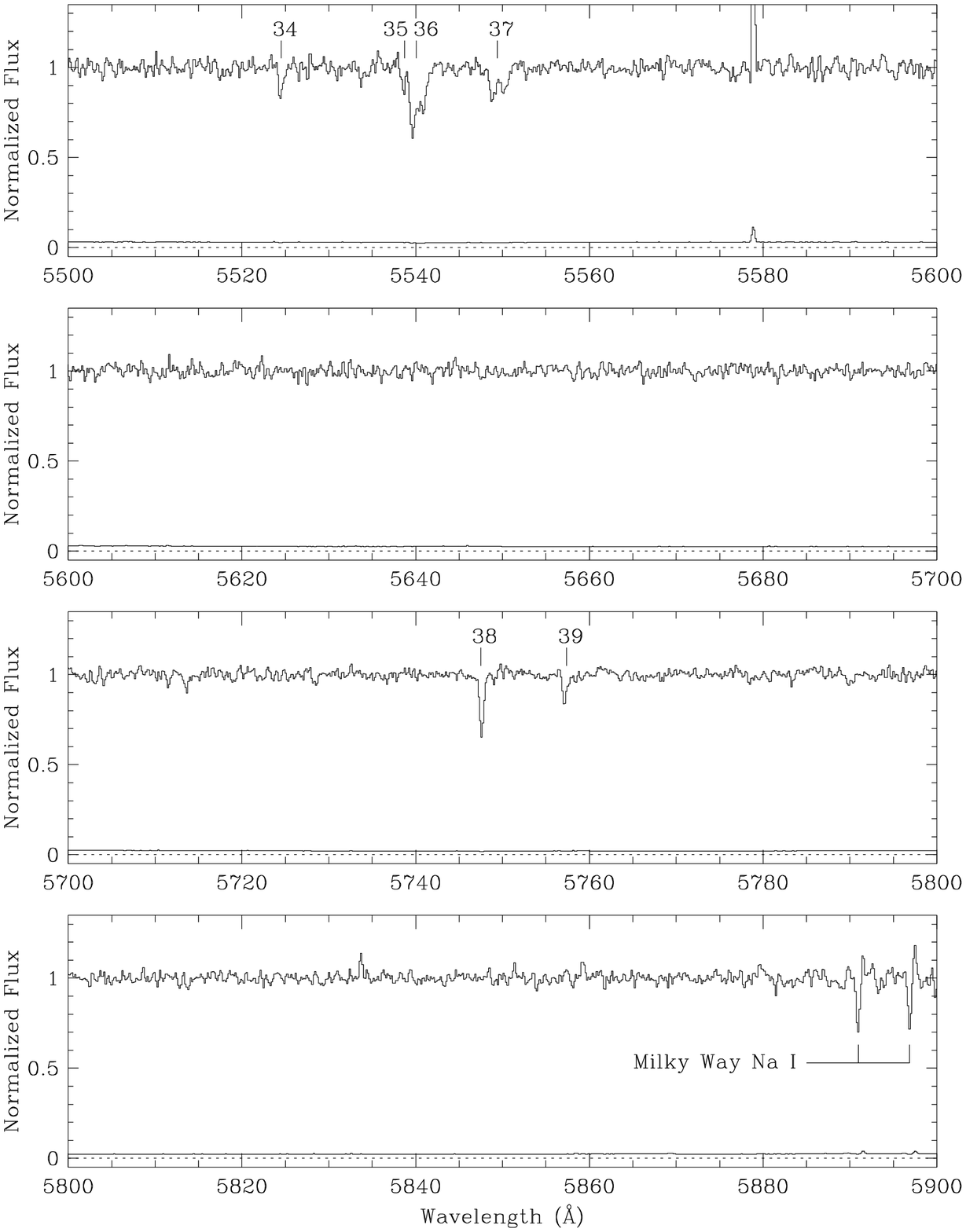}
\caption{continued}
\end{figure}
\begin{figure}
\figurenum{1}
\plotone{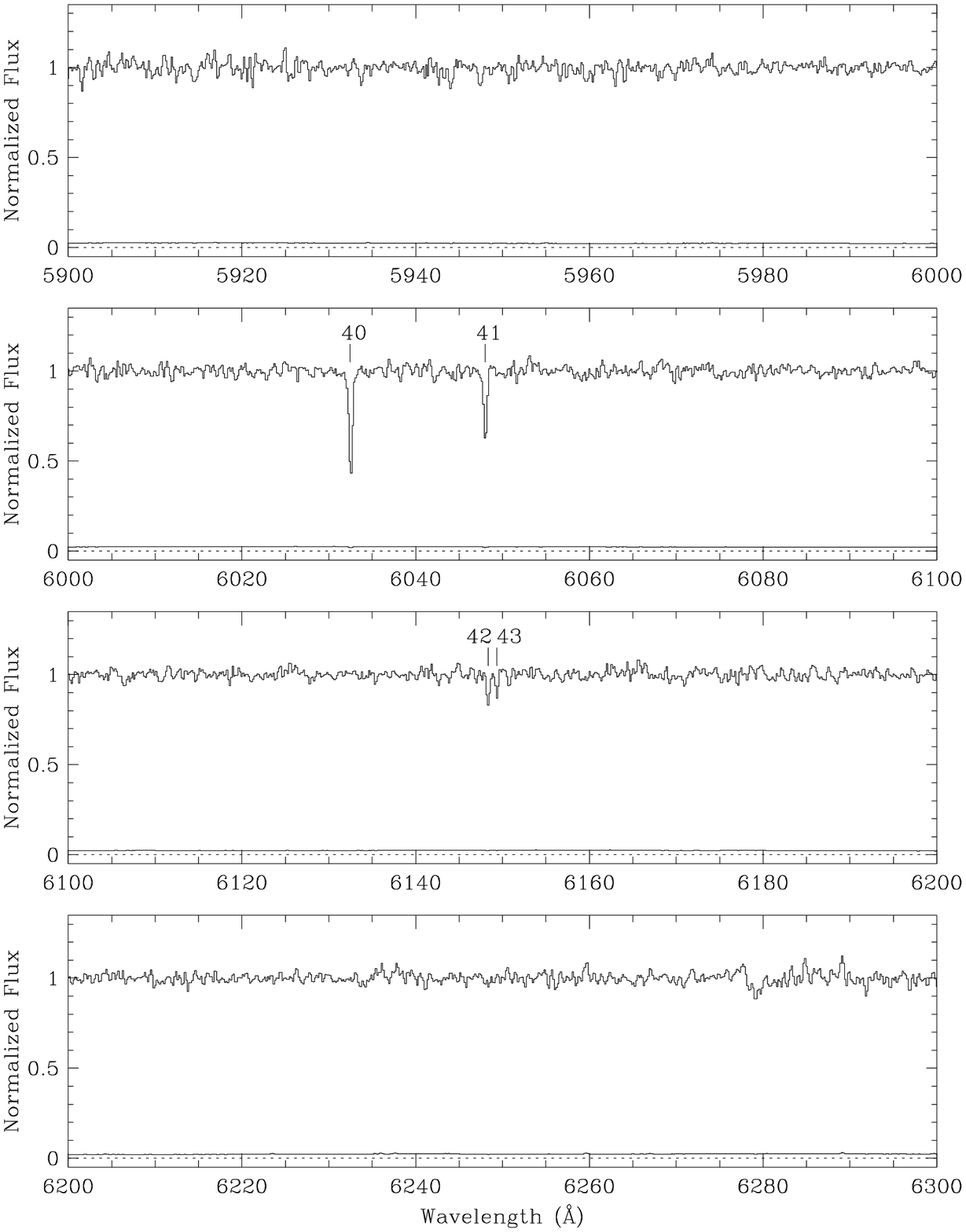}
\caption{continued}
\end{figure}
\begin{figure}
\figurenum{1}
\plotone{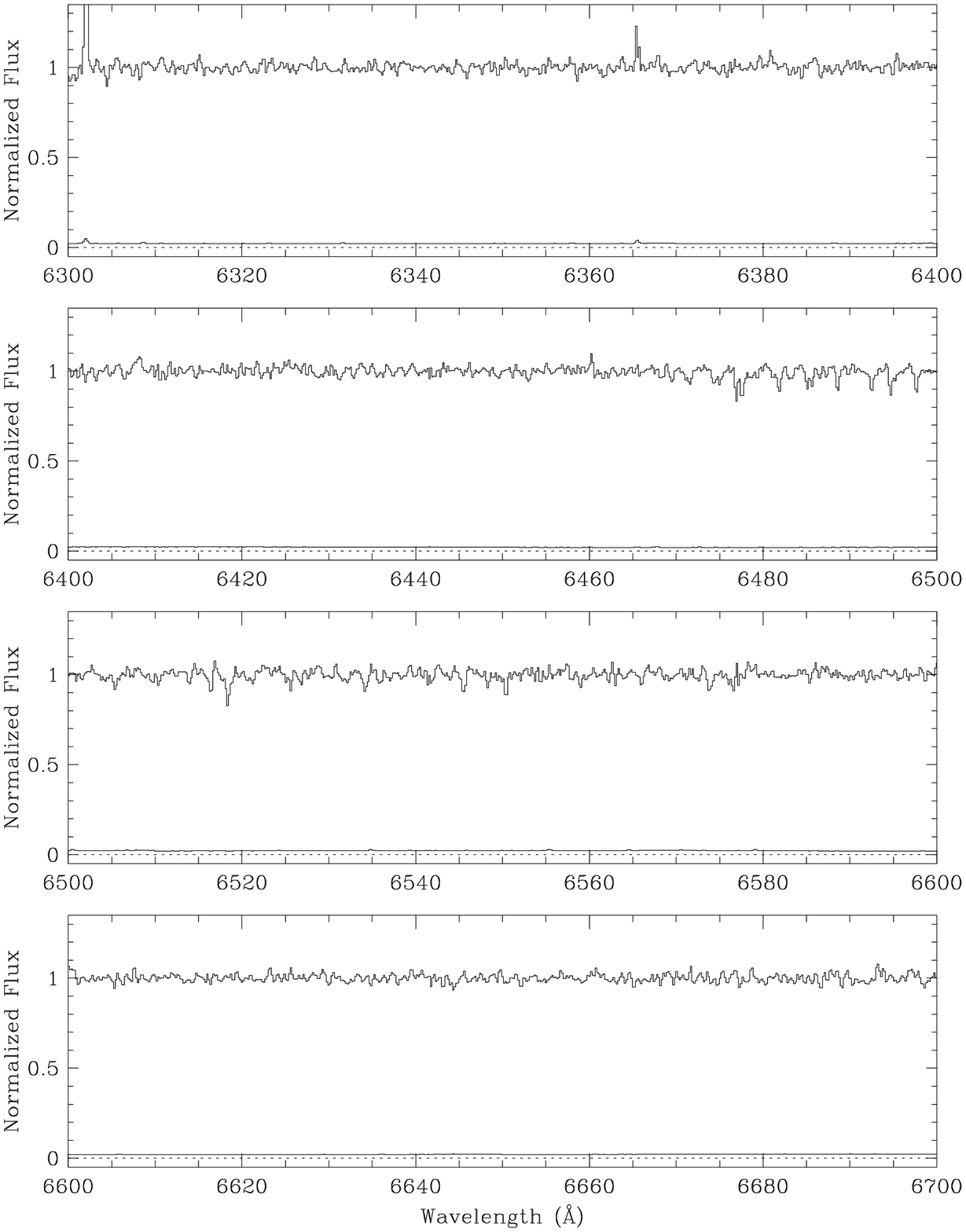}
\caption{continued}
\end{figure}
\begin{figure}
\figurenum{1}
\plotone{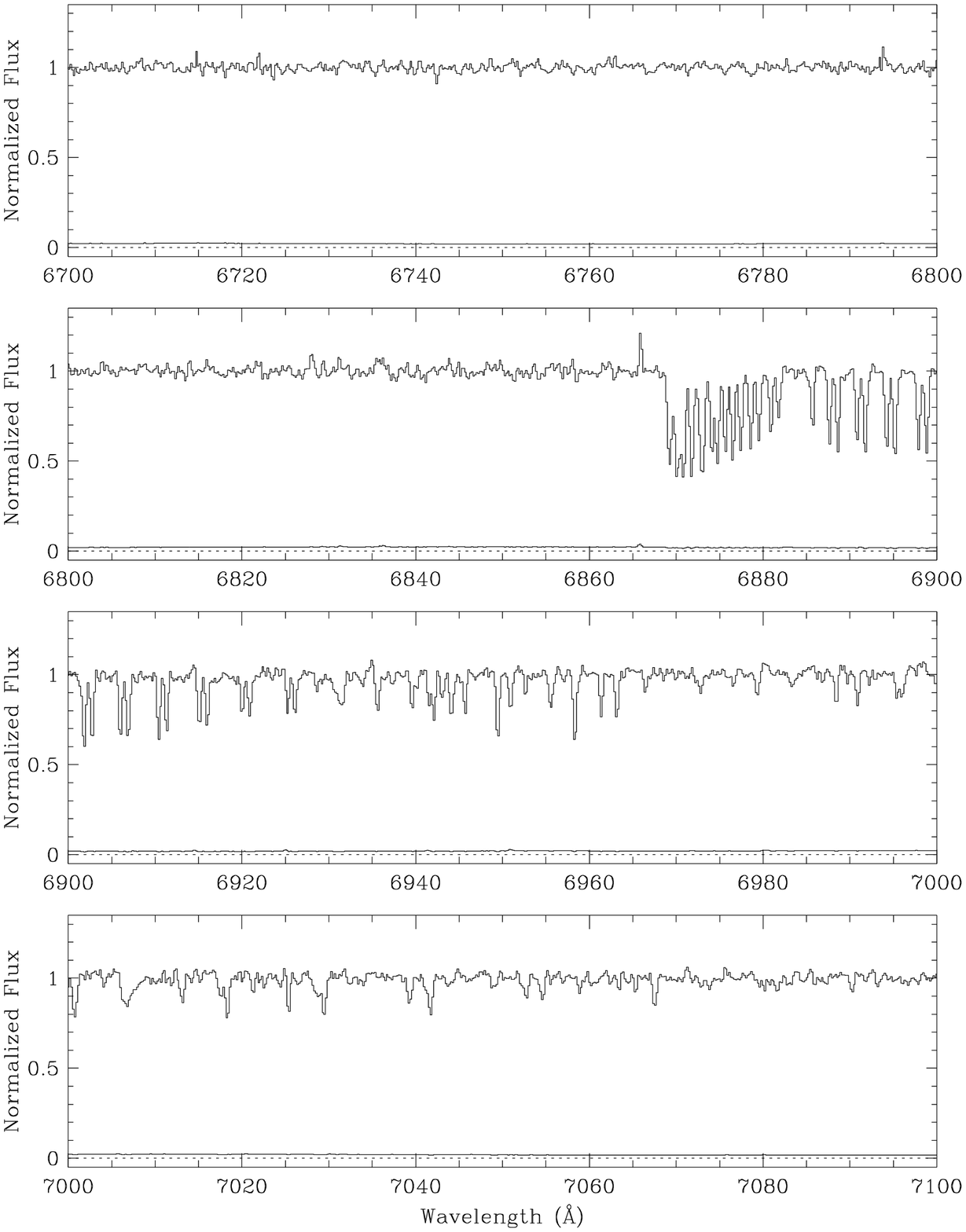}
\caption{continued}
\end{figure}
\begin{figure}
\figurenum{1}
\plotone{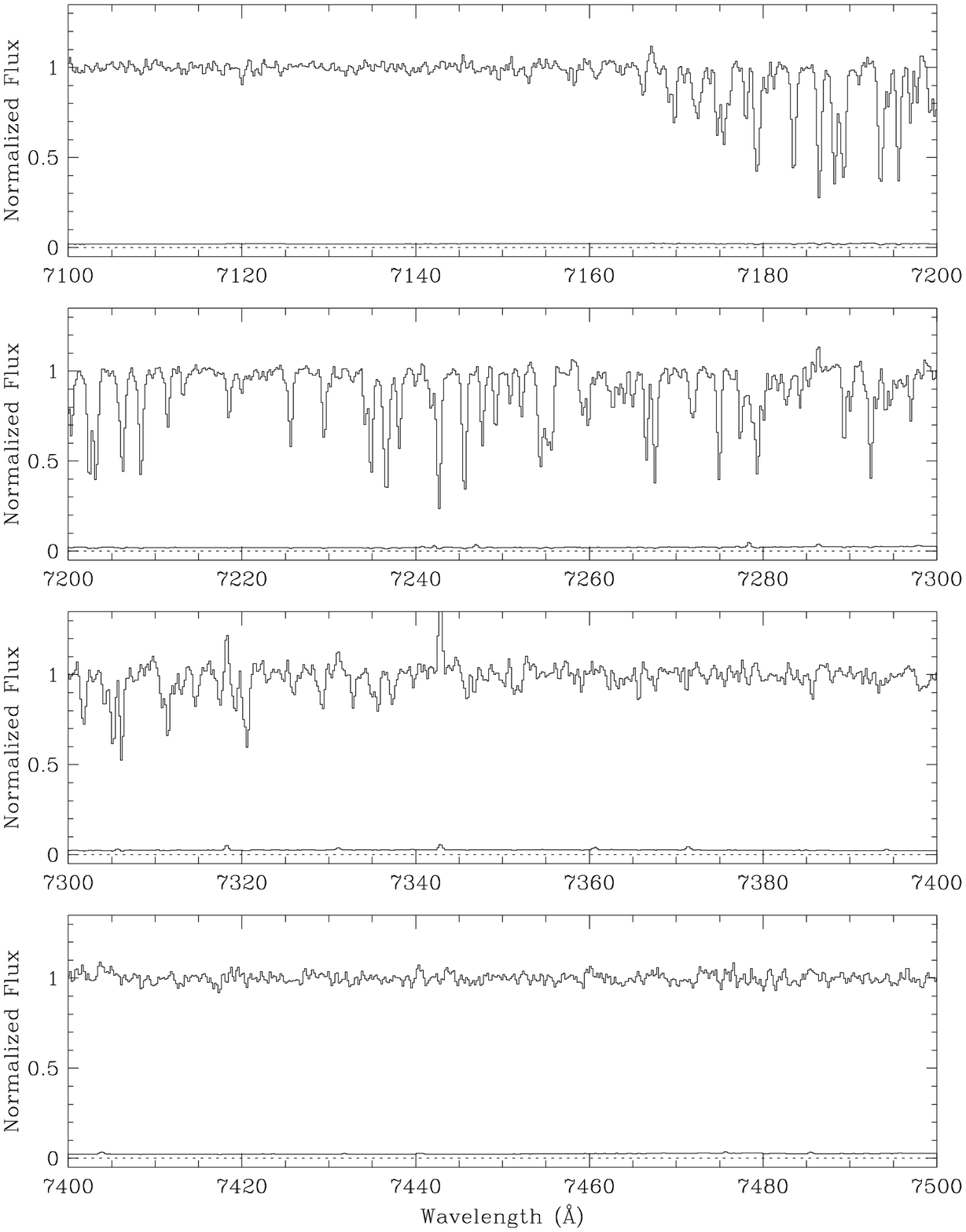}
\caption{continued}
\end{figure}
\begin{figure}
\figurenum{1}
\plotone{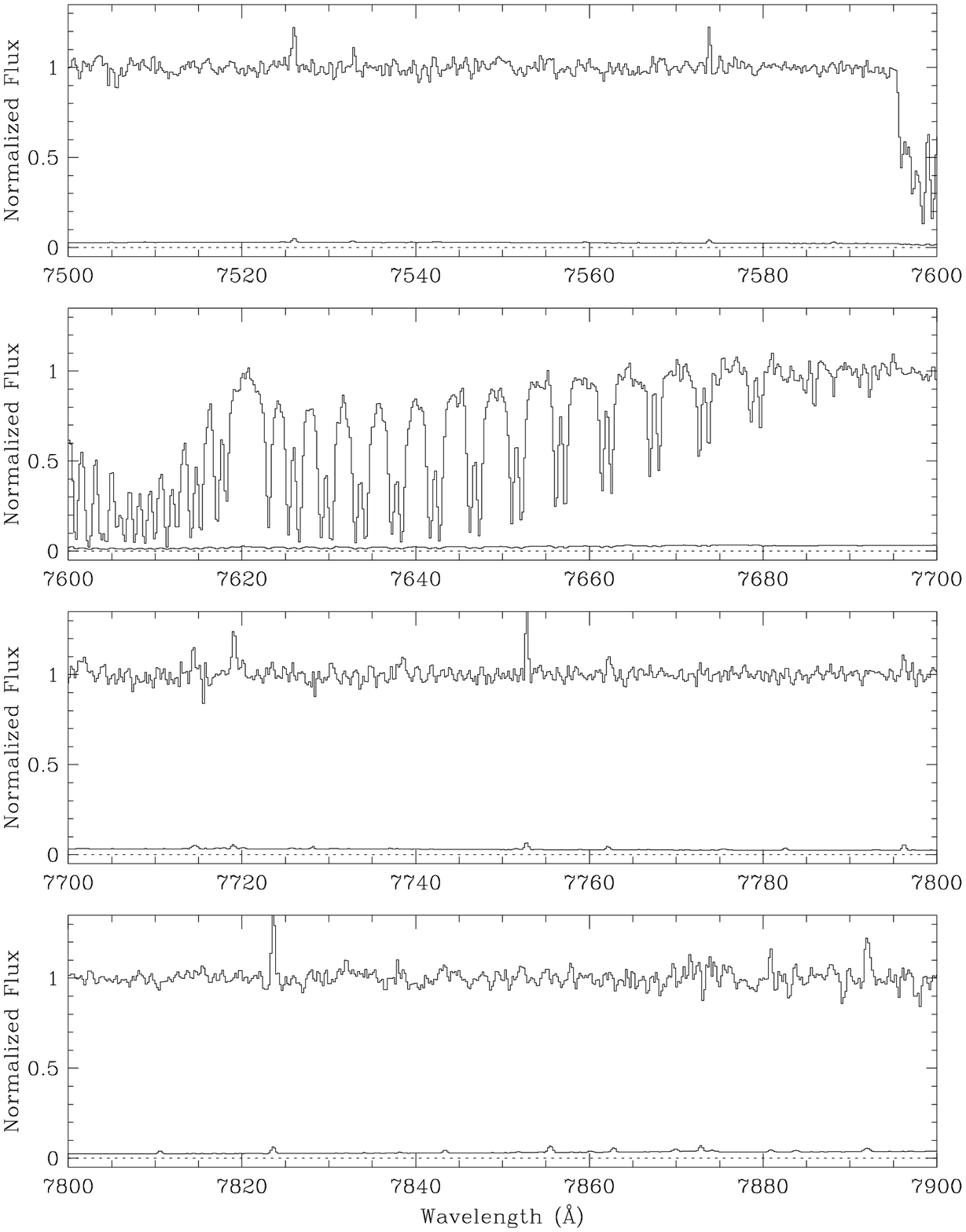}
\caption{continued}
\end{figure}
\begin{figure}
\figurenum{1}
\plotone{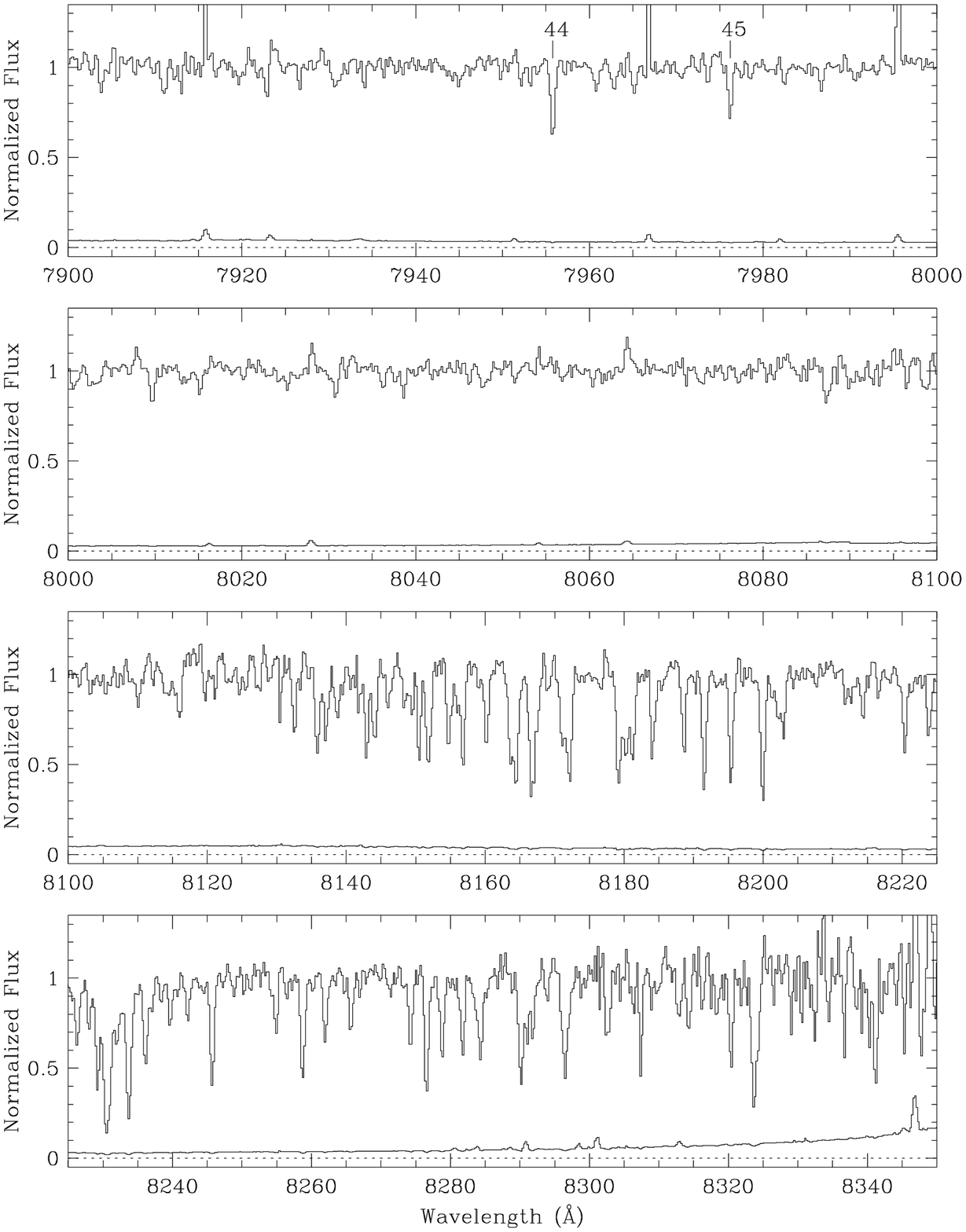}
\caption{continued}
\end{figure}

\section{Data Analysis}

\subsection{Absorption Line Selection and Identification}

Absorption lines in the HS 1700+6416 spectrum were selected for 
analysis based on the statistical significance of the observed 
equivalent width $W_{\lambda}$. Absorption line equivalent 
widths were measured using the automated method described in 
Tripp et al.\markcite{t2} (1996). Telluric absorption lines were 
removed from the line list by comparing the normalized spectrum of 
HS 1700+6416 to the normalized spectrum of HS 1946+7658 
obtained during the same observing run. Since the HS 1700+6416 
and HS 1946+7658 spectra have similar S/N levels and the same 
resolution, atmospheric absorption lines are effectively removed in a 
ratio of the two spectra (see Figure 3 in Tripp et al.\markcite{t2} 
1996). We have used this ratio of the HS 1700+6416 spectrum to the 
HS 1946+7658 spectrum to identify telluric lines {\it only}; for all 
other measurements reported in this paper we use the coadded 
spectrum shown in Figure 1 (which has not been divided by the HS 
1946+7658 spectrum).

Table 1 lists all statistically significant ($W_{\lambda } \geq \ 
4\sigma$) extragalactic absorption lines detected in the KPNO HS 
1700+6416 spectrum. Table 1 also lists the heliocentric vacuum 
wavelength of the centroid of each absorption line and a line number 
which identifies the absorption line in Figure 1. Identifications of the 
line species derived from the Morton\markcite{m8} (1991) finding 
list are tabulated in column 4 of Table 1, and Table 2 summarizes by 
redshift the heavy element systems we detect in the spectrum of this 
quasar. We have detected 13 heavy element absorbers in the 
direction of this quasar, 6 of which also produce detectable Lyman 
limits in the FOS spectrum of HS 1700+6416 obtained by Reimers 
et al.\markcite{r5} (1992). There are several absorption features in 
Table 1 for which we could not find positive identifications (labeled 
``UID'' in column 4 of Table 1). Most of these unidentified lines are 
weak ($< 5\sigma$), so they could be the \ion{C}{4} 1548.2 \AA\ 
line with the corresponding \ion{C}{4} 1550.8 \AA\ line falling 
below our detection threshold. Similarly, some of these UIDs could 
be the stronger line in the \ion{Mg}{2} doublet with the weaker line 
undetected.

\begin{deluxetable}{cccllclc}
\tablewidth{0pc}
\tablenum{1}
\tablecaption{Heavy Element Absorption Lines in the Optical Spectrum of HS 1700+6416\tablenotemark{a}}
\tablehead{
Line & $\lambda _{\rm vac}$ & $W_{\lambda}\pm \sigma _{W}$ & 
Identification & $z_{\rm abs}$ & log $N$ & \ \ \ \ $b$ & log $N_{\rm a}$ \nl
Number & (\AA ) & (\AA ) & \ & \ & (cm$^{-2}$) & (km s$^{-1}$) & (cm$^{-2}$) \nl
(1) & (2) & (3) & \ \ \ \ \ \ (4) & (5) & (6) & \ \ \ \ (7) & (8)
}
\startdata
1 & 4404.79 & 1.175$\pm$0.070 & \ion{C}{4} \ 1548.2\tablenotemark{b} & 1.84515 & 14.35$\pm$0.03 & 32.6$\pm$1.6 & $>$14.32 \nl 
2 & 4412.26 & 0.843$\pm$0.059 & \ion{C}{4} \ 1550.8\tablenotemark{b} & 1.84515 & 14.35$\pm$0.03 & 32.6$\pm$1.6 & 14.33$\pm$0.04 \nl 
3 & 4599.17 & 0.457$\pm$0.043 & \ion{N}{5} \ 1238.8 & 2.71248 & 13.86$\pm$0.02 & 25.2$\pm$1.3 & 13.78$\pm$0.03 \nl 
4 & 4613.85 & 0.253$\pm$0.028 & \ion{N}{5} \ 1242.8 & 2.71248 & 13.86$\pm$0.02 & 25.2$\pm$1.3 & 13.87$\pm$0.04 \nl 
5 & 4620.93 & 1.322$\pm$0.043 & \ion{Si}{4} \ 1393.8 & 2.31457 & 12.05$\pm$0.14 & 5.0$\pm$16.5 & 12.06$\pm$0.14 \nl 
 \ &    \    &       \         & \ \ \ & 2.31499 & 13.75$\pm$0.07 & 10.4$\pm$1.1 & 13.60$\pm$0.03 \nl 
 \ &    \    &       \         & \ \ \ & 2.31550 & 13.52$\pm$0.03 & 30.9$\pm$2.4 & 13.41$\pm$0.02 \nl 
 \ &    \    &       \         & \ \ \ & 2.31642 & 12.92$\pm$0.03 & 13.8$\pm$2.0 & 12.89$\pm$0.04 \nl
6 & 4631.32 & 0.066$\pm$0.014 & UID \ \ & \nodata & \nodata & \nodata & \nodata \nl 
7 & 4650.72 & 0.848$\pm$0.051 & \ion{Si}{4} \ 1402.8 & 2.31457 & 12.05$\pm$0.14 & 5.0$\pm$16.5 & \nodata \nl 
 \ &    \    &       \         & \ \ \ & 2.31499 & 13.75$\pm$0.07 & 10.4$\pm$1.1 & 13.68$\pm$0.03 \nl 
 \ &    \    &       \         & \ \ \ & 2.31550 & 13.52$\pm$0.03 & 30.9$\pm$2.4 & 13.40$\pm$0.04 \nl 
 \ &    \    &       \         & \ \ \ & 2.31642 & 12.92$\pm$0.03 & 13.8$\pm$2.0 & 12.89$\pm$0.10 \nl 
8 & 4679.89 & 0.057$\pm$0.013 & UID \ \ & \nodata & \nodata & \nodata & \nodata \nl 
9 & 4752.37 & 0.113$\pm$0.026 & UID \ \ & \nodata & \nodata & \nodata & \nodata \nl 
10 & 4785.28 & 0.303$\pm$0.052 & \ion{Si}{4} \ 1393.8 & 2.43339 & 13.05$\pm$0.04 & 32.7$\pm$4.6 & 13.05$\pm$0.08 \nl 
11 & 4815.90 & 0.370$\pm$0.046 & \ion{Mg}{2} \ 2796.4 & 0.72219 & 12.78$\pm$0.03 & 16.2$\pm$1.7 & 12.74$\pm$0.05 \nl 
12 & 4828.16 & 0.202$\pm$0.030 & \ion{Mg}{2} \ 2803.5 & 0.72219 & 12.78$\pm$0.03 & 16.2$\pm$1.7 & 12.80$\pm$0.08 \nl 
13 & 4842.45 & 0.094$\pm$0.020 & UID \ \ & \nodata & \nodata & \nodata & \nodata \nl 
14 & 4867.35 & 0.125$\pm$0.029 & UID \ \ & \nodata & \nodata & \nodata & \nodata \nl 
15 & 4904.41 & 0.702$\pm$0.055 & \ion{C}{4} \ 1548.2 & 2.16797 & 13.94$\pm$0.02 & 16.5$\pm$0.8 & 13.90$\pm$0.02 \nl 
16 & 4912.92 & 0.456$\pm$0.046 & \ion{C}{4} \ 1550.8 & 2.16797 & 13.94$\pm$0.02 & 16.5$\pm$0.8 & 13.94$\pm$0.04 \nl 
17 & 4952.66 & 0.162$\pm$0.024 & UID \ \ & \nodata & \nodata & \nodata & \nodata \nl 
18 & 5061.03 & 0.087$\pm$0.023 & \ion{Si}{2} \ 1526.7 & 2.31499 & 13.08$\pm$0.07* & 9.4$\pm$5.2 & 13.07$\pm$0.12 \nl 
19 & 5061.87 & 0.052$\pm$0.014 & \ion{Si}{2} \ 1526.7 & 2.31553 & \nodata & \nodata & 12.86$\pm$0.20 \nl 
20 & 5105.72 & 0.098$\pm$0.025 & UID \ \ & \nodata & \nodata & \nodata & \nodata\tablebreak \nl 
21 & 5121.89 & 0.333$\pm$0.036 & \ion{C}{4} \ 1548.2\tablenotemark{d} & 2.30812 & 13.38$\pm$0.03 & 25.0$\pm$2.2 & \nodata        \nl 
 \ &    \    &       \         & UID \ \ & \nodata & \nodata & \nodata & \nodata \nl 
22 & 5130.23 & 0.181$\pm$0.023 & \ion{C}{4} \ 1550.8\tablenotemark{d} & 2.30812 & 13.38$\pm$0.03 & 25.0$\pm$2.2 & \nodata        \nl 
23 & 5133.14 & 2.672$\pm$0.031 & \ion{C}{4} \ 1548.2 & 2.31458 & 13.19$\pm$0.11 & 14.6$\pm$5.4 \nl 
 \ &    \    &       \         & \ \ \ & 2.31505 & 15.48$\pm$1.1 & 11.3$\pm$3.8 \nl 
 \ &    \    &       \         & \ \ \ & 2.31568 & 14.57$\pm$0.02 & 28.6$\pm$1.5 \nl 
 \ &    \    &       \         & \ \ \ & 2.31638 & 16.72$\pm$0.14 & 3.9$\pm$0.2 & \raisebox{4.5ex}[0pt]{\Bigg\} $>$14.74\tablenotemark{c}}\nl 
24 & 5141.71 & 2.477$\pm$0.039 & \ion{C}{4} \ 1550.2 & 2.31458 & 13.19$\pm$0.11 & 14.6$\pm$5.4 \nl 
 \ &    \    &       \         & \ \ \ & 2.31505 & 15.48$\pm$1.1 & 11.3$\pm$3.8 \nl 
 \ &    \    &       \         & \ \ \ & 2.31568 & 14.57$\pm$0.02 & 28.6$\pm$1.5\nl 
 \ &    \    &       \         & \ \ \ & 2.31638 & 16.72$\pm$0.14 & 3.9$\pm$0.2 & \raisebox{4.5ex}[0pt]{\Bigg\} $>$14.93\tablenotemark{c}} \nl 
25 & 5213.84 & 0.177$\pm$0.027 & \ion{Mg}{2} \ 2796.4\tablenotemark{e} & 0.86431 & 12.00$\pm$0.05 & 6.0$\pm$4.8 & \nodata \nl 
 \ &    \    &       \         & UID \ \ & \nodata & \nodata & \nodata & \nodata \nl 
26 & 5232.72 & 0.156$\pm$0.030 & \ion{C}{4} \ 1548.2 & 2.37995 & 13.13$\pm$0.04 & 6.3$\pm$2.0 & 13.07$\pm$0.06 \nl 
27 & 5241.48 & 0.075$\pm$0.014 & \ion{C}{4} \ 1550.8 & 2.37995 & 13.13$\pm$0.04 & 6.3$\pm$2.0 & 13.03$\pm$0.15 \nl 
28 & 5313.49 & 0.060$\pm$0.019 & \ion{C}{4} \ 1548.2 & 2.43204 & 12.70$\pm$0.07 & 8.0$\pm$5.5 & 12.66$\pm$0.17 \nl 
29 & 5315.29 & 0.332$\pm$0.036 & \ion{C}{4} \ 1548.2 & 2.43306 & 13.16$\pm$0.19 & 4.0$\pm$3.5 \nl 
 \ &    \    &       \         & \ \ \ & 2.43329 & 13.29$\pm$0.05 & 44.5$\pm$4.9 & \raisebox{1.5ex}[0pt]{\Big\} 13.44$\pm$0.05\tablenotemark{c}} \nl 
30 & 5323.70 & 0.578$\pm$0.056 & \ion{C}{4} \ 1548.2\tablenotemark{f} & 2.43785 & 12.80$\pm$0.06 & 12.6$\pm$4.3 & \nodata        \nl 
 \ &    \    &       \         & \ \ \ & 2.43877 & 13.30$\pm$0.03 & 18.3$\pm$1.9 & \nodata        \nl 
 \ &    \    &       \         & \ion{C}{4} \ 1550.8 & 2.43204 & 12.70$\pm$0.07 & 8.0$\pm$5.5 & \nodata        \nl 
 \ &    \    &       \         & \ \ \ & 2.43306 & 13.16$\pm$0.19 & 4.0$\pm$3.5 & \nodata        \nl 
 \ &    \    &       \         & \ \ \ & 2.43329 & 13.29$\pm$0.05 & 44.5$\pm$4.9 & \nodata        \nl 
31 & 5326.70 & 0.233$\pm$0.025 & \ion{C}{4} \ 1548.2 & 2.44059 & 13.30$\pm$0.02 & 20.3$\pm$1.7 & 13.28$\pm$0.05 \nl 
32 & 5332.45 & 0.167$\pm$0.032 & \ion{C}{4} \ 1550.8 & 2.43785 & 12.80$\pm$0.06 & 12.6$\pm$4.3 & 12.70$^{+0.20}_{-0.37}$ \nl 
 \ &    \    &       \         & \ \ \ & 2.43877 & 13.30$\pm$0.03 & 18.3$\pm$1.9 & 13.35$\pm$0.10 \nl 
33 & 5335.48 & 0.136$\pm$0.028 & \ion{C}{4} \ 1550.8 & 2.44059 & 13.30$\pm$0.02 & 20.3$\pm$1.7 & 13.28$\pm$0.10 \nl 
34 & 5524.54 & 0.105$\pm$0.018 & UID \ \ & \nodata & \nodata & \nodata & \nodata \nl 
35 & 5538.74 & 0.060$\pm$0.019 & \ion{Al}{2} \ 1670.8 & 2.31498 & 11.71$\pm$0.06* & 7.9$\pm$4.6 & 11.69$\pm$0.11 \nl 
36 & 5540.05 & 0.552$\pm$0.035 & \ion{C}{4} \ 1548.2 & 2.57813 & 13.34$\pm$0.03 & 18.9$\pm$1.9 \nl 
 \ &    \    &       \         & \ \ \ & 2.57884 & 13.36$\pm$0.03 & 35.0$\pm$3.8 & \raisebox{1.5ex}[0pt]{\Big\} 16.64$\pm$0.03\tablenotemark{c}} \nl 
37 & 5549.41 & 0.311$\pm$0.043 & \ion{C}{4} \ 1550.8 & 2.57813 & 13.34$\pm$0.03 & 18.9$\pm$1.9 \nl 
 \ &    \    &       \         & \ \ \ & 2.57884 & 13.36$\pm$0.03 & 35.0$\pm$3.8 & \raisebox{1.5ex}[0pt]{\Big\} 16.63$\pm$0.06\tablenotemark{c}} \tablebreak \nl
38 & 5747.49 & 0.218$\pm$0.027 & \ion{C}{4} \ 1548.2 & 2.71245 & 13.19$\pm$0.02 & 11.4$\pm$1.1 & 13.18$\pm$0.04 \nl 
39 & 5757.38 & 0.120$\pm$0.021 & \ion{C}{4} \ 1550.8 & 2.71245 & 13.19$\pm$0.02 & 11.4$\pm$1.1 & 13.16$\pm$0.08 \nl 
40 & 6032.45 & 0.330$\pm$0.027 & \ion{Mg}{2} \ 2796.4 & 1.15729 & 12.80$\pm$0.03 & 6.7$\pm$0.6 & 12.65$\pm$0.03 \nl 
41 & 6047.98 & 0.192$\pm$0.014 & \ion{Mg}{2} \ 2803.5 & 1.15729 & 12.80$\pm$0.03 & 6.7$\pm$0.6 & 12.67$\pm$0.05 \nl 
42 & 6148.36 & 0.069$\pm$0.012 & \ion{Al}{3} \ 1854.7 & 2.31499 & 12.20$\pm$0.18* & 3.2$\pm$5.8 & 12.11$\pm$0.09 \nl 
43 & 6149.34 & 0.045$\pm$0.010 & \ion{Al}{3} \ 1854.7 & 2.31551 & 11.95$\pm$0.18* & 2.8$\pm$9.0 & 11.87$\pm$0.18 \nl 
44 & 7955.79 & 0.222$\pm$0.019 & \ion{Mg}{2} \ 2796.4 & 1.84506 & 12.65$\pm$0.10 & 3.5$\pm$0.7 & 12.31$\pm$0.06 \nl 
45 & 7976.21 & 0.142$\pm$0.015 & \ion{Mg}{2} \ 2803.5 & 1.84506 & 12.65$\pm$0.10 & 3.5$\pm$0.7 & 12.43$\pm$0.08 \nl 
\enddata
\tablenotetext{a}{NOTES: The line number in column 1 identifies
the absorption feature in Figure 1. Column 2 lists the vacuum heliocentric
wavelength of the absorption feature, and column 3 gives the observed equivalent
width. Column 4 identifies the species and lists the vacuum rest wavelength of
the line; in this column ``UID'' indicates that the line is unidentified.
Columns 5, 6, and 7 give the absorption redshift, logarithmic column density,
and Doppler parameter, respectively, determined from profile fitting,
 and column 8 lists the logarithm of the integrated apparent column
density (see \S 3.2). Most of the column densities and Doppler parameters from
profile fitting are based on the simultaneous fitting of more than one
absorption line for the ion (e.g., simultaneous fitting of \ion{C}{4} 1548.2 and
\ion{C}{4} 1550.8 \AA ). Profile fitting column densities which are based on 
fitting of a {\it single} line are marked with an asterisk in column 6.}
\tablenotetext{b}{These metal lines are found in the Ly$\alpha$ forest and thus
could be contaminated by \ion{H}{1} absorption from different redshifts (see 
\S 4.5).}
\tablenotetext{c}{The various components of this absorption feature are 
blended with adjacent components, and column (8) lists the total column density
integrated across all components.}
\tablenotetext{d}{These lines are detected at high significance levels, but
the line identifications are tentative (see \S 4.7), and due to significant
blending with unidentified absorption, integrated column densities are not
reliable.}
\tablenotetext{e}{This absorption feature is confused by blending with 
unidentified absorption, and the \ion{Mg}{2} identification is not secure (see
\S 4.2).}
\tablenotetext{f}{Due to very strong blending of the \ion{C}{4} 1548.2 and 
1550.8 \AA\ lines in this absorption feature (see Figure 4), the integrated
apparent column densities are not useful (the apparent optical depth of the
1548.2 \AA\ line cannot be distinguished from the $\tau_{\rm a}(v)$ of the
1550.8 \AA\ line).}
\end{deluxetable}
\clearpage

\subsection{Column Densities}

Column densities of identified absorption lines were measured using 
two techniques, the apparent column density method and profile 
fitting. Once again, these measurements are discussed in some detail 
in Tripp et al.\markcite{t2} (1996).

In the apparent column density method, the ``apparent'' (i.e., 
observed) optical depth per unit velocity, $\tau _{\rm a}(v)$, is used 
to calculate the apparent column density as a function of velocity 
according to the following relation:
\begin{equation}
N_{a}(v)  = \frac{m_{e}c}{\pi e^{2}} \left( \frac{\tau 
_{a}(v)}{f\lambda}\right) = 3.768\times 10^{14}  \left( \frac{\tau 
_{a}(v)}{f\lambda}\right) \ \ \ ({\rm atoms \ cm^{-2} \ (km \ s^{-
1})^{-1}}),
\end{equation}
where $f$ is the oscillator strength and $\lambda$ is the wavelength 
of the transition (in \AA ), and the other symbols have their usual 
meanings. The apparent optical depth is determined from the 
observed normalized absorption line profile, $I(v)$, according to the 
usual relation, $\tau _{\rm a}(v)$ = ln[1/$I(v)$]. If the line profile is 
fully resolved, then \nav\ provides a good estimate of the true 
column density per unit velocity. Even if the profile is not 
well-resolved, \nav\ still provides a good representation of the true 
column density, broadened by the spectral spread function, {\it if} 
the profile is not affected by unresolved saturation. However, if the 
profile contains unresolved saturated absorption, then \nav\ will be 
an underestimate of the true column density. If two or more 
resonance lines of a species differing in the quantity $f\lambda$ are 
available, then comparison of the $N_{\rm a}(v)$ profiles of the 
different lines can be used to search for unresolved saturation. If the 
absorption profile contains unresolved saturated components, then 
the \nav\ of stronger lines will be less than the \nav\ of weaker lines 
at velocities where the saturation occurs. If there is no unresolved 
saturation, the resonance lines will yield the same \nav . Finally, if 
the absorption profile is not affected by unresolved saturated 
components, then $N_{\rm a}(v)$ can be integrated to obtain a good 
measurement of the total column density, $N$(total) = $N_{\rm a}$ 
= $\int N_{\rm a}(v) dv$. The error in \nav\ due to uncertainty of 
the continuum placement is estimated as described in Tripp et 
al.\markcite{t2} (1996), and this error is added in quadrature to the 
error from photon counting and read noise to get the total 
uncertainty in \nav. Integrated apparent column densities of all 
metals detected in the HS 1700+6416 spectrum are given in column 
8 of Table 1. For further information on the apparent column density 
method, see Savage \& Sembach\markcite{s3} (1991), 
Jenkins\markcite{j4} (1996), and references therein. For most of the 
lines in Table 1 we use oscillator strengths and wavelengths from 
Morton\markcite{m8} (1991). However, for some transitions we use 
the oscillator strength revisions summarized in Tripp et 
al.\markcite{t2} (1996).

We have also used the profile fitting routine described by Lanzetta 
\& Bowen\markcite{l1} (1992) to estimate column densities by 
fitting Voigt profiles to the HS 1700+6416 absorption lines, and we 
refer the reader to their paper for a description of the minimized 
$\chi ^{2}$ profile fitting procedure. For the purposes of this paper, 
we have fitted different lines from the same ion (e.g., \ion{C}{4} 
1548.2 \AA\ and \ion{C}{4} 1550.8 \AA ) simultaneously, but we 
have fitted different elements and different ionization stages 
independently. The column densities, Doppler parameters, and 
redshifts obtained by fitting Voigt profiles in this fashion are listed 
in columns 5, 6, and 7 of Table 1. Voigt profile fitting corrects for 
saturated absorption to some degree,\footnote{However, column 
densities derived from profile fitting can have {\it large} 
uncertainties if the line is significantly saturated. These uncertainties 
may be larger than the errors formally determined in the profile 
fitting process because the true profile may be more complex than 
the assumed profile.} but direct integration of \nav\ does not correct 
for saturation effects unless special efforts are made (Savage \& 
Sembach\markcite{s3} 1991; Jenkins\markcite{j4} 1996). Therefore 
for a specific line, comparison of the column density derived from 
profile fitting to the column density given by direct integration of 
\nav\ can also be used to assess the impact of unresolved saturated 
absorption. If the column density from profile fitting is significantly 
larger than the integrated apparent column density, then the 
absorption line may be affected by unresolved saturation.

\section{Comments on Individual Systems}

In the HS 1700+6416 discovery paper, Reimers et al.\markcite{r2} 
(1989) identified four metal absorption systems at $z_{\rm abs}$ = 
2.308, 2.315, 2.433, and 2.440 on the basis of \ion{C}{4} lines 
detected in a high resolution but modest S/N optical spectrum. 
Subsequently, Reimers et al.\markcite{r5} (1992) reobserved the 
quasar with the {\it HST} FOS and detected a wide range of 
ionization stages of many heavy elements at these redshifts. They 
also reported detections of metals at $z_{\rm abs}$ = 0.8642, 
1.1572, 1.725, 1.8465, 2.1678, and 2.579. Reimers et 
al.\markcite{r5} (1992) detected \ion{H}{1} Lyman continuum 
absorption in seven of these systems, and all of the Lyman limits 
have moderate optical depth ($\tau _{\rm LL} \ \lesssim$ 1). This is 
important because with detection of flux shortward of the Lyman 
limit edge, the Lyman continuum optical depth 
$\tau_{LL}(\lambda)$ at wavelength $\lambda \ \leq$ 912 \AA\ can 
be reliably measured. Since the cross section for \ion{H}{1} 
continuum absorption is well known, it is then straightforward to 
estimate the \ion{H}{1} column density, 
\begin{equation}
N({\sc H \ i}) = 1.6 \times 10^{17} \tau_{LL}(\lambda) \left[ 
\frac{912}{\lambda}\right] ^{3}.
\end{equation}
Because of the low resolution of the FOS spectrum, this provides a 
measurement of the {\it total} \ion{H}{1} column density in the 
Lyman limit system and no information about the kinematics of the 
absorber.

More recently, HS 1700+6416 has been observed from the ground 
by Sanz et al.\markcite{s1} (1993), Rodr\'{i}guez-Pascual et 
al.\markcite{r7} (1995), and Petitjean, Riediger, \& 
Rauch\markcite{p4} (1996). Sanz et al.\markcite{s1} (1993) and 
Rodr\'{i}guez-Pascual et al.\markcite{r7} (1995) have concentrated 
their efforts on the \lya forest; they obtained high resolution spectra 
of the forest and low resolution spectra longward of the \lya 
emission line. Petitjean et al.\markcite{p4} (1995) have observed the 
quasar longward of \lya emission but with moderate resolution 
(FWHM $\sim$ 95 \kms ).

\begin{deluxetable}{clcc}
\tablewidth{0pc}
\tablenum{2}
\tablecaption{Summary of Heavy Element Systems}
\tablehead{
\colhead{$z_{{\rm abs}}$} & \colhead{Species detected} & \colhead{HST LL\tablenotemark{a}} & \colhead{log $N$(\ion{H}{1})\tablenotemark{b}} \nl
}
\startdata
0.72219  & \ion{Mg}{2} 2796.4,2803.5 & \ & \nodata \nl
0.86431  & \ion{Mg}{2} 2796.4,2803.5 & $\surd$ & 16.35 \nl
1.15729  & \ion{Mg}{2} 2796.4,2803.5 & $\surd$ & 16.85 \nl
1.725    & \nodata                   & $\surd$ & 17.05 \nl
1.84506  & \ion{Mg}{2} 2796.4,2803.5 & $\surd$ & 16.75 \nl
  \      & \ion{C}{4} 1548.2,1550.8\tablenotemark{c} & \        & \nl
2.16797  & \ion{C}{4} 1548.2,1550.8 & $\surd$ & 16.85 \nl
2.30812  & \ion{C}{4} 1548.2,1550.8 & \ & \nodata \nl
2.315,2.316 & \ion{Si}{2} 1526.7 & $\surd$ & 16.85 \nl
  \     & \ion{Al}{2} 1670.8 & \ & \nodata \nl
  \     & \ion{Al}{3} 1854.7,1862.8 & \ & \nodata \nl
  \     & \ion{Si}{4} 1393.8,1402.8 & \ & \nodata \nl
  \     & \ion{C}{4} 1548.2,1550.8 & \ & \nodata \nl
2.37995  & \ion{C}{4} 1548.2,1550.8 & \ & \nodata \nl
2.433\tablenotemark{d}   & \ion{Si}{4} 1393.8 & $\surd$ & 16.7 \nl
  \     & \ion{C}{4} 1548.2,1550.8 & \ & \nodata \nl
2.439\tablenotemark{d}   & \ion{C}{4} 1548.2,1550.8 & \ & \nodata \nl
2.44059  & \ion{C}{4} 1548.2,1550.8 & \ & \nodata \nl
2.57813,2.57884   & \ion{C}{4} 1548.2,1550.8 & \ & \nodata \nl
2.7125\tablenotemark{e}  & \ion{C}{4} 1548.2,1550.8 & \ & \nodata \nl
  \     & \ion{N}{5} 1238.8,1242.8 & \ & \nodata \nl
\enddata
\tablenotetext{a}{Absorption systems with associated Lyman continuum absorption in the {\it HST} spectrum of Reimers et al. (1992) are indicated with a check mark in this column.}
\tablenotetext{b}{Neutral hydrogen column density derived by Reimers et al. (1992) from the Lyman continuum optical depth measured in the {\it HST} FOS spectrum. Reimers et al. do not provide the uncertainties of these $N$(\ion{H}{1}) measurements.}
\tablenotetext{c}{These \ion{C}{4} lines occur in the Ly$\alpha$ forest and could be contaminated by \ion{H}{1} from different redshifts.}
\tablenotetext{d}{At least four \ion{C}{4} doublets are detected in a complex
absorption cluster at 2.432 $< \ z_{\rm abs} \ <$ 2.441 (see text \S4.10).}
\tablenotetext{e}{Associated absorption system ($z_{\rm abs} \approx z_{\rm em}$).}
\end{deluxetable}

Table 2 summarizes the heavy element systems we have identified 
in the KPNO optical spectrum shown in Figure 1. Table 2 also lists 
the \ion{H}{1} column densities determined by Reimers et 
al.\markcite{r5} (1992) from $\tau _{LL}$ as discussed above. We 
detect 13 metal systems including new systems at $z_{\rm abs}$ = 
2.37995 and 2.44059. In this section we comment on these metal 
systems including assessments of absorption saturation effects on 
the column density measurements. Our KPNO spectrum covers the 
portion of the \lya forest between $\lambda \approx$ 4300 \AA\ and 
$\lambda$ = 4520 \AA , so in principle we can measure the 
\ion{H}{1} column densities of the metal systems at $z_{\rm abs}$ 
= 2.57813, 2.57884 and \zabs\ = 2.7125 from the \lya absorption 
profile. However, the \ion{H}{1} lines at 2.57813 and 2.57884 are 
badly blended and saturated, so the values of $N$(\ion{H}{1}) are 
not usefully constrained by our spectrum. The highest redshift metal 
absorber is found at \zabs\ = 2.7125, but there are at least five 
additional \ion{H}{1} \dla\ lines at 2.7125 $<$ \zabs\ $\leq$ 
2.74427 which are not detected in any metal lines despite high S/N 
in spectral regions where these metal lines would be observed.

\subsection{$\bf z_{\bf abs}$ = 0.72219}

Reimers et al.\markcite{r5} (1992) did not recognize this metal 
system in their initial analysis of the {\it HST} FOS spectrum, but it 
was identified in the reanalysis by Vogel \& Reimers\markcite{v3} 
(1995). Confirmation of this system was provided by Petitjean et 
al.\markcite{p4} (1996) who detected the \ion{Mg}{2} lines at 
2796.4 and 2803.5 \AA . We also detect this \ion{Mg}{2} doublet 
(lines 11 and 12 in Figure 1), and comparison of the apparent 
column densities of the \ion{Mg}{2} lines indicates that there is 
some unresolved saturation in the core of the \ion{Mg}{2} 2796.4 
\AA\ profile. The \ion{Mg}{2} 2796.4 \AA\ line (the stronger 
member of the doublet) is mildly blended with \ion{Si}{4} 1402.8 
\AA\ absorption from the \zabs\ = 2.433 system; this blending could 
mask some saturation in the \ion{Mg}{2} profile because the excess 
optical depth from the \ion{Si}{4} line artificially boosts the 
\ion{Mg}{2} 2796.4 \AA\ \nav . However, the weaker \ion{Mg}{2} 
2803.5 \AA\ line yields an integrated column density in good 
agreement with the column density from profile fitting (see Table 1), 
so this magnesium doublet is only weakly saturated. Reimers et 
al.\markcite{r5} (1992) do not report detection of Lyman continuum 
absorption at this redshift, but the Lyman limit occurs in a region of 
the FOS spectrum which has very low S/N.

\subsection{$\bf z_{\bf abs}$ = 0.86431}

This system produces partial Lyman continuum absorption in the {\it 
HST} FOS spectrum, and from the Lyman limit optical depth 
Reimers et al. (1992) estimate that log $N$(\ion{H}{1}) = 16.35 
[unfortunately, Reimers et al.\markcite{r5} (1992) do not provide 
the uncertainties in their measurements of $N$(\ion{H}{1})]. We 
detect a 6.6$\sigma$ absorption feature (line 25 in Fig. 1) at the 
expected wavelength of \ion{Mg}{2} 2796.4 \AA\ at this redshift. 
We consider this to be a probable, but not definite, identification. 
We detect a 2.3$\sigma$ absorption line at the expected wavelength 
(5226.65\AA ) of \ion{Mg}{2} 2803.5 \AA , but this weak line does 
not line up well with the stronger line: the stronger line is best-fit 
with \zabs\ = 0.86452 while the weaker line indicates that \zabs\ = 
0.86431. However, the stronger line shows evidence of two 
components, and one of the components is reasonably fit with \zabs\ 
= 0.86431. On this basis, we tentatively conclude that the stronger 
\ion{Mg}{2} line is blended with an unidentified absorption feature. 
The \nav\ profiles of these \ion{Mg}{2} lines are rather noisy so it is 
difficult to assess the impact of unresolved saturation, but these lines 
are weak and unlikely to be saturated.

\subsection{$\bf z_{\bf abs}$ = 1.15729}

This metal system produces obvious Lyman continuum absorption 
in the {\it HST} spectrum, and Reimers et al.\markcite{r5} (1992) 
find that log $N$(\ion{H}{1}) = 16.85. The \ion{Mg}{2} 2796.4 
and 2803.5 \AA\ lines at this redshift are recorded at high 
significance levels in the KPNO spectrum (see lines 40 and 41 in 
Figure 1 and Table 1). The integrated apparent column densities of 
these \ion{Mg}{2} lines are in agreement, but inspection of the 
\nav\ profiles suggests that they are undersampled, and the higher 
column density from profile fitting indicates that the lines may be 
moderately affected by unresolved saturation.

\subsection{$\bf z_{\bf abs}$ = 1.725}

No metals are unambiguously detected in this absorption system, 
which is identified on the basis of obvious Lyman limit absorption 
in the {\it HST} spectrum with log $N$(\ion{H}{1}) = 17.05 
(Reimers et al.\markcite{r5} 1992). The redshift is not favorable for 
clear detection of many metal lines from the ground; only the 
\ion{Al}{2}, \ion{Al}{3}, \ion{Fe}{2}, and \ion{Mg}{2} 
resonance lines are sufficiently redshifted to escape from the \lya 
forest. Unfortunately, the \ion{Mg}{2} 2796.4, 2803.5 \AA\ doublet 
at this redshift falls within the O$_{2}$ atmospheric A band (see 
Figure 1), but after removal of the A band using the HS 1946+7658 
spectrum as a template (see \S 3.1), we still do not see any 
convincing absorption which can be attributed to \ion{Mg}{2}. The 
strongest \ion{Fe}{2} line at 2382.8 \AA\ is redshifted into a cleaner 
region of the optical spectrum; the weak telluric lines near the 
expected wavelength of this Fe line are effectively removed after 
division by the template, and the strong \ion{Fe}{2} line is not 
evident after removal of the atmospheric lines. Petitjean et 
al.\markcite{p4} (1996) point out that the strong and broad line at 
4551.8 \AA\ could be \ion{Al}{2} 1670.8 \AA\ at \zabs\ = 1.7243, 
but this line occurs at the edge of the \lya forest and Petitjean et 
al.\markcite{p4} acknowledge that it could be \ion{H}{1} \lya at 
\zabs\ = 2.744 instead. We strongly prefer the \ion{H}{1} 
identification for this line because such a strong and broad 
\ion{Al}{2} line\footnote{If the line at 4551.8 \AA\ is indeed Al II 
1670.8 \AA , then from profile fitting with one component we find 
log $N$(Al II) = 13.1 with $b$ = 41.9 \kms .} should be 
accompanied by \ion{Mg}{2} absorption which is easily detected 
(even within the telluric A band), but strong \ion{Mg}{2} 
absorption is not seen. We cannot confidently place upper limits on 
the column densities of metals we do not detect in this system 
because the redshift is not adequately constrained by the {\it HST} 
spectrum (see \S 7). The \ion{Mg}{2} 2796.4 and 2803.5 \AA\ lines 
are among the strongest resonance lines typically detected in neutral 
interstellar gas in the Milky Way (see, for example, \S 3.4 in 
Lockman \& Savage\markcite{l3} 1995), so the absence of these 
lines in absorption suggests that this Lyman limit system is highly 
ionized.

\subsection{$\bf z_{\bf abs}$ = 1.84506}

Reimers et al.\markcite{r5} (1992) and Vogel \& 
Reimers\markcite{v3} (1995) identify a Lyman limit system at 
\zabs\ = 1.8465 in the {\it HST} FOS spectrum of HS 1700+6416, 
and from the measured $\tau _{\rm LL}$ they estimate that log 
$N$(\ion{H}{1}) = 16.75. We do not detect any metals at \zabs\ = 
1.8465, but we do detect the \ion{Mg}{2} 2796.4, 2803.5 \AA\ 
doublet at \zabs\ = 1.84506 (lines 44 and 45 in Figure 1) at high 
significance levels. Our KPNO spectrum covers the \ion{C}{4} 
doublet at this redshift, and two strong absorption features are 
present at the expected wavelengths of the 1548.2 and 1550.8 \AA\ 
transitions (see lines 1 and 2 in Figure 1). These \ion{C}{4} lines 
occur within the \lya forest and thus are prone to confusion due to 
blending with \ion{H}{1} absorption from different redshifts; 
however, the velocity extents, centroids, and even the \nav\ profiles 
derived from lines 1 and 2 are in reasonable agreement, and we 
believe these absorption features are primarily due to \ion{C}{4} at 
this redshift.

Comparison of the \ion{Mg}{2} \nav\ profiles indicates that these 
magnesium lines are significantly affected by unresolved saturated 
absorption. Similarly, comparison of the \ion{C}{4} \nav\ profiles 
suggests that the \ion{C}{4} column density may be slightly 
underestimated due to unresolved saturation, but it is important to 
bear in mind that the \ion{C}{4} column might also be 
overestimated due to contamination by \lya forest lines.

\subsection{$\bf z_{\bf abs}$ = 2.16797}

The \ion{C}{4} 1548.2 and 1550.8 \AA\ lines are clearly detected at 
this redshift (lines 15 and 16 in Figure 1), and Reimers et 
al.\markcite{r5} (1992) report Lyman limit absorption as well with 
log $N$(\ion{H}{1}) = 16.85. Inspection of the \nav\ profiles 
indicates that these \ion{C}{4} lines are not significantly affected by 
saturation, and this conclusion is supported by the good agreement 
of the column densities from profile fitting and \nav\ integration (see 
Table 1). Therefore $N$(\ion{C}{4})/$N$(\ion{H}{1}) = 1.23 
$\times \ 10^{-3}$. This column density ratio is similar to the ratio 
measured in lower $N$(\ion{H}{1}) \lya clouds by Cowie et 
al.\markcite{c1} (1995). However, we report this ratio with some 
trepidation because the {\it HST} Lyman limit provides an 
estimation of the {\it total} \ion{H}{1} column density, and while 
some of this \ion{H}{1} absorption is likely to be associated with 
the observed \ion{C}{4} lines, there may be additional \ion{H}{1} 
gas which does not produce \ion{C}{4} absorption.

\subsection{$\bf z_{\bf abs}$ = 2.30812}

Reimers et al.\markcite{r5} (1992) identified this system based on 
detection of Lyman series lines as well as various metal lines. At 
first glance, lines 21 and 22 in Figure 1 appear to be at the expected 
wavelengths of the \ion{C}{4} 1548.2, 1550.8 \AA\ doublet at this 
redshift. However, closer inspection of these lines casts some doubt 
on this identification. First, line 21 shows two components while 
line 22 shows only one. Line 22 is close to the strong and complex 
\ion{C}{4} 1548.2 \AA\ absorption at \zabs\ = 2.315, but it is 
separated enough to at least partially show two components and the 
second component is not apparent. The second component of line 21 
is not telluric, but it could be an unidentified line from a different 
redshift. However, line 22 is significantly broader than the stronger 
component of line 21, and most importantly, lines 21 and 22 do not 
line up as well: profile fitting gives \zabs\ = 2.30805 for the stronger 
component of line 21 and \zabs\ = 2.30817 for line 22. We were 
unable to find a better identification for these lines, so we tentatively 
suggest that these are the \ion{C}{4} lines, substantially confused 
by blending, at the redshift of the {\it HST} system. Line 22 could 
be an additional component of the multicomponent \ion{C}{4} 
1548.2 \AA\ profile at \zabs\ = 2.315, but in this case a 
corresponding line should be detected in the \ion{C}{4} 1550.8 
\AA\ profile at \zabs\ = 2.315, and nothing is apparent at the 
expected wavelength (see Figure 2).

\subsection{$\bf z_{\bf abs}$ = 2.3150, 2.3155}

\begin{figure}
\plotone{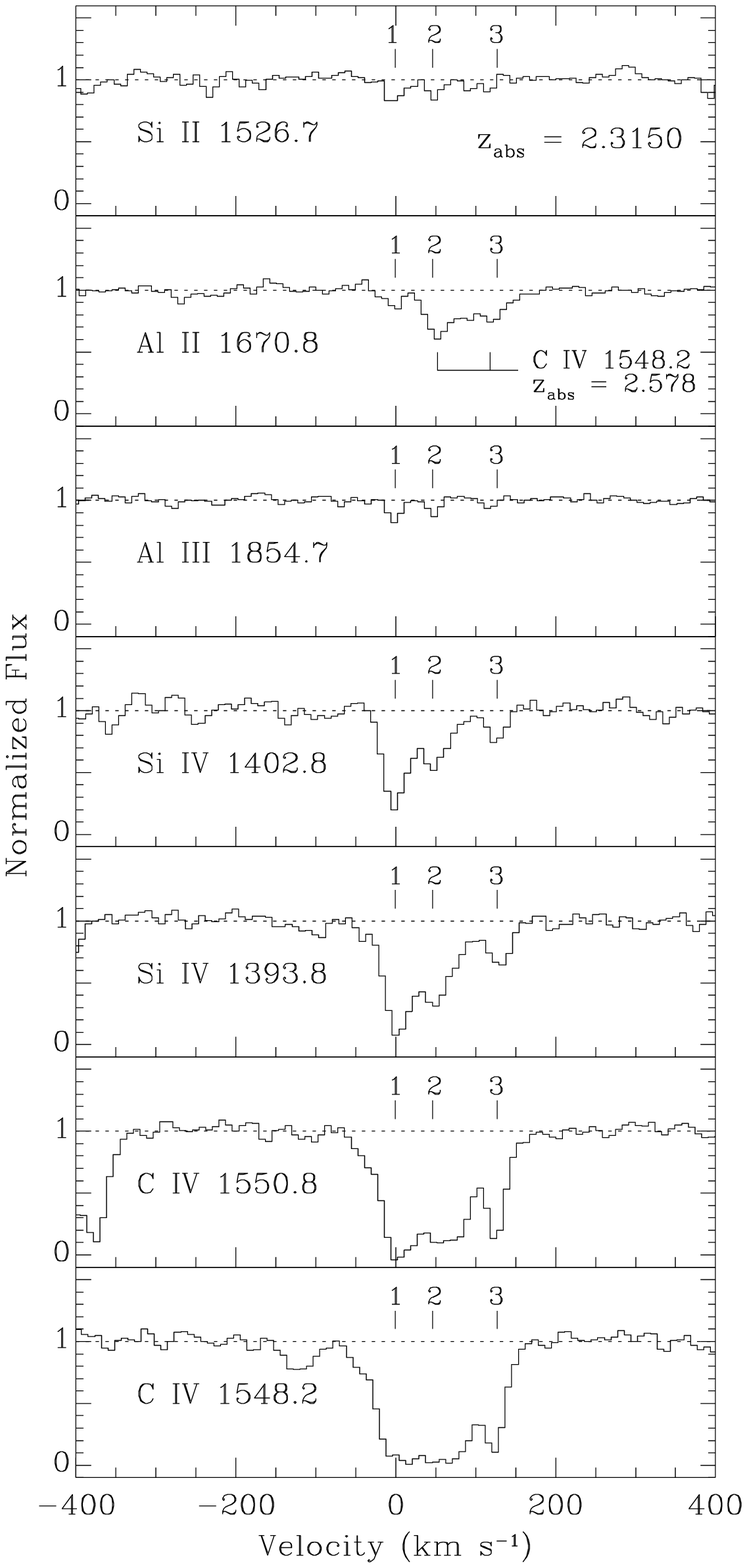}
\vspace{-1.5cm}
\caption{Normalized absorption profiles of heavy elements 
associated with the Lyman limit system at \zabs\ = 2.3150. The 
profiles are plotted versus restframe velocity with $v$ = 0 \kms\ at 
$z$ = 2.315. For convenience, we refer to the three resolved 
components as components 1-3 as marked in each panel. The 
absorption near components 2 and 3 in the \ion{Al}{2} 1670.8 \AA\ 
profile is predominantly due to \ion{C}{4} at a different redshift 
(see text \S 4.8). The absorption feature at $v \approx$ --120 \kms\ 
in the \ion{C}{4} 1548.2 \AA\ profile is also probably a line from a 
different redshift system, although the identification of this line is 
not secure (\S 4.7).}
\end{figure}

This absorption system, which is detected in \ion{Si}{2}, 
\ion{Si}{4}, \ion{C}{4}, \ion{Al}{2}, and \ion{Al}{3}, is the most 
prominent heavy element system in our optical spectrum of HS 
1700+6416. This system also shows Lyman limit absorption in the 
FOS spectrum, and from this Reimers et al.\markcite{r5} (1992) 
measure log $N$(\ion{H}{1}) = 16.85. Figure 2 shows the 
normalized absorption profiles of all metals we detect in this system 
(excluding profiles which occur in the \lya forest). Two components 
are apparent in the \ion{Si}{2} and \ion{Al}{3} absorption 
profiles\footnote{The absorption lines at \zabs\ = 2.3150 and 2.3155 
are detected at the 5.7$\sigma$ and 4.5$\sigma$ levels, respectively, 
in the Al III 1854.7 \AA\ transition. Given the weakness of these 
lines, it is not surprising that they are not significantly detected in 
the weaker Al III 1862.8 \AA\ line of this doublet.}, and at least 
three components are clearly present in the profiles of \ion{Si}{4} 
and \ion{C}{4}. We shall refer to these main components at \zabs\ = 
2.3150, 2.3155, and 2.3164 as components 1, 2, and 3 respectively 
(as labeled in Figure 2). In addition, the \ion{C}{4} 1548.2 and 
1550.8 \AA\ profiles both show absorption at \zabs\ $\approx$ 
2.3146 (i.e., at $v \approx$ --40 \kms\ in Figure 2), and very weak 
absorption is present at this redshift in the \ion{Si}{4} 1393.8 \AA\ 
profile. There are some indications of further component structure in 
the profiles (see below), but the combined S/N and resolution of 
these data do not warrant addition of more components to the profile 
fits.

Unfortunately, the \ion{Al}{2} 1670.8 \AA\ absorption occurs close 
to the \ion{C}{4} 1548.2 \AA\ absorption line at \zabs\ = 2.578. 
Furthermore, the \zabs\ = 2.578 absorber shows multiple 
components (\zabs\ = 2.57813 and \zabs\ = 2.57884), so the 
identification of the absorption line at 5538.7 \AA\ is ambiguous; 
this absorption could be due to \ion{Al}{2} in component 1 (see 
Figure 2), or it could be a third component of the \ion{C}{4} 
absorber at \zabs\ = 2.57746. The \ion{C}{4} 1550.8 \AA\ profile at 
\zabs\ = 2.578 does not show a third component at \zabs\ = 2.57746, 
so the line at 5538.7 \AA\ is probably due to \ion{Al}{2}. However, 
we cannot rule out the \ion{C}{4} identification because the line is 
weak, and it may fall below our detection threshold in the weaker 
\ion{C}{4} 1550.8 \AA\ profile.

\begin{figure}
\plotone{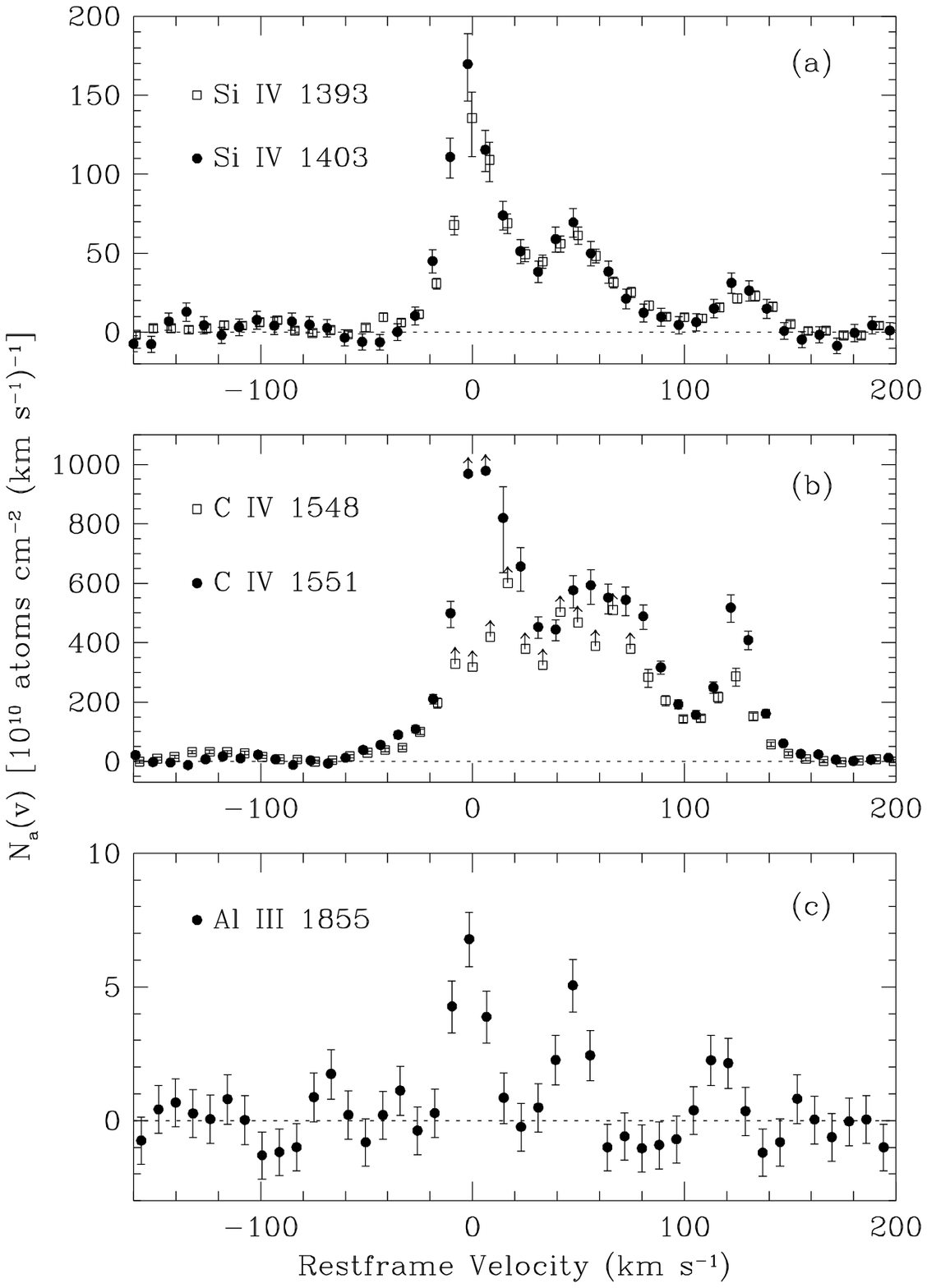}
\caption{Apparent column density profiles (see text \S 3.2) of heavy 
elements in the Lyman limit system at \zabs\ = 2.3150: (a) the 
\ion{Si}{4} 1393.8 and 1402.8 \AA\ lines, (b) the \ion{C}{4} 
1548.2 and 1550.8 \AA\ lines, and (c) the \ion{Al}{3} 1854.7 \AA\ 
line. In panels (a) and (b) the stronger member of the doublet is 
plotted with an open square and the weaker line is indicated with a 
filled circle.}
\end{figure}

In Figure 3 we compare the apparent column density profiles of the 
\ion{Si}{4} 1393.8 and 1402.8 \AA\ lines (panel a) as well as the 
\ion{C}{4} 1548.2 and 1550.8 \AA\ lines (panel b) and the 
\ion{Al}{3} 1854.7 \AA\ line. Not surprisingly, this shows that the 
\ion{C}{4} profile is badly saturated in components 1 and 2. The 
\ion{C}{4} profile is also significantly saturated in component 3; 
the weaker line in the \ion{C}{4} doublet (filled circles in Figure 3) 
yields a significantly higher \nav\ than the strong line in this 
component. This high degree of \ion{C}{4} saturation indicates that 
component 3 is intrinsically rather narrow, and this in turn indicates 
that the gas is relatively cool. The \ion{Si}{4} profile is less 
affected by saturation. From Figure 3 we see that the \ion{Si}{4} 
weak and strong line \nav\ profiles are in good agreement in 
components 2 and 3, which indicates that the \ion{Si}{4} column 
density is reliably measured in these components. In component 1, 
on the other hand, the weak \ion{Si}{4} line has a higher apparent 
column density than the strong line, so this component is affected by 
unresolved saturation. However, Figure 3 indicates that \ion{Si}{4} 
is only mildly saturated in component 1, so a correction can be 
applied to the weak line apparent column density to obtain a reliable 
estimate of $N$(\ion{Si}{4}) [Savage \& Sembach 1991; Jenkins 
1996]. Using the method of Savage \& Sembach (1991), we estimate 
that the correction required for the integrated \nav\ of the 
\ion{Si}{4} 1402.8 \AA\ line is 0.08 dex. Therefore log $N_{\rm 
a}$(\ion{Si}{4}) = 13.68 + 0.08 = 13.76 in component 1. This 
corrected column density is in excellent agreement with the column 
density from profile fitting, log $N$(\ion{Si}{4}) = 
13.75$\pm$0.07, which gives us confidence in the measurement. For 
the \ion{Si}{2}, \ion{Al}{2}, and \ion{Al}{3} profiles we do not 
have the benefit of additional transitions for \nav\ comparison, but 
these lines are all rather weak so saturation is not likely to be a 
problem. From Figure 3 we see that a very weak absorption feature 
is present at $v \approx$ +118 \kms\ in the \ion{Al}{3} profile. This 
is close to component 3 but is shifted slightly blueward compared to 
\ion{Si}{4} and \ion{C}{4}. Similarly, from Figure 2 we see that 
there may be weak \ion{Si}{2} absorption near but slightly 
blueward of component 3.

In \S 5 we discuss the physical conditions in this Lyman limit 
absorber, and we use standard photoionization models to examine 
the heavy element abundances.

\subsection{$\bf z_{\bf abs}$ = 2.37995}

This new metal system is detected only in the \ion{C}{4} 1548.2, 
1550.8 \AA\ doublet (lines 26 and 27 in Figure 1 and Table 1). 
Comparison of the \nav\ profiles indicates that these \ion{C}{4} 
lines are not substantially affected by unresolved saturated 
component structure. Similarly, the \ion{C}{4} column densities 
from profile fitting and \nav\ integration agree within the noise, so 
saturation is not a problem.

\subsection{The C IV complex at 2.432 $< \ \bf z_{\bf abs} \ <$ 
2.441}

A prominent complex of \ion{C}{4} doublets is detected at 5313 
$<$ \zabs\ $<$ 5336 \AA\ (lines 28--33 in Figure 1), which 
corresponds to 2.432 $<$ \zabs\ $<$ 2.441. Figure 4 shows an 
expanded plot of this region of the KPNO spectrum along with the 
component fit which gives the profile parameters listed in columns 
5-7 in Table 1. To show the clustering of \ion{C}{4} lines, we also 
plot at the top of Figure 4 the redshifts at which we detect 
\ion{C}{4} absorbers in the optical spectrum of HS 1700+6416. 
This absorption complex has two interesting features:

(1) There is a very high density of absorption lines in this small 
redshift interval. Inspection of the spectrum in Figure 4 reveals that 
there are at least 4 \ion{C}{4} doublets in this complex within the 
redshift range $\Delta z$ = 0.00855, and this implies that the number 
of \ion{C}{4} absorbers per unit redshift is $dN/dz \gtrsim$ 450. In 
addition, profile fitting requires a broad component with a 
substantial column density at \zabs\ = 2.43329; this broad 
component is not well constrained by the present data and may split 
up into subcomponents when observed at higher resolution. Also, 
the strength of the line at 5322.4 \AA\ indicates that another narrow 
component is present (see below). In contrast, Petitjean et 
al.\markcite{p3} (1994) find $dN/dz \approx$ 7.7 and Tripp et 
al.\markcite{t2} (1996) derive $dN/dz$ = 7.1$\pm$1.7 using 
samples of intervening \ion{C}{4} systems with much larger 
redshift intervals of $\Delta z$ = 1.3 and $\Delta z$ = 2.4, 
respectively. The Petitjean et al.\markcite{p3} and Tripp et 
al.\markcite{t2} samples include {\it all} \ion{C}{4} lines with rest 
equivalent width $W_{\rm r} \ >$ 0.03 \AA\ along multiple sight 
lines, and therefore these samples provide a measure of the {\it 
average} $dN/dz$ of \ion{C}{4} systems at moderately high 
redshifts. However, Petitjean et al.\markcite{p3} and Tripp et 
al.\markcite{t2} treat any \ion{C}{4} doublets separated by less 
than 200 \kms\ as a single absorber. Applying this treatment to the 
data in Figure 4, we still find that there are 2 or 3 \ion{C}{4} 
systems in this small redshift range implying that $dN/dz >$ 200. 
The high $dN/dz$ measured in this \ion{C}{4} cluster only 
indicates that the {\it local} density is very high and is not in conflict 
with previous estimates of $dN/dz$ over broad redshift intervals.

\begin{figure}
\caption[]{(opposite page) The normalized absorption profiles of the \ion{C}{4} 
complex at 2.432 $<$ \zabs\ $<$ 2.441 (lines 28-33 in Figure 1 and 
Table 1), plotted versus vacuum Heliocentric wavelength (bottom 
axis) and velocity in the \zabs\ = 2.43877 rest frame (top axis). The 
thin histogram shows the observed optical spectrum, and the thick 
line shows the profile fit which yields the component parameters 
listed in Table 1. To show the clustering of the \ion{C}{4} 
absorption systems, and in particular to show the high density of 
\ion{C}{4} doublets in this redshift range, we plot the redshifts of 
all detected \ion{C}{4} doublets (not including lines in the \lya 
forest) at the top of the figure. Here each \ion{C}{4} doublet is 
indicated with a single thin vertical line at the measured redshift. 
The line identification numbers from Table 1 are shown at the 
bottom of the lower panel.}
\end{figure}
\begin{figure}
\figurenum{4}
\plotone{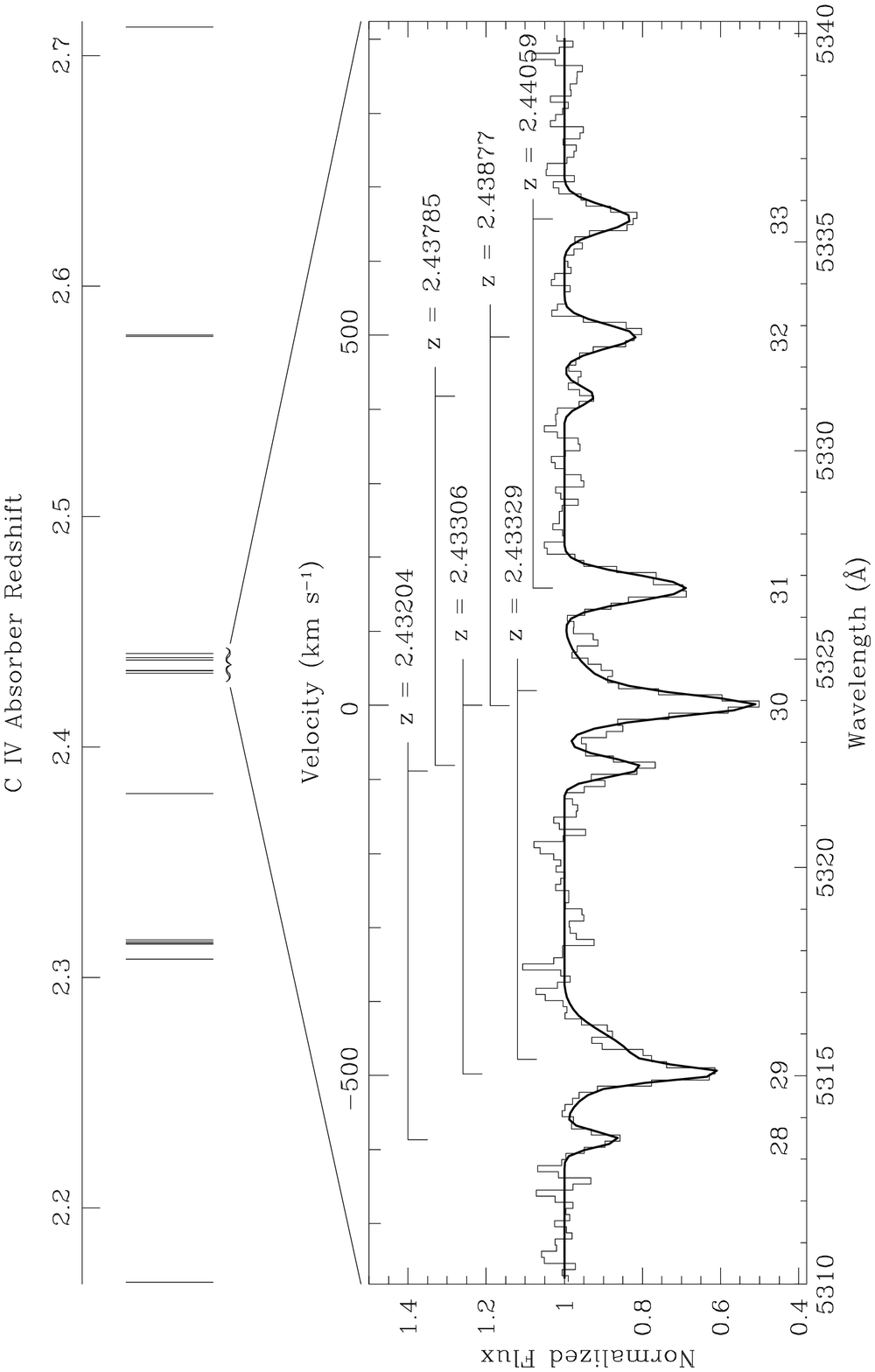}
\caption{continued}
\end{figure}

(2) The weak line (1550.8 \AA ) of the \ion{C}{4} doublet at \zabs\ 
= 2.43306 is remarkably well-aligned with the strong line of the 
\ion{C}{4} doublet at \zabs\ = 2.43877 (see Figure 4). To within 
what velocity tolerance are these \ion{C}{4} lines aligned? 
Traditionally the quantity $R = (1 + z_{2})/(1 + z_{1})$ has been 
used to search for alignments of absorbers at redshifts $z_{1}$ and 
$z_{2}$; if the absorbers are aligned in some doublet with 
wavelengths $\lambda_{i}$ and $\lambda_{j}$, then $R = 
\lambda_{i}/\lambda_{j}$. Thus for the \ion{C}{4} doublet we 
expect $R$ = 1550.770/1548.195 = 1.001663, and with $z_{1}$ = 
2.433064$\pm$0.000013 and $z_{2}$ = 2.438767$\pm$0.000013 
we obtain $R$ = 1.001661$\pm$0.000005. These redshifts and 
uncertainties were obtained by fitting the profiles of all lines shown 
in Figure 4 simultaneously, and the uncertainties in the profile 
parameters may be greater than the uncertainties formally estimated 
by the fitting program due to the complexity of the profiles. 
However, the 1548.2 \AA\ line of the $z_{1}$ system and the 
1550.8 \AA\ line of the $z_{2}$ system are free from confusion 
from this alignment, so independently fitting these lines provides an 
estimate of the redshifts which is not affected by the severe blending 
due to the redshift overlap. The $z_{1}$ 1548.2 \AA\ line is blended 
with a broad component at $z \approx$ 2.4333 (see Figure 4), and it 
is necessary to fit the broad component and $z_{1}$ simultaneously 
to obtain an acceptable fit, but the core of the $z_{1}$ line is narrow 
and deep enough that $z_{1}$ is well-constrained despite the blend. 
The $z_{2}$ 1550.8 \AA\ line is not blended. From this independent 
fitting exercise, we obtain $z_{1}$ = 2.433055$\pm$0.000016 and 
$z_{2}$ = 2.438787$\pm$0.000019. From these redshifts we 
estimate that $z_{1}$ and $z_{2}$ are aligned to within 10 \kms\ at 
the 3$\sigma$ level.

There is another pair of weak \ion{C}{4} doublets in Figure 4 which 
show the weak line of one absorber aligned with the strong line of 
another, although this case is not as strong. The line at 5313.49 \AA\ 
is attributed to \ion{C}{4} 1548.2 at \zabs\ = 2.43204. The 
corresponding weak (1550.8 \AA ) line is detected at the expected 
wavelength (5322.3 \AA ), but it is considerably stronger than the 
1548.2 \AA\ line, and fitting the profiles of the 5313.5 and 5322.3 
\AA\ lines with only one component yields a very poor fit. Since the 
line at 5313.5 only allows one component, we attribute the line at 
5322.3 \AA\ to the 1550.8 \AA\ line at \zabs\ = 2.43204 blended 
with a \ion{C}{4} 1548.2 \AA\ line at \zabs\ = 2.43785, and we find 
that there is a weak absorption line at the expected wavelength of 
\ion{C}{4} 1550.8 \AA\ at \zabs\ = 2.43785. Since these doublets 
are weaker than the aligned doublets discussed in the previous 
paragraph, we cannot constrain the velocity alignment as tightly. We 
estimate that the weak line of the \zabs\ = 2.43204 absorber is 
aligned to within $\sim$ 25 \kms\ with the strong line of the \zabs\ = 
2.43785 absorber.

This alignment of the weak \ion{C}{4} line from one redshift with 
the strong \ion{C}{4} line from a different redshift is reminiscent of 
the absorption ``line locking'' phenomenon which was once a topic 
of considerable interest and controversy (Perry, Burbidge, \& 
Burbidge\markcite{p1} 1978; Weymann, Carswell, \& 
Smith\markcite{w3} 1981 and references therein). Some interesting 
examples of aligned weak and strong doublet lines from different 
redshift systems have appeared in more recent literature (e.g., Foltz 
et al.\markcite{f2} 1987; Wampler\markcite{w1} 1991; Wampler, 
Bergeron, \& Petitjean\markcite{w2} 1993). In some cases these 
occur in broad absorption line (BAL) and ``associated'' absorption 
systems with \zabs\ $\approx$ $z_{\rm em}$, which is perhaps not 
surprising since radiative acceleration provides a natural explanation 
of these alignments, and the presumably close proximity of the 
\zabs\ $\approx$ $z_{\rm em}$ absorbers to the QSO would subject 
them to an extreme non-isotropic radiation field. Several models of 
BAL outflows have recently been proposed in which radiation 
pressure plays a critical role (e.g., Arav, Li, \& 
Begelman\markcite{a2} 1994; Murray et al.\markcite{m9} 1995; 
Scoville \& Norman\markcite{s6} 1995), and Arav et 
al.\markcite{a1} (1995) have shown that a double-trough spectral 
signature predicted by their model is convincingly detected in the 
types of QSOs where it is expected to be present. However, the 
\ion{C}{4} complex shown in Figure 4 is definitely not a BAL 
system, and it is displaced from $z_{\rm em}$ by $\sim$24000 \kms 
, so it usually would not be considered an associated system either. 
BAL outflows do attain velocities this large (Turnshek\markcite{t3} 
1988), so perhaps this \ion{C}{4} complex is close to the QSO and 
was accelerated in an earlier BAL outflow that has since mostly 
dissipated (a scenario proposed by Turnshek\markcite{t3} 1988). 
Alternatively, this may be a BAL outflow which is just beginning to 
form with the high velocity absorption showing up first to be 
followed later by lower velocity absorption with \zabs\ closer to 
$z_{\rm em}$.

Before pursuing these speculations, it is important to ascertain 
whether or not the doublet alignments in Figure 4 could simply be 
coincidental. If we consider randomly placing some number of 
\ion{C}{4} absorbers in a specified velocity interval, then we can 
use the binomial distribution to estimate the probability that the 
weak line of one absorber could by chance be aligned with the 
strong line of a different absorber to within some velocity tolerance. 
It is not immediately obvious what velocity interval or number of 
absorbers should be assumed for this calculation, however, and these 
assumptions significantly affect the probability. The {\it minimum} 
velocity interval (which translates to an {\it upper limit} on the 
probability) is the difference between the velocity of the strong line 
of the highest redshift doublet in the cluster and the velocity of the 
strong line of the lowest redshift doublet (if the velocity interval 
were less than this, then the \ion{C}{4} complex shown in Figure 4 
would never occur because some of the doublets would be outside of 
the velocity interval). This corresponds to a wavelength interval of 
$\Delta \lambda$ = 13.2 \AA\ or a velocity interval of $\Delta v$ = 
746 \kms . There are four well-detected absorbers (\zabs\ = 2.43204, 
2.43306 2.43877, and 2.44059) in this $\lambda$ range, and as 
noted above, it is likely that there is another at \zabs\ = 2.43785, and 
there is a broad component at \zabs\ = 2.43329 which could be due 
to multiple blended \ion{C}{4} lines. With four absorbers and 
$\Delta v$ = 746 \kms , the probability of one weak/strong line 
alignment to within 10 \kms\ is 15\% . If we include the second 
apparent weak/strong line alignment, then we must relax the velocity 
overlap tolerance to 25 \kms , and we must randomly place at least 
five absorbers in the velocity interval, so in this case the probability 
of two chance alignments is 12\% . If we increase the number of 
absorbers to, say, 7, then the probability of two alignments to within 
25 \kms\ increases to 28\% .

Evidently it is possible that these apparently line locked systems are 
just chance coincidences in an overdense \ion{C}{4} complex. 
Nevertheless, since similar highly displaced line alignments have 
been reported elsewhere (Wampler et al.\markcite{w2} 1993), it is 
worthwhile to make a note of this. If highly displaced line 
alignments show up in future high resolution QSO absorber studies, 
then this may reveal an interesting new QSO phenomenon or 
evolutionary stage. It is also interesting to note that Sargent, 
Boksenberg, \& Steidel\markcite{s2} (1988), Hamann, Barlow, \& 
Junkkarinen\markcite{h2} (1997a), and Jannuzi et al.\markcite{j2} 
(1996) have reported evidence of a new class of QSO absorption 
systems which are similar to BALs but highly displaced from the 
QSO systemic redshift and lacking the \zabs\ $\approx \ z_{\rm 
em}$ absorption usually detected in BALs. These displaced BALs 
may be related to the \ion{C}{4} complex discussed in this section. 
It may be possible to establish that this \ion{C}{4} complex is close 
to the QSO by monitoring the absorption profile; if the profile shows 
temporal variability, then the absorbing gas must be close to the 
QSO.

\subsection{$\bf z_{\bf abs}$ = 2.57813, 2.57884}

Reimers et al.\markcite{r5} (1992) report detection of this system in 
several metals, but they do not detect Lyman limit absorption. 
However, the spectral coverage of the FOS observation barely 
reaches the expected wavelength of this Lyman limit, and their S/N 
drops off precipitously in this region (see Figure 1 in Vogel \& 
Reimers\markcite{v3} 1995), so the FOS spectrum is not sensitive 
to Lyman continuum absorption at this redshift. We detect the 
\ion{C}{4} 1548.2, 1550.8 \AA\ doublet in this system, and two 
components are obvious in the profiles of both lines (see lines 36 
and 37 in Figure 1). Comparison of the \nav\ profiles of the 
\ion{C}{4} doublet does not show unresolved saturation in either 
component. The \ion{C}{4} 1548.2 \AA\ profile shows a third 
component, but there is no evidence of this component in the 1550.8 
\AA\ profile and, as discussed above, this third 
component\footnote{If the third component is due to C IV at \zabs\ 
= 2.57746 rather than Al II, then from profile fitting we obtain $b$ = 
7.3 \kms\ and log N(C IV) = 12.66.} may be \ion{Al}{2} 1670.8 at 
\zabs\ = 2.3150. The \zabs\ = 2.3150 system shows complex 
component structure (see Figure 2), so the \ion{C}{4} 1548.2 
profile could be contaminated with \ion{Al}{2} absorption from 
\zabs\ = 2.3155.

\subsection{$\bf z_{\bf abs}$ = 2.7125}

Reimers et al.\markcite{r5} (1992) and Vogel \& 
Reimers\markcite{v3} (1995) do not report detections of any metals 
at this redshift, but as discussed by Petitjean et al.\markcite{p4} 
(1996), some of the heavy element lines at this redshift in the FOS 
spectrum may have been misidentified by Vogel \& 
Reimers\markcite{v3}. For example, Petitjean et al.\markcite{p4} 
point out that the line Vogel \& Reimers\markcite{v3} attribute to 
\ion{O}{2} 833.3 \AA\ at \zabs\ = 2.433 could instead be 
\ion{Ne}{8} 770.4 \AA\ at \zabs\ = 2.712. In the KPNO optical 
spectrum we detect the \ion{N}{5} 1238.8, 1242.8 \AA\ and 
\ion{C}{4} 1548.2, 1550.8 \AA\ doublets at this redshift (see lines 
3-4 and 38-39, respectively, in Table 1 and Figure 1). A 4.7$\sigma$ 
absorption line is present at 4842.45 \AA , the expected wavelength 
of \ion{Si}{2} 1304.4 \AA\ at \zabs\ = 2.71249. However, this 
identification probably is not correct because other \ion{Si}{2} lines 
should be detected as well, and they are not. For example, the 
\ion{Si}{2} 1260.4 \AA\ line is considerably stronger than the 
1304.4 \AA\ line, but the 1260.4 \AA\ line is not detected. There is a 
weak unidentified feature recorded at 4679.89 \AA\ (line 8 in Figure 
1), but this is shifted redward from the expected wavelength of the 
\ion{Si}{2} 1260.4 \AA\ line by $\Delta z$ = +0.00047, and 
furthermore this unidentified line is much weaker than the expected 
\ion{Si}{2} 1260.4 \AA\ line (based on the strength of the line at 
4842.45, assuming it is due to \ion{Si}{2} 1304.4 \AA ). Similarly, 
the \ion{Si}{2} 1526.7 \AA\ line should be easily detected, but no 
line is apparent at the expected wavelength (see Figure 1).

\begin{figure}
\plotone{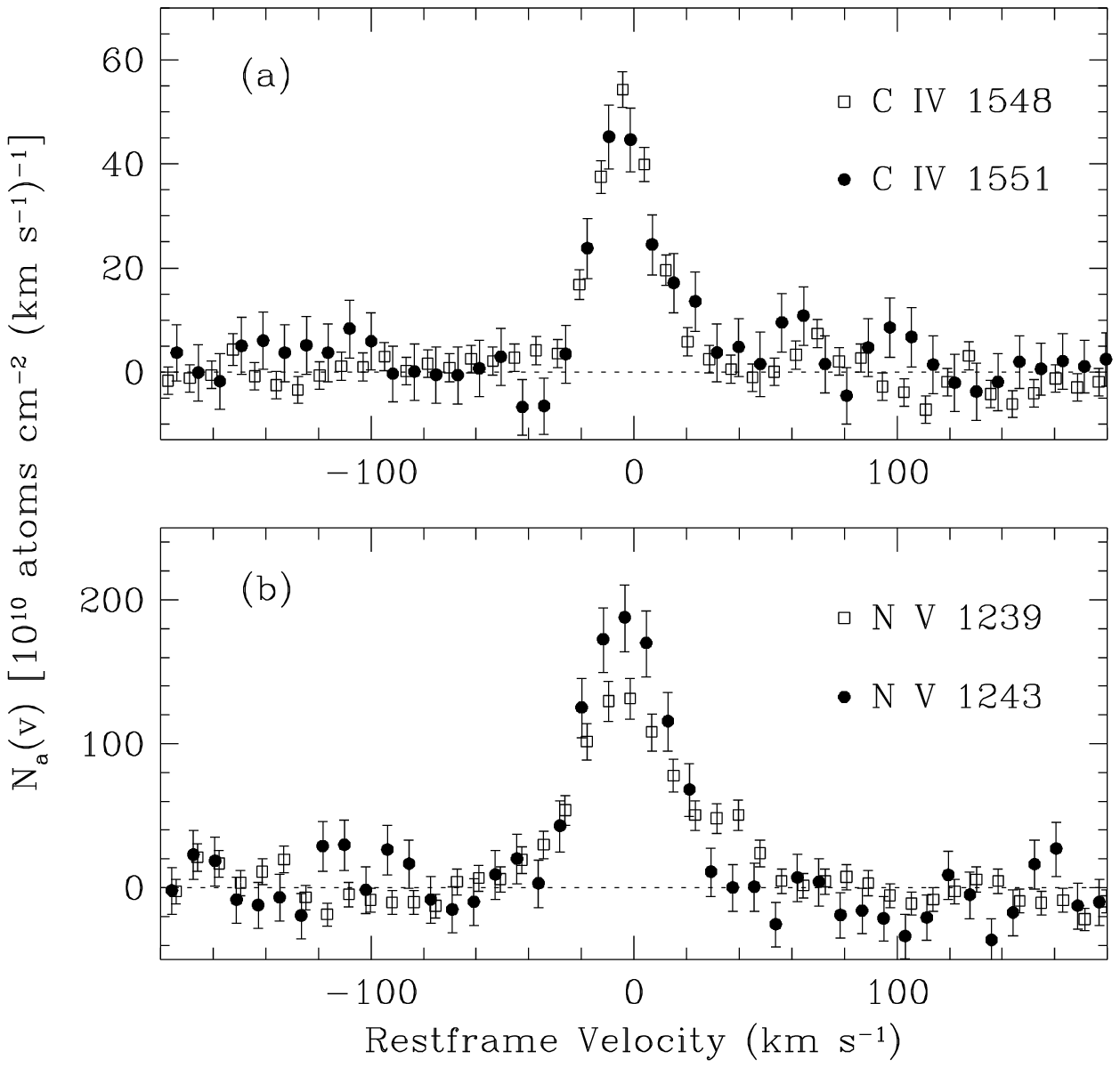}
\vspace{-5cm}
\caption[]{Apparent column density profiles (\S 3.2) of (a) the 
\ion{C}{4} 1548.2 and 1550.8 \AA\ lines and (b) the \ion{N}{5} 
1238.8 and 1242.8 \AA\ lines of the associated absorber at \zabs\ = 
2.7125. The stronger line of each doublet is plotted with an open 
square and the weaker line with a filled circle. The \nav\ profiles of 
the weak and strong \ion{C}{4} lines are in good agreement, which 
indicates that \ion{C}{4} is not affected by unresolved saturation in 
this absorber. However, in the core of the profile, the weaker 
\ion{N}{5} line yields a significantly higher \nav\ than the strong 
line. This indicates that either the \ion{N}{5} profile is affected by 
unresolved saturated absorption, or the absorbing gas does not 
completely cover the QSO flux source.}
\end{figure}

The apparent column density profiles of the \ion{C}{4} and 
\ion{N}{5} doublets at \zabs\ = 2.7125 are shown in Figure 5. This 
figure suggests that the \ion{N}{5} absorption profile is affected by 
unresolved saturation; the weak \ion{N}{5} line (1242.8 \AA ) has a 
significantly higher \nav\ than the strong line in the core of the 
profile, and the integrated \ion{N}{5} 1242.8 \AA\ column density, 
log $N$(\ion{N}{5}) = 13.87, is 0.09 dex greater than the integrated 
column of the 1238.8 \AA\ line. Alternatively, Figure 5 may indicate 
that the \ion{N}{5} absorbing gas does not completely cover the 
flux source (see below). The column density from profile fitting 
agrees with the integrated 1242.8 \AA\ column density, but 
inspection of the fitted profiles shows that the fit is not very good 
and the observed optical depth in the core of the 1242.8 \AA\ line 
exceeds the profile fit, so the true column density could be 
somewhat higher. From Figure 5 we see that \ion{C}{4}, on the 
other hand, is not significantly affected by unresolved saturation (or 
incomplete covering) at this redshift; the \ion{C}{4} weak and 
strong line \nav\ profiles are in good agreement as are the column 
densities from profile fitting and \nav\ integration (see Table 1).

This absorption system occurs within 5000 \kms\ of the QSO 
emission redshift and thus qualifies as an ``associated'' absorber. 
There are some indications that these are a special class of QSO 
absorbers which may be intrinsically related to the QSO nucleus or 
host galaxy, but there is also evidence that some absorption systems 
within 5000 \kms\ of $z_{\rm em}$ are at least a few hundred kpc 
away from the QSO (Turnshek, Weymann, \& 
Williams\markcite{t3} 1979; Morris et al.\markcite{m7} 1986; 
Tripp et al.\markcite{t2} 1996). Probably this class of absorbers 
includes a mix of intrinsic and intervening absorption systems. In \S 
5.2.3 we use photoionization models to study the abundances and 
ionization in this absorption system. Recently it has been suggested 
that the gas in some associated absorbers does not completely cover 
the QSO flux source (Wampler et al.\markcite{w2} 1993; Petitjean 
et al.\markcite{p3} 1994; Hamann et al.\markcite{h2} 
1997a,b\markcite{ha2}; Barlow \& Sargent\markcite{b1} 1997). If 
this is true, then these associated absorbers must be close to the QSO 
nucleus. The case presented by Barlow \& Sargent\markcite{b1} 
(1997) is particularly convincing because the \ion{H}{1} \lya 
profile has an obvious flat bottom indicative of strong saturation, but 
the flux in the flat bottom of the profile is well-detected and is not 
zero. While quite interesting, this complicates the column density 
measurements because even if the lines are unsaturated, the 
incomplete covering will introduce errors in column densities from 
profile fitting and \nav\ integration. The discrepancy between the 
\ion{N}{5} 1238.8 and 1242.8 \AA\ \nav\ profiles shown in Figure 
5 could indicate that the absorbing gas does not completely cover 
the flux source. In principle, one can attempt to correct the  
absorption profiles for partial covering (c.f., Hamann et 
al.\markcite{ha2} 1997b; Barlow \& Sargent\markcite{b1} 1997), 
but in this case this is difficult since the absorption lines are strong 
enough that a mix of partial covering and saturation could affect the 
measurements. For the \ion{N}{5} column densities in the \zabs\ = 
2.7125 absorber, we adopt the conservative approach of simply 
treating the integrated column density of the weak line as a lower 
limit on $N$(\ion{N}{5}).

\section{Physical Conditions and Abundances}

In this section we examine the physical conditions and abundances 
in the metal systems of HS 1700+6416. We begin with simple 
inferences drawn from the absorption line measurements, and then 
we apply standard photoionization models to estimate abundances. 
However, we find some indications that the standard single-zone 
photoionization models are too simple to explain the data and 
multi-zone models may be required (see below).

\subsection{Temperature and Ionization}

The observed line widths provide one of the most straightforward 
probes of physical conditions. If the absorption profile is dominated 
by Doppler broadening (i.e., thermal motions), then the temperature 
$T$ of the gas can be calculated from the measured Doppler 
parameter, $b$ = $\sqrt{2}\sigma$ = 0.129$\sqrt{T/A}$, where $A$ 
is the atomic weight of the element and $\sigma$ is the radial 
velocity dispersion of the gas (Spitzer\markcite{s10} 1978). In QSO 
absorbers, the gas is often cool enough that other factors such as 
bulk motions, turbulence, or multiple unresolved components can 
significantly broaden the line, so the temperature derived from this 
equation must be treated as an upper limit\footnote{However, in a 
recent Keck study of a large number of \ion{C}{4} QSO absorbers, 
Rauch et al.\markcite{ra1} 1996 conclude that most \ion{C}{4} 
profiles are dominated by thermal motions.}. Since $b$ goes as 
$A^{-1/2}$, lighter elements like carbon are more useful for 
constraining the temperature of the gas. In Table 3 we list upper 
limits on the gas temperature derived from the Doppler 
parameters\footnote{Doppler parameter uncertainties could be larger 
than the formal fitting errors suggest since the true absorption 
profiles may be kinematically more complex than the simple one 
component Voigt profile assumed. This is particularly true for 
strongly blended adjacent components and narrow components with 
$b$ substantially smaller than the 20 \kms\ resolution.} of all 
\ion{C}{4} lines detected in the KPNO optical spectrum.

\begin{deluxetable}{ccc}
\tablewidth{0pc}
\tablenum{3}
\tablecaption{Temperature Upper Limits from C IV Lines\tablenotemark{a}}
\tablehead{
\colhead{$z_{{\rm abs}}$} & \colhead{$b$ (km s$^{-1}$)} & \colhead{$T$ (K)}
}
\startdata
2.16797                    & 16.5$\pm$0.8      & $<$2.0$\times 10^{5}$ \nl
2.30812                    & 25.0$\pm$2.2      & $<$4.7$\times 10^{5}$ \nl
2.31458\tablenotemark{b,c} & 14.6$\pm$5.4      & $<$1.5$\times 10^{5}$ \nl
2.31505\tablenotemark{b,c} & 11.3$\pm$3.8      & $<$9.2$\times 10^{4}$ \nl
2.31568\tablenotemark{b,c} & 28.6$\pm$1.5      & $<$5.9$\times 10^{5}$ \nl
2.31638\tablenotemark{b,c} &  3.9$\pm$0.2      & $<$1.1$\times 10^{4}$ \nl
2.37995                    &  6.3$\pm$2.0      & $<$2.9$\times 10^{4}$ \nl
2.43204\tablenotemark{b}   &  8.0$\pm$5.5      & $<$4.6$\times 10^{4}$ \nl
2.43306\tablenotemark{b}   &  4.0$\pm$3.5      & $<$1.2$\times 10^{4}$ \nl
2.43329\tablenotemark{b}   & 44.5$\pm$4.9      & $<$1.4$\times 10^{6}$ \nl
2.43785\tablenotemark{b}   & 12.6$\pm$4.3      & $<$1.1$\times 10^{5}$ \nl
2.43877\tablenotemark{b}   & 18.3$\pm$1.9      & $<$2.4$\times 10^{5}$ \nl
2.44059                    & 20.3$\pm$1.7      & $<$3.0$\times 10^{5}$ \nl
2.57813\tablenotemark{b}   & 18.9$\pm$1.9      & $<$2.6$\times 10^{5}$ \nl
2.57884\tablenotemark{b}   & 35.0$\pm$3.8      & $<$8.8$\times 10^{5}$ \nl
2.71245                    & 11.4$\pm$1.1      & $<$9.4$\times 10^{4}$ \nl
\enddata
\tablenotetext{a}{Temperature upper limits derived from the Doppler parameter, $b = 0.129 \sqrt{T/A}$. The atomic weight of carbon is $A$ = 12.0111. The
temperature estimates assume the absorption lines are well represented by
single component Voigt profiles unless otherwise noted.}
\tablenotetext{b}{Blended absorption line.}
\tablenotetext{c}{Saturated absorption line.}
\end{deluxetable}

An issue of considerable interest is the question of the ionization 
mechanism(s) in the various types of QSO absorbers. Are 
intervening QSO absorbers photoionized by the UV background 
radiation from QSOs and AGNs, or do local photoionizing sources 
and/or collisional ionization play a role? From Table 3 we see that 
four of the \ion{C}{4} absorption lines in the HS 1700+6416 optical 
spectrum have $T <$ 50000 K. At these temperatures very little 
\ion{C}{4} is present in gas in collisional ionization equilibrium 
(Sutherland \& Dopita\markcite{s15} 1993), so these systems are 
probably photoionized. However, it is also possible that the narrow 
\ion{C}{4} lines occur in nonequilibrium gas which has cooled 
more rapidly than it has recombined; this is predicted in some 
renditions of galactic fountain theory (Edgar \& 
Chevalier\markcite{e1} 1986; Benjamin \& Shapiro\markcite{b4} 
1997). It is important to note that in some cases these lines are 
moderately weak and noise could cause the profiles to appear 
artificially narrow. On the other hand, some of the \ion{C}{4} lines 
are so broad as to indicate that the line profiles are probably {\it not} 
dominated by thermal Doppler broadening. For example, the 
\ion{C}{4} absorber at \zabs\ = 2.57884 has $b = 35.0\pm3.8$ \kms 
, which implies that $T \approx 10^{5.9}$K if the profile is 
dominated by thermal motions; at this temperature carbon is 
predominantly in higher ionization stages than \ion{C}{4} in 
collisional ionization equilibrium (Sutherland \& Dopita 1993).

The high ion column density ratios in the Lyman limit absorber at 
\zabs\ = 2.3150, 2.3155 indicate that the ionization conditions differ 
significantly from the conditions in the gaseous halo of the Milky 
Way. From Figure 3 we see that $N$(\ion{C}{4})/$N$(\ion{Si}{4}) 
$\geq$ 10 in components 2 and 3 of this absorber and could be 
significantly larger since the \ion{C}{4} profile is affected by strong 
saturation. For comparison, Sembach \& Savage (1992) measure 
$N$(\ion{C}{4})/$N$(\ion{Si}{4}) = 3.6$\pm$1.3 in the Milky 
Way gaseous halo. More recently Sembach, Savage, \& Tripp 
(1997) have compiled the best measurements made with the {\it 
HST} and {\it International Ultraviolet Explorer (IUE)} of high ion 
column densities in the Galactic ISM, and they find 
$N$(\ion{C}{4})/$N$(\ion{Si}{4}) = 3.8$\pm$1.9 with the vast 
majority of the ISM \ion{C}{4}/\ion{Si}{4} ionic ratios being well 
below 10 (see their Figure 8). This high ionic ratio in the LL 
absorber is not expected on the basis of cooling galactic fountain 
theory but could be produced in supernova remnants, conductive 
interfaces between warm and hot gas, or ``turbulent mixing layers'' 
(see \S 8 in Sembach et al. 1997 and references therein). This higher 
ratio could also be due to the significantly increased intensity of the 
extragalactic UV background radiation at $z \ \approx$ 2 compared 
to $z \ \approx$ 0 (c.f., Bechtold 1995 and references therein).

Since the temperature of the gas producing the \ion{C}{4} 
absorption line in this LL absorber at \zabs\ = 2.31638 (i.e., 
component 3, see Figure 2) is evidently less than roughly 11000 K 
(Table 3), it seems likely that this system is photoionized. 
Components 1 and 2 may or may not be photoionized (the line 
widths are large enough to allow collisional ionization), but given 
the small total \ion{H}{1} column density for all components 
derived from $\tau_{LL}$ by Reimers et al.\markcite{r5} (1992) 
[log N(\ion{H}{1}) =  16.85], photoionization by the extragalactic 
UV background radiation is likely to be significant in all three 
components (c.f., Viegas\markcite{v2} 1995), so modeling is 
required to estimate elemental abundances.

\subsection{Photoionization Models}

Following the seminal paper of Bergeron \& 
Stasi\'{n}ska\markcite{b5} (1986), it has become customary to use 
photoionization models to infer abundances in moderate column 
density QSO absorbers, and the physical conditions discussed above 
indicate that this is appropriate. In this section we use models 
constructed with the photoionization code CLOUDY (version 84.12; 
Ferland\markcite{fe1} 1993) to estimate the abundances in the 
Lyman limit and associated absorption systems. We model the LL 
absorbers as a constant density plane parallel slab photoionized 
predominantly by the Madau\markcite{m1} (1992) extragalactic 
background radiation from QSOs/AGNs, which includes the effects 
of H and He opacity in intervening \lya clouds and Lyman limit 
systems. The slab is illuminated on one side. We assume the mean 
intensity at the hydrogen Lyman limit is $J_{\nu}$(LL) = 
10$^{-21}$ ergs s$^{-1}$ cm$^{-2}$ Hz$^{-1}$ sr$^{-1}$. For the 
associated absorber, the model assumes the gas is predominantly 
photoionized by the flux from HS 1700+6416 itself (see \S 5.2.3). 
Once the radiation field has been specified, other parameters which 
can be modified to fit the observed column densities are the overall 
metallicity [M/H], the relative heavy element abundances, the 
ionization parameter $\Gamma \ (\equiv n_{\gamma}/n_{\rm H}$ = 
H ionizing photon density/total hydrogen number density), and the 
\ion{H}{1} column density. We express the logarithmic abundance 
of element X relative to element Y using the usual notation, [X/Y] = 
log (X/Y) - log (X/Y)$_{\odot}$, and we use [M/H] to indicate the 
overall abundance of the metals relative to solar. Throughout this 
paper we use the solar abundances from Grevesse \& 
Anders\markcite{g2} (1989). 

We begin by placing lower limits on the heavy element abundances, 
and then for absorbers detected in several species and ionization 
stages, we use the models to more tightly constrain the metallicities. 
As discussed by Hamann (1997), even if we have measured column 
densities of {\it only one} ionization stage of metal X (e.g., 
\ion{Mg}{2}) and \ion{H}{1}, we can still place a lower limit on 
[X/H] in the context of the photoionization model. The abundance 
[X/H] can be expressed as
\begin{equation}
\left[ \frac{\rm X}{\rm H}\right] = {\rm log}\left( \frac{N({\rm 
X}_{i})}{N({\rm H \ I)}}\right) + {\rm log}\left( \frac{f({\rm H \ 
I})}{f({\rm X}_{i})}\right) - {\rm log}\left( \frac{\rm X}{\rm 
H}\right) _{\odot}
\end{equation}
where ${\rm X}_{i}$ is element X in ionization stage $i$, $N$ is 
the column density, $f$ is the ionization fraction, and 
(X/H)$_{\odot}$ is the abundance ratio in the sun. By choosing the 
model ionization parameter which minimizes $f({\rm H \ I})/f({\rm 
X}_{i})$, we can derive a lower limit on [X/H]; any other value of 
$\Gamma$ will lead to higher [X/H]. In practice our method is to 
calculate $N$(X$_{i}$) vs. $\Gamma$ for a series of models with 
decreasing values of [X/H] and fixed $N$(\ion{H}{1}); eventually 
[X/H] drops low enough so that the peak of the $N$(X$_{i}$) vs. 
$\Gamma$ curve is lower than the observed column density of 
X$_{i}$. For several of the HS 1700+6416 LL absorbers, Table 4 
summarizes the lower limits on [X/H] which agree with the 
measured column densities, to within the 2$\sigma$ uncertainties, at 
the peak of the $N$(X$_{i}$) vs. $\Gamma$ curve calculated by the 
model described above.

\begin{deluxetable}{cccccc}
\tablenum{4}
\tablewidth{0pc}
\tablecaption{Lower Limits on Heavy Element Abundances\tablenotemark{a}}
\tablehead{ \ & \multicolumn{5}{c}{\underline{Lyman Limit Absorbers}} \nl
 \ & \ & \ & \ & \ & \nl
$z_{\rm abs}$ & $\left[\frac{\rm Mg}{\rm H}\right]$ & 
$\left[\frac{\rm C}{\rm H}\right]$ & $\left[\frac{\rm N}{\rm H} \right]$ 
& $\left[\frac{\rm Si}{\rm H} \right]$ & $\left[\frac{\rm Al}{\rm H}\right]$ 
}
\startdata
1.15729 & $\geq$ --0.22 & \nodata & \nodata & \nodata & \nodata \nl
1.84506 & $\geq$ --0.45 & $\geq$ --2.26 & \nodata & \nodata & \nodata \nl
2.16797 & \nodata & $\geq$ --2.81 & \nodata & \nodata \nl
2.31499 & \nodata & \nodata & \nodata & $\geq$ --0.95\tablenotemark{b} & $\geq$ --0.96\tablenotemark{c} \nl
2.31551 & \nodata & \nodata & \nodata & $\geq$ --1.35\tablenotemark{b} & $\geq$ --1.40\tablenotemark{c} \nl
 \ & \ & \ & \ & \ & \nl
 \ & \multicolumn{5}{c}{\underline{Associated Absorber -- Model 1}\tablenotemark{d}} \nl
2.7125  & \nodata & $\geq$ --0.82 & $\geq$ --0.65  & \nodata & \nodata \nl
 \ & \ & \ & \ & \ & \nl
 \ & \multicolumn{5}{c}{\underline{Associated Absorber -- Model 2}\tablenotemark{d}} \nl
2.7125  & \nodata & $\geq$ --0.25 & $\geq$ +0.19  & \nodata & \nodata \nl
\enddata
\tablenotetext{a}{These abundance lower limits were derived using equation 3 as discussed in the text, \S 5.2.}
\tablenotetext{b}{Lower limit derived from $N$(\ion{Si}{2}).}
\tablenotetext{c}{Lower limit derived from $N$(\ion{Al}{3}).}
\tablenotetext{d}{Two models were used to place lower limits on the associated absorber abundances. Model 1 used the observed QSO spectral energy distribution
to set the model continuum shape while Model 2 used continuum shape `A' from Hamann (1997) with log $T_{c}$ = 5.7.}
\end{deluxetable}

\subsubsection{Lyman Limit Absorber at $z_{abs}$ = 2.3150, 
2.3155}

The abundance lower limits derived using equation 3 as discussed 
above are [Si/H] $\geq$ --0.95 and [Al/H] $\geq$ --0.96 in 
component 1 (\zabs\ = 2.3150) and [Si/H] $\geq$ --1.35 and [Al/H] 
$\geq$ --1.40 in component 2 (\zabs\ = 2.3155). These lower limits 
were calculated assuming log $N$(\ion{H}{1}) = 16.85 {\it for each 
component}. There are at least three absorption components in this 
LL system (see Figure 2), and log $N$(\ion{H}{1}) = 16.85 is the 
{\it total} \ion{H}{1} column (i.e., integrated across all 
components) derived from the Lyman limit optical depth. Therefore 
these are rather conservative lower limits; $N$(\ion{H}{1}) in the 
{\it individual} components is almost certainly lower than the 
assumed value so the abundances are almost certainly higher than 
these lower limts.

In this absorber we detect absorption lines from several species 
including multiple ionization stages of Si and Al (see Table 2), so 
we may be able to more accurately pin down the abundances using 
photoionization models which simultaneously match all of the 
measured heavy element column densities. However, with these 
models one must bear in mind that different ionization stages such 
as \ion{Si}{2} and \ion{Si}{4} may not occur in the same gas, i.e., 
the absorber may be a multiphase medium. We will start with the 
single zone model in which all of the absorption lines are assumed to 
arise in the same gas because it is the simplest model. With this 
caveat, consider component 1 (\zabs\ = 2.3150). Based on the heavy 
element column densities, we estimate that log $N$(\ion{H}{1}) 
$\approx$ 16.6 for component 1 assuming components 1 and 2 have 
roughly the same metallicity and contain the bulk of the neutral 
hydrogen. Our best estimations of the metal column densities in this 
component are log $N$(\ion{Si}{2}) = 13.07$\pm$0.12, log 
$N$(\ion{Al}{2}) = 11.69$\pm$0.11, log $N$(\ion{Al}{3}) = 
12.11$\pm$0.09, and log $N$(\ion{Si}{4}) = 13.75$\pm$0.07 (see 
\S 4.8). A very crude lower limit of log $N$(\ion{C}{4}) $>$ 14.6 
is obtained by integrating the apparent column density of 
\ion{C}{4} 1550.8 \AA\ over the velocity range of component 1.

\begin{figure}
\plotone{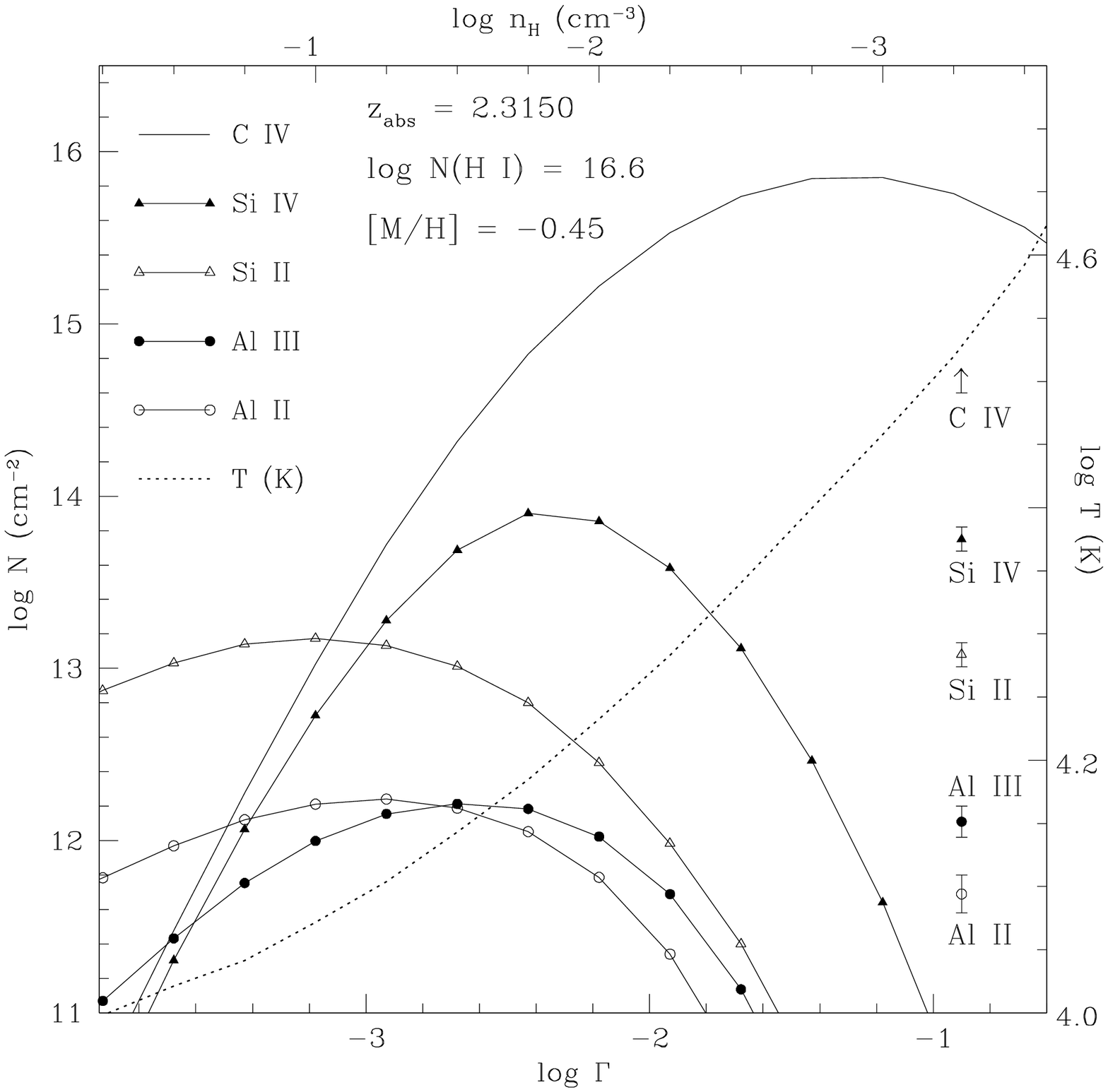}
\caption{Column densities predicted by a photoionized model of the 
Lyman limit absorber at \zabs\ = 2.3150 (see text \S 5.2.1) as a 
function of ionization parameter $\Gamma$ (bottom axis) and total 
hydrogen number density $n_{\rm H}$ (top axis). The column 
densities of heavy elements detected in component 1 of the Lyman 
limit system (\zabs\ = 2.3150) are shown at right with 1$\sigma$ 
error bars; \ion{C}{4} is strongly affected by absorption saturation. 
The dotted line shows the model temperature vs. $\Gamma$ with 
the temperature scale on the right axis.}
\end{figure}

To simultaneously produce enough \ion{Si}{2} and \ion{Al}{2} in 
component 1 with log $N$(\ion{H}{1}) = 16.6 (ignoring for the 
moment the higher ionization stages), the model must have [M/H] 
$\geq$ --0.73. Since \ion{Si}{2} and \ion{Al}{2} are the dominant 
ionization stages of Si and Al in interstellar \ion{H}{1} regions, this 
lower limit holds even if the higher ion absorption does not occur in 
the same gas. However to simultaneously match $N$(\ion{Si}{2}) 
and $N$(\ion{Si}{4}), the model must have a higher metallicity. To 
search for a model which agrees with {\it all} of the observed 
column densities, we have carried out a series of CLOUDY 
calculations with metallicities ranging from [M/H] = --0.60 to [M/H] 
= --0.30. Figure 6 shows a sample CLOUDY calculation, and Figure 
7 summarizes the results of these modeling efforts. Figure 6 shows 
the column densities of these species, as a function of $\Gamma$, 
calculated by a photoionization model with [M/H] = --0.45. We 
present this particular CLOUDY calculation as a sample because 
this model provides the single best fit. From the figure we see that 
the model is consistent with the observed column densities of 
\ion{Si}{2}, \ion{Al}{3}, and \ion{Si}{4}, to within the 2$\sigma$ 
uncertainties, when --2.73 $\leq$ log $\Gamma \ \leq$ --2.46, and it 
is consistent with the lower limit on $N$(\ion{C}{4}) when the 
ionization parameter exceeds --2.54. The dashed line in Figure 6 
shows the mean temperature of the model as a function of 
$\Gamma$, and we see that in this ionization parameter range the 
model is also in agreement with the observational constraints on the 
gas temperature. However, over this ionization parameter range the 
calculation predicts too much \ion{Al}{2} by 0.4--0.5 dex, which is 
greater than the 3$\sigma$ uncertainty of $N$(\ion{Al}{2}). This 
problem is aggravated by the possible contamination of the 
\ion{Al}{2} absorption from \ion{C}{4} 1548.2 \AA\ at \zabs\ = 
2.57746 (see \S 4.8 and Figure 2); if this absorption is partly or 
mostly due to \ion{C}{4} at a different redshift, then the 
discrepancy between the model and the observed $N$(\ion{Al}{2}) 
is even greater.

\begin{figure}
\plotone{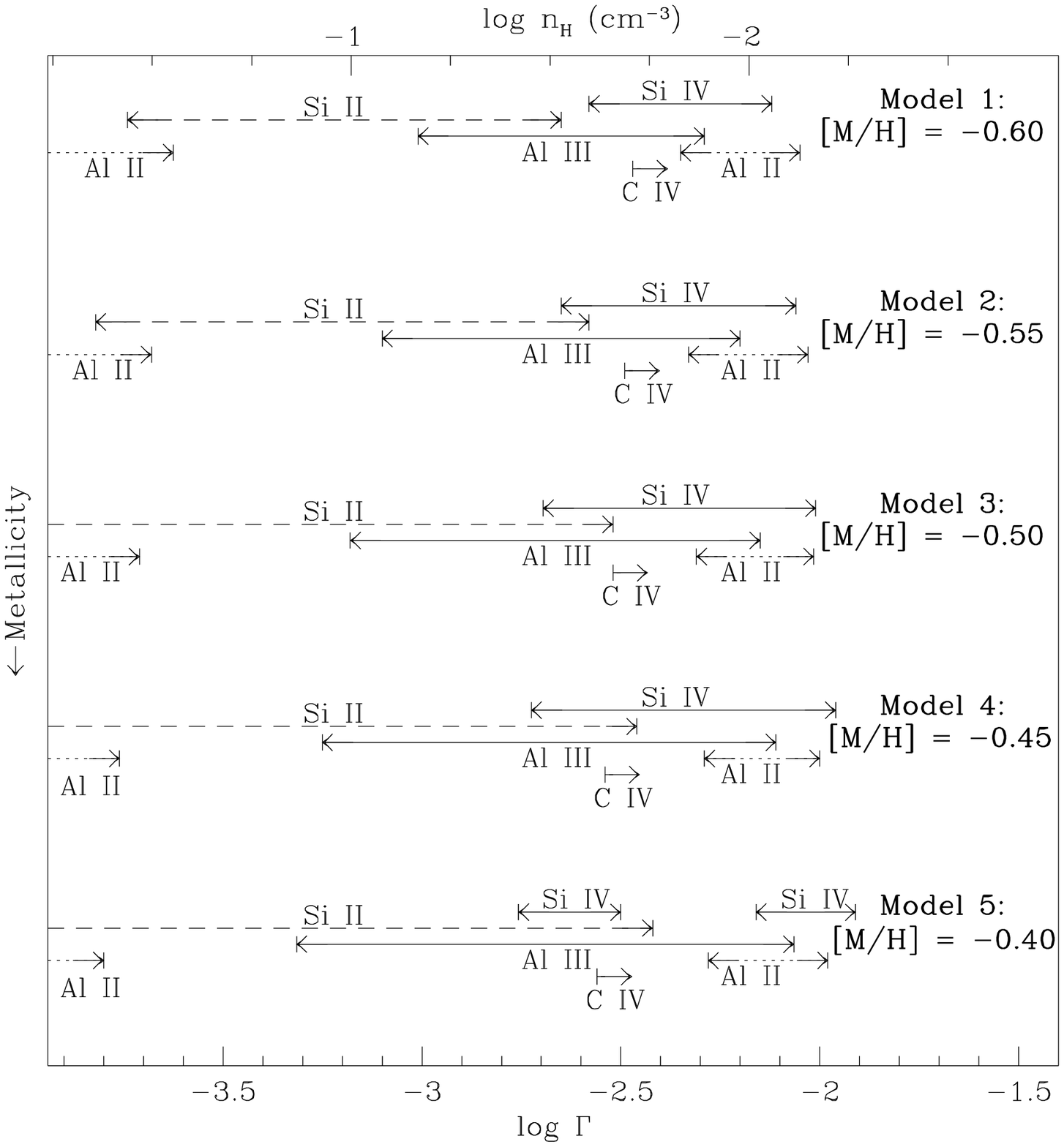}
\caption[]{Summary of photoionization modeling results for the 
Lyman limit absorber at \zabs\ = 2.3150, 2.3155. Five models are 
shown with metallicities ranging from [M/H] = --0.60 to [M/H] = 
--0.40. For each observed species, a horizontal bar indicates the 
range of ionization parameter $\Gamma$ over which the model is in 
agreement with the measured column density to within the 
2$\sigma$ uncertainties. For convenience in distinguishing the 
various species, the ionization parameter range is indicated with a 
dashed line for \ion{Si}{2} and a dotted line for \ion{Al}{2}. Note 
that for some species (e.g., \ion{Al}{2}) there are multiple ranges of 
$\Gamma$ which are in agreement with the measured column 
densities. None of these models which assume solar relative 
abundances are able to simultaneously fit all of the observed column 
densities for any value of the ionization parameter.}
\end{figure}

Figure 7 summarizes the results from five photoionization models 
with --0.60 $\leq$ log $\Gamma \ \leq$ --0.40. For each species, the 
labeled horizontal bar in Figure 7 shows the ionization parameter 
range which yields model column densities in 2$\sigma$ agreement 
with the measured values. From Figure 7 we see that {\it none} of 
the models are able to fit all of the observed column densities for 
any value of $\Gamma$. Models with --0.50 $\lesssim$ [M/H] 
$\lesssim$ --0.40 can fit all of the species except \ion{Al}{2}. When 
[M/H] exceeds --0.40, the model fails to fit several of the observed 
column densities at any given value of $\Gamma$. There are several 
possible explanations for the difficulty in fitting \ion{Al}{2} 
including the following. (1) Aluminum could be underabundant in 
the gas-phase relative to Si and C, due either to depletion onto dust 
grains or relative metal abundances which are intrinsically different 
from solar. In the ISM of the Milky Way, aluminum is strongly 
depleted from the gas-phase by incorporation into dust grains 
(Jenkins\markcite{j3} 1987) while Si and C are moderately or 
lightly depleted (Sofia, Cardelli, \& Savage\markcite{s9} 1994), and 
therefore depletion could cause Al to be underabundant. 
Alternatively, the intrinsic Al/Si and Al/C abundances could be 
different from solar. In low metallicity stars in the Milky Way halo, 
Si is overabundant relative to aluminum, and this is attributed to 
different nucleosynthesis time scales for Si and Al 
(McWilliam\markcite{m3} et al. 1995; Ryan, Norris, \& 
Beers\markcite{r9} 1996; Wheeler, Sneden, \& 
Truran\markcite{w4} 1989 and references therein). Either of these 
phenomena could alleviate the problem with \ion{Al}{2} in the 
model. If Al is underabundant by $\sim$0.2 dex relative to Si and C, 
for example, then the model shown in Figure 6 could fit all of the 
metal column densities at log $\Gamma$ = --2.54. (2) The 
absorption lines could arise in a multiphase medium in which the 
low ionization species are associated with the \ion{H}{1} but some 
or all of the higher ion absorption occurs in a physically distinct 
region. In this case the single-zone models of Figures 6 and 7 are not 
expected to succeed; more complex multi-zone models are required. 
(3) There may be some problems with the Al atomic data in the 
model, especially the Al dielectronic recombination rates (see \S 4.1 
in Petitjean et al. 1994). (4) We may be deceived by noise since the 
\ion{Al}{2} line is marginally detected.

\begin{figure}
\plotone{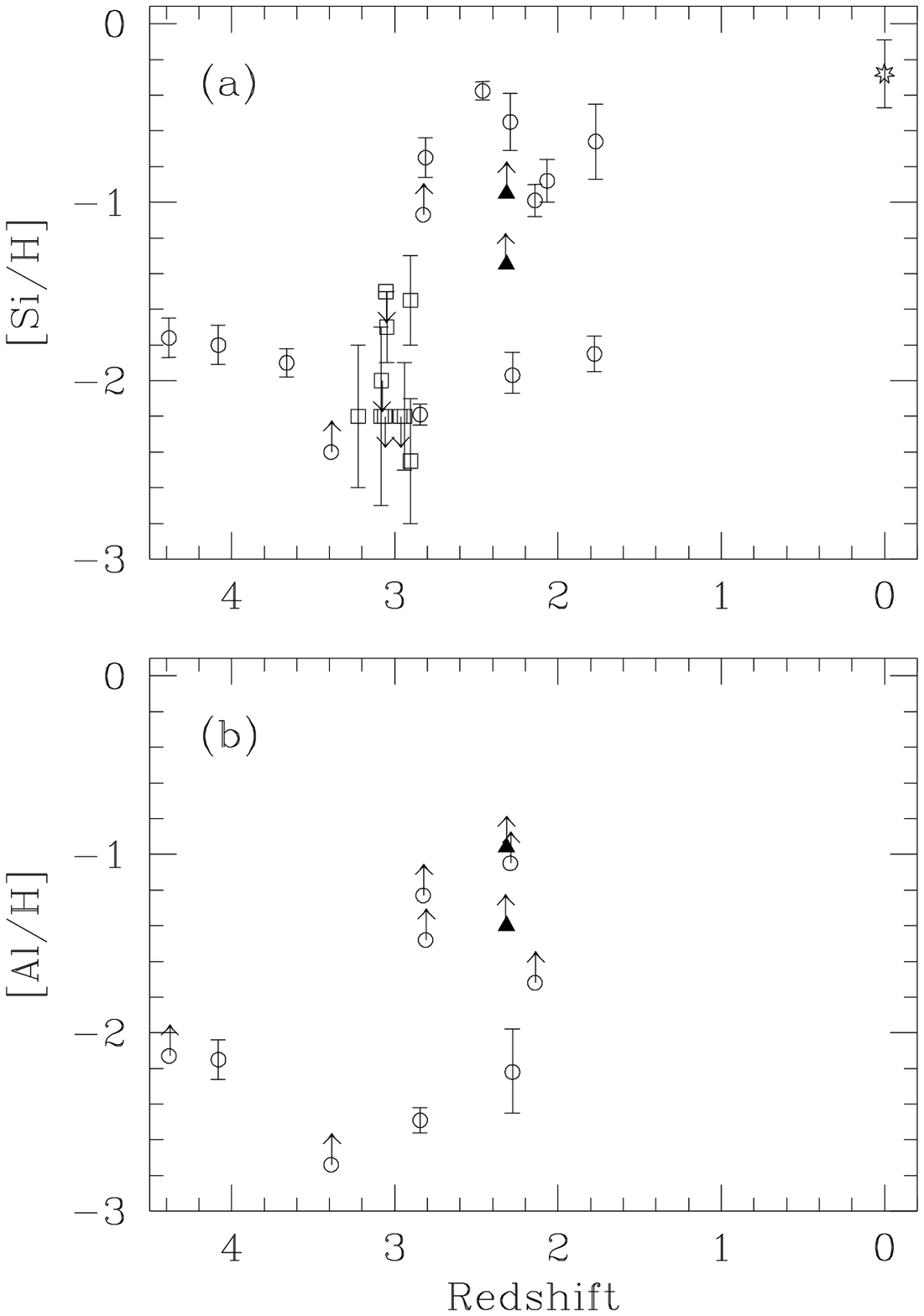}
\caption[]{Metal abundances measured in damped \lya QSO absorbers 
(open circles) as a function of redshift from Lu et al. (1996) and 
references therein for (a) silicon and (b) aluminum. In (a) we also 
show the abundances derived by Steidel (1990) for several high 
redshift Lyman limit absorbers (open squares). Limits are plotted 
with arrows. The conservative lower limits on [Si/H] and [Al/H] 
inferred in this paper for the Lyman limit absorber at \zabs\ = 
2.3150, 2.3155 (see text \S 5.2.1) are indicated with filled triangles 
with arrows. Evidently, the silicon abundance in this LL absorber is 
comparable to the silicon abundance measured in several damped 
\lya absorbers at similar redshifts. In panel (a), the open star at $z$ = 
0 shows the typical gas-phase abundance of Si observed in the 
Milky Way halo (Savage \& Sembach 1996).}
\end{figure}

The presence of multiple components in this absorber suggests that 
it may not be appropriate to model the total column densities as a 
single constant density slab; the multiple components indicate that 
the density changes along the line of sight. However, some previous 
studies of abundances in Lyman limit absorbers (e.g., Vogel \& 
Reimers 1995) are based on total column densities due to the low 
resolution of the instruments used, so if we wish to compare our 
abundances to these studies (see \S 7), we should start with total 
column densities. The column densities integrated across the 
velocity range of the main low ion absorption (components 1 and 2) 
in this absorber are log $N$(\ion{Si}{2}) = 13.28$\pm$0.12, log 
$N$(\ion{Al}{3}) = 12.31$\pm$0.10, and log $N$(\ion{Si}{4}) = 
13.95$\pm$0.05. In this case, the \ion{H}{1} column density 
appropriate for the model is the total, log $N$(\ion{H}{1}) = 16.85. 
With these column densities, we find that the photoionization model 
agrees with the measurements to within the 2$\sigma$ uncertainties 
when --0.62 $\leq$ [M/H] $\leq$ --0.43.

Throughout the rest of this section we will use [M/H] $\approx$ 
--0.45 as a rough estimate of the metallicity in this absorber since 
this value for [M/H] provides the best fit to the component 1 column 
densities and agrees with the constraints set in the previous 
paragraph from the integrated column densities. How does the 
metallicity derived for this LL absorber compare to other 
intervening QSO absorption systems at similar redshifts? At first 
glance, [M/H] $\approx$ --0.45 seems like a high metallicity for 
\zabs\ = 2.315. However, it may not be. Recently, several high 
quality abundance measurements for intervening damped \lya QSO 
absorbers have appeared in the literature (Lu et al.\markcite{l5} 
1996 and references therein, see their Table 16). In Figure 8 we 
show the Si and Al abundances measured in these damped \lya 
absorbers with open circles. The filled triangles in Figure 8 indicate 
the {\it conservative} lower limits on [Si/H] and [Al/H] at \zabs\ = 
2.3150 and 2.3155 from Table 4. We also show in Figure 8a the Si 
abundances derived by Steidel (1990) in a study of several high 
redshift LL absorbers (open squares). From this figure we see that 
the gas-phase Si and Al abundances in this LL absorber are 
comparable to the gas-phase abundances in a significant fraction of 
the damped \lya absorbers at similar redshifts. The open star at $z$ = 
0 in Figure 8a is the diffuse cloud gas-phase abundance of Si 
observed in the Galactic halo (Savage \& Sembach\markcite{s4} 
1996). It is interesting to note that at \zabs\ $\approx$ 2, [Si/H] in 
many intervening QSO absorbers is already approaching the value 
observed in the gaseous halo of the Milky Way. It is also interesting 
to note that [Si/H] appears to increase significantly when \zabs\ 
$\lesssim$ 3. Lu et al. (1996) suggest that the sudden increase in 
metallicity at $z \ \approx$ 3 is due to chemical enrichment from the 
first wide-spread star formation in high redshift galaxies. However 
the sample is small, and more abundance measurements are needed.

Some additional properties of this LL absorber can be derived from 
the models. Since we have assumed a value for the mean intensity at 
the Lyman limit $J_{\nu}$(LL), the hydrogen number density can 
be determined from the ionization parameter. The top axis of Figure 
6 indicates how $n_{\rm H}$ changes with $\Gamma$. The models 
also specify the thickness $t$ of the absorber, and this in turn can be 
used to estimate the cloud mass, assuming the absorber is spherical 
(see \S VIc in Steidel\markcite{s11} 1990). The model with [M/H] 
= --0.45 and log $\Gamma$ = --2.6 has $n_{\rm H}$ = 0.02 
cm$^{-3}$, $t$ = 240 pc, and cloud mass $M_{\rm c}$ = 1.0 
$\times$ 10$^{5} M_{\odot}$. These rough estimates of $t$ and 
$M_{\rm c}$ are close to the values predicted by the recent 
two-phase gaseous halo model of Mo \& 
Miralda-Escud\'{e}\markcite{m5} (1996) for a cold photoionized 
phase embedded in a hot halo.

\subsubsection{Other Lyman Limit Absorbers}

We only detect \ion{C}{4} or \ion{Mg}{2} absorption lines 
associated with the rest of the LL absorbers (see Table 2), so 
photoionization models cannot tightly constrain the elemental 
abundances (the level of ionization is unknown). Nevertheless, we 
can use CLOUDY models to place lower limits on the metallicities 
of these absorbers using equation 3 as discussed above. Applying 
this method with the Madau\markcite{m1} (1992) extragalactic 
radiation field, we derive [C/H] $\geq$ --2.26 for the LL absorber at 
\zabs\ = 1.84506 and [C/H] $\geq$ --2.81 for the LL absorber at 
\zabs\ = 2.16797. However, for the \zabs\ = 1.84506 system we also 
detect \ion{Mg}{2}, and we derive a lower limit of [Mg/H] $\geq$ 
--0.45. If we assume that the \ion{H}{1}, \ion{Mg}{2}, and 
\ion{C}{4} absorption lines at \zabs\ = 1.84506 all originate in the 
same gas, then the photoionization model requires [Mg/H] = [C/H] = 
--0.43. This must also be treated as a lower limit on [M/H] since 
both \ion{Mg}{2} and \ion{C}{4} are affected by absorption 
saturation at this redshift (see \S 4.5). Finally, applying this same 
method to the \ion{Mg}{2} absorber at \zabs\ = 1.15729, we derive 
[Mg/H] $\geq$ --0.22. Once again we see indications that high 
metallicities are already present in moderate redshift intervening 
absorbers.

\subsubsection{Associated Absorber at $z_{abs}$ = 2.7125}

We detect \ion{C}{4} and rather strong \ion{N}{5} absorption at 
\zabs\ = 2.7125 (lines 38-39 and 3-4 in Figure 1). We can also place 
good upper limits on the column densities of \ion{O}{1}, 
\ion{C}{2}, \ion{Al}{2}, \ion{Si}{2}, \ion{Fe}{2}, and 
\ion{Si}{4} at this redshift since these lines are not in the \lya forest. 
No absorption lines are apparent at the expected wavelengths; upper 
limits on these species along with the \ion{C}{4} and \ion{N}{5} 
column densities are summarized in Table 5. The \ion{C}{4} and 
\ion{N}{5} absorption lines are displaced by $\sim$600 \kms\ from 
the emission redshift of the QSO and thus belong in the ``associated'' 
absorber class. Recent high resolution studies have shown that 
associated absorbers are often characterized by strong high ion 
absorption (especially \ion{N}{5}) and overall metallicities 
comparable to or greater than solar (Wampler\markcite{w1} 1991; 
Petitjean et al.\markcite{p3} 1994; Savaglio et al.\markcite{s5} 
1994; Hamann et al.\markcite{h1} 1995; Tripp et al. 1996; 
Hamann\markcite{ha1} 1997). The ionization level of the HS 
1700+6416 associated system inferred from the high ion column 
density ratios is similar to other associated absorbers and rather 
different from the ISM of the Milky Way. The \zabs\ = 2.7125 ionic 
ratios are $N$(\ion{C}{4})/$N$(\ion{N}{5}) $<$ 0.2 and 
$N$(\ion{C}{4})/$N$(\ion{Si}{4}) $>$ 5.2, whereas Sembach et 
al.\markcite{s8} (1997) find that the median ratios in the Galactic 
ISM are $N$(\ion{C}{4})/$N$(\ion{N}{5}) = 4.0$\pm$2.4 and 
$N$(\ion{C}{4})/$N$(\ion{Si}{4}) = 3.8$\pm$1.9 based on 40 
sight lines studied with {\it HST} and the {\it IUE}. For comparison, 
we note that Petitjean et al.\markcite{p3} (1994) report 
$N$(\ion{C}{4})/$N$(\ion{N}{5}) $\sim$ 0.3 and 0.4 for 
associated absorbers of PKS 0424-131 and Q 0450-131 respectively. 
This high degree of ionization is perhaps not surprising since 
associated absorbers are probably physically close to rather UV 
luminous QSOs, either somehow associated with the QSO 
nucleus/host galaxy or possibly occuring in the halos of galaxies in a 
cluster near the quasar, and the gas is predominantly photoionized 
by the QSO. Given the observed \ion{H}{1} column density (see 
below), ionization corrections in the HS 1700+6416 associated 
system are certain to be large, so once again photoionization 
modeling is required to infer abundances.

\begin{deluxetable}{lcc}
\tablewidth{0pc}
\tablenum{5}
\tablecaption{Associated System Column Densities}
\tablehead{ \colhead{Species} & \colhead{Wavelength (\AA )\tablenotemark{a}} & \colhead{log $N$ (cm$^{-2}$)}}
\startdata
\ion{H}{1}...... & 1215.670 & 13.60$\pm$0.06 \nl
\ion{O}{1}.....  & 1302.168 & $<$13.74\tablenotemark{b} \nl
\ion{C}{2}.....  & 1334.532 & $<$13.28\tablenotemark{b} \nl
\ion{Al}{2}....  & 1670.787 & $<$11.68\tablenotemark{b} \nl
\ion{Si}{2}....  & 1526.707 & $<$12.97\tablenotemark{b} \nl
\ion{Fe}{2}....  & 1608.451 & $<$13.21\tablenotemark{b} \nl
\ion{Si}{4}....  & 1393.755 & $<$12.46\tablenotemark{b} \nl
\ion{C}{4}....   & 1548.195, 1550.770 & 13.18$\pm$0.04 \nl
\ion{N}{5}.....  & 1238.821, 1242.804 & $\geq$13.87 \nl
\enddata
\tablenotetext{a}{Vacuum wavelength from Morton (1991).}
\tablenotetext{b}{Column density upper limit derived from the 4$\sigma$ upper
limit on the rest equivalent width, assuming the linear part of the curve of
growth applies.}
\end{deluxetable}

In this associated absorber, we do not have the benefit of optically 
thin Lyman continuum absorption for the determination of 
$N$(\ion{H}{1}), so we have used the \ion{H}{1} \lya absorption 
profile (the line at 4513.2 \AA\ in Figure 1) to constrain the 
\ion{H}{1} column density. Direct integration of the \ion{H}{1} 
apparent column density profile at this redshift yields log 
$N$(\ion{H}{1}) = 13.60$\pm$0.06. For the purpose of this 
photoionization modeling effort, we assume that this \ion{H}{1} 
column is not underestimated due to unresolved absorption 
saturation, but at any rate this provides a firm lower limit. If the 
\ion{H}{1} absorption occurs in the same gas that produces the 
\ion{C}{4} and \ion{N}{5} absorption (which is assumed when 
abundances are derived from a constant density single slab 
photoionized model), then at the likely gas temperatures the 
\ion{H}{1} profile is well enough resolved so that any saturation 
should be apparent. Instead, the \ion{H}{1} profile is broad and the 
apparent optical depth is only $\tau _{\rm a}$ = 0.7 in the deepest 
part of the profile. Furthermore, if this \ion{H}{1} profile contains 
one or more components which are narrow enough to cause 
unresolved saturation, then these cold components might be 
detectable in low ionization stages such as \ion{C}{2}, and they are 
not apparent (see Figure 1 and Table 5). Therefore, we shall proceed 
with the photoionization modeling using log $N$(\ion{H}{1}) = 
13.6. However, we noted in \S4.12 that the \ion{N}{5} \nav\ 
profiles show evidence of unresolved saturation or incomplete 
covering of the QSO flux source. If the \ion{H}{1} absorbing gas 
doesn't completely cover the QSO flux source, then the column 
density from direct integration of \nav\ may require a correction. 
This would increase the measured column density, and in the case of 
\ion{H}{1}, this might decrease the metallicity required by the 
photoionization model to match the heavy element column densities. 
However, the heavy element column densities may be 
underestimated as well, so the corrections could push the model to 
higher or lower metallicities.

\begin{figure}
\plotone{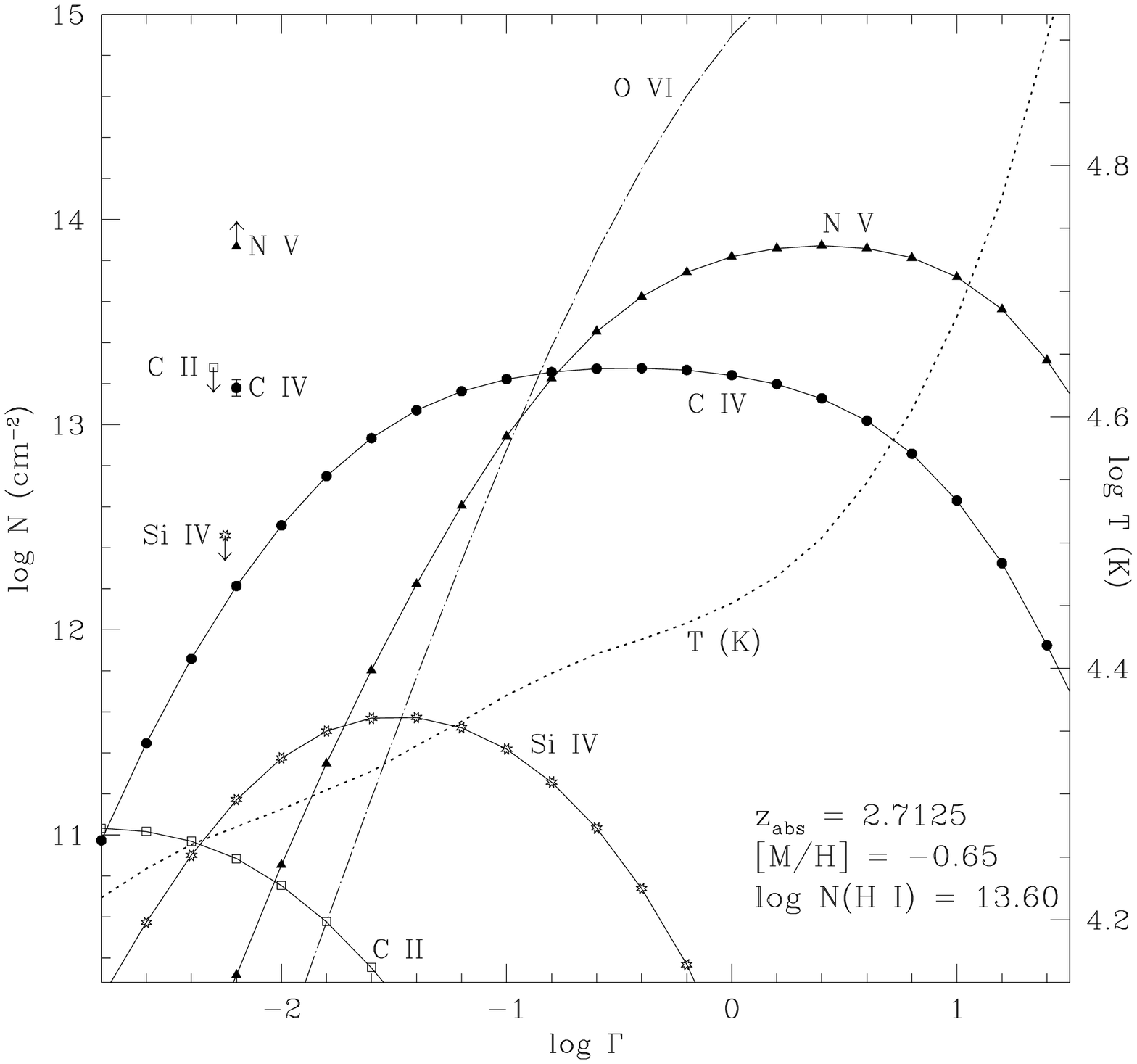}
\caption{Column densities calculated by a photoionized model of 
the associated absorber of HS 1700+6416 at \zabs\ = 2.7125 (\S 
5.2.3) as a function of ionization parameter $\Gamma$. The mean 
temperature of the model is plotted vs. $\Gamma$ with a dotted line 
using the scale on the right axis. At the left side of the figure 
(arbitrarily placed near log $\Gamma$ = --2.2) the observed 
\ion{C}{4} column density (with 1$\sigma$ error bars) and the 
limits on $N$(\ion{N}{5}), $N$(\ion{Si}{4}) and $N$(\ion{C}{2}) 
are indicated. The radiation field used for this calculation is the 
spectral energy distribution of the QSO assembled by 
Siemiginowska et al. (1996).}
\end{figure}

Since the QSO itself is likely to be the dominant source of 
ionization, we have used the observed spectral energy distribution 
(SED) of HS 1700+6416 for the input radiation field for CLOUDY 
abundance estimations. The QSO continuum is {\it detected} from 
$\lambda _{\rm r}$ = 912 \AA\ (in the QSO frame) down to 
$\lambda _{\rm r} \ \approx$ 320 \AA , which is the region of the 
spectrum which governs the ionization of most low and intermediate 
ionization stages and is close to the critical region of the spectrum 
where \ion{C}{3} -- \ion{C}{4} and \ion{N}{4} -- \ion{N}{5} are 
photoionized, i.e., 125 $< \lambda _{\rm r} \ <$ 300 \AA . 
However, since the observed QSO flux is attenuated by at 
least\footnote{The \zabs\ = 0.72219 absorber may be an eighth LL 
system with the Lyman continuum absorption undetected because it 
occurs in a region of the FOS spectrum with low S/N. The 
\ion{Mg}{2} lines detected at this redshift are strong and saturated 
(see \S 4.1), so this absorber is likely to have an appreciable 
\ion{H}{1} column.} seven optically thin Lyman limit absorbers 
(see Figure 1 in Reimers et al.\markcite{r5} 1992), the true 
photoionizing flux impinging on the associated absorber gas is 
certainly greater than the observed flux. This true flux is difficult to 
estimate because the observed flux must be corrected for (1) the 
seven optically thin LL absorbers, (2) the net reddening introduced 
by any dust in intervening absorbers (which is smeared by the 
various redshifts of the intervening systems), and (3) attenuation by 
the \lya forest and the `\lya valley', which includes the cumulative 
Lyman continuum absorption due to all of the intervening absorbers 
(M\o ller \& Jakobsen\markcite{m6} 1990). The correction for the 
seven LL systems is straightforward (see equation 2), but we have 
little information on the magnitude and wavelength dependence of 
the other corrections. Several recent high resolution studies have 
suggested that there may be very little dust in intervening damped 
\lya absorbers (Lu et al.\markcite{l6} 1995; Meyer, Lanzetta, \& 
Wolfe\markcite{m4} 1995; Prochaska \& Wolfe\markcite{p8} 
1996; Lu et al.\markcite{l5} 1996). However, this conclusion has 
been contested by Pettini et al.\markcite{p5} (1997) who argue that 
the observed abundances are consistent with a modest amount of 
dust in the QSO absorbers. To derive a first-order estimation of the 
associated system metallicity, we have used the observed SED of 
HS 1700+6416 from Siemiginowska et al.\markcite{sie1} (1996, see 
their Figure 2) for the CLOUDY radiation field. This SED is based 
on the {\it HST} FOS data from Reimers et al.\markcite{r5} (1992) 
combined with ground-based optical spectrophotometry, the 
ROSAT PSPC data from Reimers et al.\markcite{r1} (1995a), and 
an IRAS detection of the QSO at 60 $\mu$m. Siemiginowska et 
al.\markcite{sie1} have corrected the {\it HST} UV spectrum for the 
seven LL absorbers, but they have not applied any corrections for 
intervening dust or \lya forest/valley attenuation. Therefore the 
Siemiginowska et al. SED provides a firm lower limit on the 
photoionizing flux; various corrections could make the true flux 
greater, but the flux cannot be less than the Siemiginowska et al. 
SED.

We show in Figure 9 the column densities calculated by this model 
as a function of the ionization parameter $\Gamma$ for [M/H] = 
--0.65 with solar relative abundances and log $N$(\ion{H}{1}) = 
13.6. We show this particular model because it provides a lower 
limit on [N/H] (see below). We also show in this figure how the 
mean temperature (dotted line) of the model changes with 
$\Gamma$, and we plot the observed \ion{C}{4} column density as 
well as the lower limit on $N$(\ion{N}{5}) and the upper limits on 
$N$(\ion{C}{2}) and $N$(\ion{Si}{4}) near (arbitrarily) log 
$\Gamma$ = --2.2. The constraints on the low ions and \ion{Si}{4} 
are easily satisfied by the model, and the model temperature is 
compatible with the upper limit set by the \ion{C}{4} line width 
(see Table 3). We see from Figure 9 that the model agrees with the 
observed \ion{C}{4} column density to within 2$\sigma$ over a 
broad range in ionization parameter, --1.4 $\leq$ log $\Gamma \ 
\leq$ +0.4. However, over most of this ionization parameter range, 
the model falls far short of the lower limit on $N$(\ion{N}{5}). To 
produce enough \ion{N}{5}, the model requires log $\Gamma$ = 
+0.4, and even then the model \ion{N}{5} column is just barely 
large enough, which sets a lower limit of [N/H] $\geq$ --0.65. For 
carbon the model requires [C/H] $\geq$ --0.82. However, the true 
\ion{N}{5} column density may be significantly larger than the 
lower limit in Table 5, and to obtain a higher 
\ion{N}{5}/\ion{C}{4} ratio, $\Gamma$ must be increased (see 
Figure 9). At higher values of $\Gamma$ the \ion{C}{4} and 
\ion{N}{5} column densities predicted by the model decrease, so 
the metallicity must in turn be increased as well to agree with the 
measured column densities. These limits assume solar N/C relative 
abundances. Hamann \& Ferland\markcite{h4} (1993) have argued, 
on the basis of QSO {\it emission} line intensities, that nitrogen is 
overabundant relative to carbon in the immediate vicinity of a QSO. 
This may provide another means to boost the 
\ion{N}{5}/\ion{C}{4} ratio in this associated absorber {\it if} the 
absorbing gas is close to the quasar, but in this case models with 
lower ionization parameters and lower metallicities can be brought 
into agreement with the measured column densities.

Associated system abundance measurements based on \ion{C}{4} 
and \ion{N}{5} are sensitive to the shape of the ionizing continuum 
assumed for the model. Hamann (1997) has recently assembled a 
suite of photoionization models for the study of abundances in 
associated and broad absorption lines using a variety of continuum 
shapes, and we can use his work to illustrate this sensitivity to 
continuum shape. His continuum shape `A' with a UV cutoff 
temperature $T_{c} = 10^{5.7}$ K (see Appendix in Hamann 1997) 
is consistent with the observed SED of HS 1700+6416 from 
Siemiginowska et al. (1996) in the FUV but has significantly higher 
flux in the ionizing ultraviolet where Siemiginowska et al. did not 
measure the QSO flux. Using this continuum shape `A', we derive 
[C/H] $\geq$ --0.25 and [N/H] $\geq$ +0.19 for the associated 
system of HS 1700+6416. The 
continuum shape we used in the previous paragraph and in Figure 9 
(based on Siemiginowska et al.) provides firm lower limits on [C/H] 
and [N/H] because this continuum is the lower limit on the ionizing 
flux. There are various corrections which could make the SED harder 
and hence increase the lower limits on [C/H] and [N/H].

Finally, we comment that these photoionization models are highly 
simplified, and there are some indications that the real absorbers are 
more complicated than the assumed models. Consider the Doppler 
parameters derived from profile fitting in Table 1. These $b$-values 
can be expressed as a sum with a term due to thermal motions 
($b_{t}$) and a term which includes all 
non-thermal motions ($b_{nt}$) such as bulk motions and 
turbulence,
\begin{equation}
b^{2} = b_{nt}^{2} + b_{t}^{2} = b_{nt}^{2} + \frac{2kT}{m} = 
b_{nt}^{2} + \frac{(0.129)^{2}T}{A},
\end{equation}
where $m$ is the mass and $A$ is the atomic weight of the element 
and $b$ is measured in \kms . For the photoionized models 
presented above, we assumed that all of the absorption lines of 
various species in a given component arise in a constant density gas 
slab. Therefore one might expect different elements to have the 
same values for $b_{nt}$ and $T$, and higher mass elements should 
have lower $b$-values. In principle one can solve equation 4 for 
$b_{nt}$ and $T$ given $b$-values for two or more elements. 
However, in this associated absorber profile fitting yields $b$ = 
25.2$\pm$1.3 \kms\ for \ion{N}{5} and $b$ = 11.4$\pm$1.1 \kms\ 
for \ion{C}{4}. This is not consistent with the assumption that 
$b_{nt}$ and $T$ are the same for the \ion{C}{4} and \ion{N}{5} 
absorption lines because this would require an unphysical negative 
temperature. This suggests that (1) there is some problem with the 
Doppler parameter measurements or uncertainties, or (2) the 
properties of the gas causing the \ion{C}{4} absorption are different 
from the properties of the gas producing the \ion{N}{5} absorption. 
The disparity between $b$(\ion{N}{5}) and $b$(\ion{C}{4}) 
cannot be blamed on the profile fitting procedure; attempts to fit the 
\ion{N}{5} and \ion{C}{4} profiles with compatible $b$-values, 
assuming $b_{nt}$ and $T$ are the same, yield unacceptably bad 
fits. The nitrogen line would be broader than the \ion{C}{4} line if 
\ion{N}{5} is strongly saturated, but profile fitting should account 
for such saturation and yield consistent $b$-values. Therefore in this 
associated absorber, the \ion{C}{4} gas and \ion{N}{5} gas do not 
have the same non-thermal motions and temperature. It is possible 
that the \ion{C}{4} absorption and {\it some} of the \ion{N}{5} 
absorption occurs in the same gas cloud with additional \ion{N}{5} 
absorption arising in a separate cloud which does not produce 
\ion{C}{4} absorption. Some support for this suggestion is provided 
by the fact that the \ion{N}{5} profiles are affected by unresolved 
saturation and/or incomplete covering of the QSO flux source, while 
the \ion{C}{4} profiles show no evidence of these effects (see 
Figure 5). Future studies of this type of QSO absorber may require 
more intricate models.

\section{Weak \ion{Mg}{2} Systems}

Based on a survey of \ion{Mg}{2} absorption lines in the spectra of 
103 QSOs, Steidel \& Sargent\markcite{s14} (1992) conclude that 
there is paucity of {\it weak} \ion{Mg}{2} lines, and they assert that 
``at least 80\% of the \ion{Mg}{2} absorbers are accounted for in a 
sample sensitive to $W_{\rm r}^{\rm min}$ = 0.3 \AA '' ($W_{\rm 
r}^{\rm min}$ is the minimum rest equivalent width in the sample). 
We draw a different (albeit tentative) conclusion from our high 
resolution high S/N spectra of HS 1700+6416 and HS 1946+7658 
(Tripp et al.\markcite{t2} 1996). The sensitivity of our spectra of 
these QSOs is adequate for detection of \ion{Mg}{2} lines with 
restframe equivalent width $W_{\rm r}$ as small as 0.03 \AA\ (at 
the 4$\sigma$ level) over a total redshift path $\Delta z$ = 2.56. We 
detect six definite \ion{Mg}{2} absorbers in the spectra of these 
QSOs, and only one of these \ion{Mg}{2} lines has $W_{\rm r} 
\geq$ 0.3\AA . To expand the sample size, we can combine our data 
with the similar quality spectra of PKS 0424--131 ($z_{\rm em}$ = 
2.1657) and Q 0450--131 ($z_{\rm em}$ = 2.2535) obtained by 
Petitjean et al.\markcite{p3} (1994)\footnote{The spectra obtained 
by Petitjean et al.\markcite{p3} (1994) have a resolution of 
$\sim$20 \kms\ and sufficient S/N for detection of all lines with 
$W_{\rm r} \geq$ 0.03\AA .}. From this combined sample of four 
quasars, we derive a mean number of \ion{Mg}{2} absorbers per 
unit redshift of $dN/dz$ = 2.3$\pm$0.8. From their sample of 
\ion{Mg}{2} absorbers with $W_{\rm r} \geq$ 0.3\AA , Steidel \& 
Sargent\markcite{s14} (1992) derive $dN/dz$ = 0.97$\pm$0.10. It 
is important to emphasize that $dN/dz$ derived from our sample of 
four quasars is highly uncertain due to the small sample. However, 
Womble\markcite{w5} (1995) has recently detected a similar excess 
of {\it weak} \ion{Mg}{2} systems ($W_{\rm r} \geq 0.015$ \AA ) 
in a high S/N high resolution Keck spectrum of PG 1634+706 
($z_{\rm em}$ = 1.334). Furthermore, Steidel \& 
Sargent\markcite{s14} find that $dN/dz$ increases smoothly as the 
minimum $W_{\rm r}$ in the sample decreases from 1.0 \AA\ to 
0.3 \AA\ (see their Figure 5). This also suggests that $dN/dz$ is 
dominated by weak \ion{Mg}{2} systems. This excess of weak 
\ion{Mg}{2} lines may indicate that the cross section of the 
\ion{Mg}{2} absorbers is large; Womble\markcite{w5} estimates 
that the required radius is $\sim$90 $h^{-1}$ kpc ($h$ = 
H$_{0}$/100 \kms\ Mpc$^{-1}$) for lines with $W_{\rm r} \geq 
0.015$ \AA . An excess of weak \ion{Mg}{2} absorbers may also 
have implications for the study of galaxies {\it selected} because 
they produce \ion{Mg}{2} absorption in the spectrum of the 
background QSO. For example, Steidel, Dickinson, \& 
Persson\markcite{s12} (1994) find that their sample of \ion{Mg}{2} 
absorbers are mostly associated with normal luminous galaxies, but 
their sample may be biased against underluminous galaxies because 
of the lower equivalent width cutoff (they select \ion{Mg}{2} 
systems with $W_{\rm r} \geq$ 0.3 \AA\ for follow-up imaging and 
redshift measurements). With the echelle spectrograph on the Keck 
telescope, it should be straightforward to confirm or refute this 
excess of weak \ion{Mg}{2} absorbers with a larger sample.

As discussed by Petitjean \& Bergeron\markcite{p2} (1990) and 
Steidel \& Sargent\markcite{s14} (1992), there are some indications 
that the fractions of \ion{Mg}{2} absorbers which are weak and 
strong may evolve with redshift in the sense that there are more 
strong \ion{Mg}{2} lines at higher redshift. This probably is not the 
reason that the $dN/dz$ obtained by Steidel \& Sargent 
\markcite{s14} is lower than the $dN/dz$ we have derived because 
the redshift range probed by Steidel \& Sargent\markcite{s14}, 0.2 
$\leq z_{\rm abs} \leq$ 2.2, is similar to the redshift range probed by 
our sample, 0.4 $\leq z_{\rm abs} \leq$ 2.1.

\section{Comments on {\it HST} FOS Abundances}

The {\it HST} FOS observation of HS 1700+6416 obtained by 
Reimers et al.\markcite{r5} (1992) is ground-breaking work, and the 
spectrum has provided an excellent reconnaissance of this 
extraordinary sight line. However, the data have some serious 
limitations due to the low resolution (FWHM $\sim$300 \kms ) of 
the FOS, and we suspect that the abundances derived from the FOS 
data by Vogel \& Reimers\markcite{v3} (1995) are unreliable. 
Several problems are introduced by the low resolution of the data 
combined with the high redshift of the QSO (which leads to a very 
high line density in the FOS spectral region including metal lines, 
He lines, and many \ion{H}{1} Lyman series lines): (1) In many 
cases, the absorption lines are strongly blended with several lines of 
different species at different redshifts (see Figure 4 in Vogel \& 
Reimers\markcite{v3} 1995), and extracting accurate column 
densities from the blends is rather difficult. (2) As Lipman, Pettini, 
\& Hunstead\markcite{l2} (1995) have noted (see their Figure 1), 
narrow lines are unlikely to be detected in the FOS spectrum, and 
the FOS line list will be biased toward strongly saturated lines which 
do not provide reliable column densities. (3) Quasar absorption line 
profiles frequently show multiple components which will be 
completely unresolved at the FOS resolution, and consequently 
absorption by low and high ionization species which appear to be at 
the same redshift in the FOS data may not actually be associated 
(i.e., the low and high ion absorption occur in different unresolved 
components). Confusion due to multiple unresolved components 
will be especially severe in the absorption line complex at 2.432 $<$ 
\zabs\ $<$ 2.441 (see \S 4.10).

This paper confirms that the Vogel \& Reimers\markcite{v3} (1995) 
abundances may be problematic. For example, we estimate that the 
strongest component of the LL absorber at \zabs\ = 2.3150 has 
[Si/H] $\geq$ --0.95 and [Al/H] $\geq$ --0.96 and is best fit with 
[M/H] $\approx$ --0.45 (see \S 5.2.1), while Vogel \& 
Reimers\markcite{v3} report [N/H] $<$ --1.04 and [O/H] = --1.52. 
The low resolution of the FOS also yields redshifts which may not 
be accurate enough to facilitate line identifications. Consider the LL 
system for which Vogel \& Reimers\markcite{v3} give \zabs\ = 
1.8465. We detect the \ion{Mg}{2} doublet at high significance 
levels at \zabs\ = 1.84506 (lines 44 and 45 in Figure 1), which 
differs from the Vogel \& Reimers\markcite{v3} redshift by 
$\sim$150 \kms . This amount of uncertainty in $z$ can make line 
identification difficult given the high density of lines, and indeed 
Petitjean et al.\markcite{p4} (1996) have pointed out several lines 
that Vogel \& Reimers\markcite{v3} may have incorrectly identified 
because they were not aware of the associated absorption system of 
HS 1700+6416. Similarly, the extended X-ray sources near the sight 
line (see \S 1) may produce absorption lines, at \zabs\ = 0.225 in the 
case of Abell 2246, and these low redshift absorption lines could 
have been misidentified as well. These problems are generic 
drawbacks of low resolution data which should be greatly alleviated 
when HS 1700+6416 is observed with the Space Telescope Imaging 
Spectrograph.

\section{Summary}

The paper is summarized as follows.

(1) We have obtained a high S/N high resolution (20 \kms\ FWHM) 
spectrum of HS 1700+6416 ($z_{\rm em}$ = 2.72) with complete 
spectral coverage from 4300 to 8350 \AA . Thirteen absorption 
systems with metals are detected in this optical spectrum, including 
six absorbers which cause optically thin Lyman limit absorption in 
the {\it HST} spectrum obtained by Reimers et al.\markcite{r5} 
(1992).

(2) Four \ion{C}{4} doublets are narrow ($b \ <$ 8 \kms ) which 
indicates that these absorbers are relatively cool ($T \ <$ 50000 K) 
and probably photoionized.

(3) A dense cluster of \ion{C}{4} doublets is detected at 2.432 $<$ 
\zabs\ $<$ 2.441, which corresponds to a displacement of 
$\sim$24000 \kms\ from the QSO emission redshift. Despite the 
large displacement from $z_{\rm em}$, two pairs of \ion{C}{4} 
doublets are apparently line locked, i.e., the strong line of one 
\ion{C}{4} absorber is aligned with the weak line of a different 
\ion{C}{4} doublet, to within 10 \kms\ in one case and to within 25 
\kms\ in the other case. This may be the remnant (or precursor) of a 
BAL outflow. However, it is possible that these alignments are 
coincidental rather than true line locking.

(4) The column density ratios of high ionization stages in the Lyman 
limit absorber at \zabs\ = 2.3150 differ significantly from the ratios 
observed in the gaseous halo of the Milky Way. For example, 
$N$(\ion{C}{4})/$N$(\ion{Si}{4}) $\geq$ 10 in two components 
of this absorber. This high ionic ratio in the LL absorber is not 
expected in cooling galactic fountain gas but could be produced in 
supernova remnants, conductive interfaces between warm and hot 
gas, turbulent mixing layers, or from photoionization by the diffuse 
extragalactic background radiation.

(5) Photoionization modeling of the Lyman limit absorber at \zabs\ = 
2.3150 places conservative lower limits on [Si/H] and [Al/H]. The 
model which provides the best fit has [M/H] $\approx$ --0.45. This 
is substantially larger than the metallicity derived by Vogel \& 
Reimers\markcite{v3} (1995) from the {\it HST} FOS spectrum. 
The discrepancy is probably due to the low resolution of the FOS 
data which causes severe blending and saturation problems.

(6) An associated absorber at \zabs\ = 2.7125 shows \ion{C}{4} and 
strong \ion{N}{5} lines. The \ion{C}{4} lines are unsaturated, but 
the \ion{N}{5} apparent column density profiles indicate that 
\ion{N}{5} is affected by unresolved saturation, or the \ion{N}{5} 
absorbing gas does not completely cover the QSO flux source.

(7) Despite coverage of several low ionization stages of abundant 
elements with high $f\lambda$ values, no low ion absorption is 
evident in the associated absorber. Also, the high ion column density 
ratios differ significantly from the ratios observed in the Milky Way 
ISM. This indicates that the absorber is highly ionized, presumably 
by the QSO. Photoionization modeling using the QSO spectral 
energy distribution for the input radiation field requires [N/H] 
$\geq$ --0.65 and [C/H] $\geq$ --0.82. These are firm lower limits 
which could be increased by plausible corrections to the ionizing 
continuum shape. However, profile fitting yields $b$(\ion{N}{5}) = 
25.2$\pm$1.3 \kms\ and $b$(\ion{C}{4}) = 11.4$\pm$1.1, and 
therefore the gas which produces the \ion{N}{5} absorption does 
not have the same temperature and/or non-thermal motions as the 
\ion{C}{4} gas, and a more complex photoionization model may be 
required for this absorber.

(8) We briefly examine the number of \ion{Mg}{2} absorbers 
detected per unit redshift. The line density appears to be dominated 
by weak \ion{Mg}{2} lines with $W_{\rm r} \ <$ 0.3 \AA , but the 
sample suffers from small number statistics.

\acknowledgements
It is a pleasure to thank Fred Hamann for several excellent 
comments which significantly improved this paper. We 
acknowledge Gary Ferland for use of his photoionization code 
CLOUDY, and we thank Ken Lanzetta for sharing software which 
was used for profile fitting. Kathy Romer and David Cohen 
provided helpful comments on the ROSAT data and the nature of 
the extended X-ray sources near HS 1700+6416. T.M.T. received 
support from NASA grant NGT51003. This research is also 
supported by NASA through grant HF1062.01-94A for L.L. from 
the Space Telescope Science Institute, which is operated by the 
Association of Universities for Research in Astronomy, Inc., for 
NASA under contract NAS5-26555. B.D.S. acknowledges support 
from NASA grant NAG5-1852.


\begin{references}

\reference{a1} Arav, N., Korista, K. T., Barlow, T. A., \& 
Begelman, M. C. 1995, Nature, 376, 576
\reference{a2} Arav, N., Li, Z.-Y., \& Begelman, M. C. 1994, ApJ, 
432, 62
\reference{b1} Barlow, T. A., \& Sargent, W. L. W. 1997, AJ, 113, 
136
\reference{b2} Beaver, E. A., et al. 1991, ApJ, 377, L1
\reference{b3} Bechtold, J. 1995 in QSO Absorption Lines, ed. G. 
Meylan, (Berlin:Springer), 299
\reference{b4} Benjamin, R. A., \& Shapiro, P. R. 1997, ApJS, 
submitted
\reference{b5} Bergeron, J., \& Stasi\'{n}ska, G. 1986, A\&A, 169, 
1
\reference{c1} Cowie, L. L., Songaila, A., Kim, T.-S., \& Hu, E. M. 
1995, AJ, 109, 1522
\reference{e1} Edgar, R. J., \& Chevalier, R. A. 1986, ApJ, 310, 
L27
\reference{fe1} Ferland, G. J. 1993, University of Kentucky 
Department of Physics and Astronomy Internal Report
\reference{f1} Foltz, C. B., Chaffee, F. H., Weymann, R. J., \& 
Anderson, S. F. 1988, in Quasar Absorption Lines: Probing the 
Universe, ed. J. C. Blades, D. A. Turnshek, \& C. A. Norman 
(Cambridge: Cambridge University Press), 53
\reference{f2} Foltz, C. B., Weymann, R. J., Morris, S. L., \& 
Turnshek, D. A. 1987, ApJ, 317, 450
\reference{g2} Grevesse, N., \& Anders, E. 1989, in Cosmic 
Abundances of Matter, ed. C. J. Waddington, (New York: American 
Institute of Physics), 1
\reference{ha1} Hamann, F. 1997, ApJS, in press
\reference{h1} Hamann, F., Barlow, T. A., Beaver, E. A., Burbidge, 
E. M., Cohen, R. D., Junkkarinen, V., \& Lyons, R. 1995, ApJ, 443, 
606
\reference{h2} Hamann, F., Barlow, T. A., \& Junkkarinen, V. 
1997a, ApJ, in press
\reference{ha2} Hamann, F., Barlow, T. A., Junkkarinen, V., \& 
Burbidge, E. M. 1997b, ApJ, in press
\reference{h4} Hamann, F., \& Ferland, G. 1993, ApJ, 418, 11
\reference{h5} Horne, K. 1986, PASP, 98, 609
\reference{j1} Jakobsen, P., Boksenberg, A., Deharveng, J. M., 
Greenfield, P., Jedrzejewski, R., \& Paresce, F. 1994, Nature, 370, 
35
\reference{j2} Jannuzi, B. T., et al. 1996, ApJ, 470, L11
\reference{j3} Jenkins, E. B. 1987, in Interstellar Processes, ed. D. J. 
Hollenbach \& H. A. Thronson (Dordrecht: Reidel), 533
\reference{j4} Jenkins, E. B. 1996, ApJ, 471, 292
\reference{l1} Lanzetta, K. M., \& Bowen, D. V. 1992, ApJ, 391, 48
\reference{l2} Lipman, K., Pettini, M., \& Hunstead, R. W. 1995, in 
QSO Absorption Lines, ed. G. Meylan, (Berlin:Springer), 89
\reference{l3} Lockman, F. J., \& Savage, B. D. 1995, ApJS, 97, 1
\reference{l4} Lyons, R. W., Cohen, R. D., Junkkarinen, V. T., 
Burbidge, E. M., \& Beaver, E. A. 1995, AJ, 110, 1544
\reference{l5} Lu, L., Sargent, W. L. W., \& Barlow, T. A., 
Churchill, C. W., \& Vogt, S. 1996, ApJS, 107, 475
\reference{l6} Lu, L, Savage, B. D., Tripp, T. M., \& Meyer, D. M. 
1995, ApJ, 447, 597
\reference{m1} Madau, P. 1992, ApJ, 389, L1
\reference{m3} McWilliam, A., Preston, G., Sneden, C., \& Searle, 
L. 1995, AJ, 109, 2757
\reference{m4} Meyer, D. M., Lanzetta, K. M., \& Wolfe, A. M. 
1995, ApJ, 451, L13
\reference{m5} Mo, H. J., \& Miralda-Escud\'{e}, J. 1996, ApJ, 469, 
589
\reference{m6} M\o ller, P., \& Jakobsen, P. 1990, A\&A, 228, 299
\reference{m7} Morris, S. L., Weymann, R. J., Foltz, C. B., 
Turnshek, D. A., Shectman, S., Price, C., \& Boroson, T. A. 1986, 
ApJ, 310, 40
\reference{m8} Morton, D. C. 1991, ApJS, 77, 119
\reference{m9} Murray, N., Chaing, J., Grossman, S. A., \& Voit, G. 
M. 1995, ApJ, 451, 498
\reference{p1} Perry, J. J., Burbidge, E. M., \& Burbidge, G. R. 
1978, PASP, 90, 337
\reference{p2} Petitjean, P., \& Bergeron, J. 1990, 231, 309
\reference{p3} Petitjean, P., Rauch, M., \& Carswell, R. F., 1994, 
A\&A, 291, 29
\reference{p4} Petitjean, P., Riediger, R., \& Rauch, M. 1996, 
A\&A, 307, 417
\reference{p5} Pettini, M., King, D. L., Smith, L. J., \& Hunstead, 
R. W. 1997, ApJ, submitted
\reference{p6} Pettini, M., Smith, L. J., Hunstead, R. W., \& King, 
D. L. 1994, ApJ, 426, 79
\reference{p7} Picard, A., \& Jakobsen, P. 1993, A\&A, 276, 331
\reference{p8} Prochaska, J. X., \& Wolfe, A. M. 1996, ApJ, 470, 
474, 140
\reference{ra1} Rauch, M., Sargent, W. L. W., Womble, D. S., \& 
Barlow, T. A. 1996, ApJ, 467, L5
\reference{r1} Reimers, D., Bade, N., Schartel, N., Hagen, H.-J., 
Engles, D., \& Toussaint, F. 1995a, A\&A, 296, L49
\reference{r2} Reimers, D., Clavel, J., Groote, D., Engels, H. J., 
Naylor, T., Wamsteker, W., \& Hopp, U. 1989, A\&A, 218, 71
\reference{r4} Reimers, D., Rodriguez-Pascual, P., Hagen, H.-J., \& 
Wisotzki, L. 1995b, A\&A, 293, L21
\reference{r5} Reimers, D., Vogel, S., Hagen, H.-J., Engels, D., 
Groote, D., Wamsteker, W., Clavel, J., \& Rosa, M. R. 1992, Nature, 
360, 561
\reference{r6} Robertson, J. G. 1986, PASP, 98, 1220
\reference{r7} Rodr\'{i}guez-Pascual, P. M., de la Fuente, A., Sanz, 
J. L., Recondo, M. C., Clavel, J., Santos-Lle\'{o}, M., \& 
Wamsteker, W. 1995, ApJ, 448, 575
\reference{ro} Romer, A. K. 1996a, private communication
\reference{r8} Romer, A. K., Ulmer, M. P., Nichol, B. C., Holden, 
B. P., Burke, D., \& Collins, C. A. 1996b, BAAS, 28, 853
\reference{r9} Ryan, S. G., Norris, J. E., \& Beers, T. C. 1996, ApJ, 
471, 254
\reference{s1} Sanz, J. L., Clavel, J., Naylor, T., \& Wamsteker, W. 
1993, MNRAS, 260, 468
\reference{s2} Sargent, W. L. W., Boksenberg, A., \& Steidel, C. C. 
1988, ApJS, 68, 539
\reference{s3} Savage, B. D., \& Sembach, K. R. 1991, ApJ, 379, 
245
\reference{s4} Savage, B. D., \& Sembach, K. R. 1996, \araa, 34, 
279
\reference{s5} Savaglio, S., D'Odorico, S., \& M\o ller, P. 1994, 
A\&A, 281, 331
\reference{s6} Scoville, N., \& Norman, C. 1995, ApJ, 451, 510
\reference{s7} Sembach, K. R., \& Savage, B. D. 1992, ApJS, 83, 
147
\reference{s8} Sembach, K. R., Savage, B. D., \& Tripp, T. M. 
1997, ApJ, in press
\reference{sie1} Siemiginowska, A., Bechtold, J., Tran, K.-V., \& 
Dobrzycki, A. 1996, in Science with the Hubble Space Telescope - 
II, ed. P. Benvenuti, F. D. Macchetto, \& E. J. Schreier, 
STScI/ST-ECF Workshop, 181
\reference{s9} Sofia, U. J., Cardelli, J. A., \& Savage, B. D. 1994, 
ApJ, 430, 650
\reference{s10} Spitzer, L. 1978, Physical Processes in the 
Interstellar Medium, (New York: John Wiley \& Sons)
\reference{s11} Steidel, C. C., 1990, ApJS, 74, 37
\reference{s12} Steidel, C. C., Dickinson, M., \& Persson, S. E. 
1994, ApJ, 437, L75
\reference{s13} Steidel, C. C., \& Sargent, W. L. W. 1989, ApJ, 
343, L33
\reference{s14} Steidel, C. C., \& Sargent, W. L. W. 1992, ApJS, 
80, 1
\reference{s15} Sutherland, R. S., \& Dopita, M. A. 1993, ApJS, 88, 
253
\reference{t1} Tripp, T. M., Bechtold, J., \& Green, R. F. 1994, ApJ, 
433, 533
\reference{t2} Tripp, T. M., Lu, L., \& Savage, B. D. 1996, ApJS, 
102, 239
\reference{tu1} Turnshek, D. A. 1988 in Quasar Absorption Lines: 
Probing the Universe, ed. J. C. Blades, D. A. Turnshek, \& C. A. 
Norman (Cambridge: Cambridge University Press), 17
\reference{t3} Turnshek, D. A., Weymann, R. J., \& Williams, R. E. 
1979, ApJ, 230, 330
\reference{t4} Tytler, D., Fan, X.-M., Burles, S., Cottrell, L., Davis, 
C., Kirkman, D., \& Zuo, L. 1995, in QSO Absorption Lines, ed. G. 
Meylan, (Berlin:Springer), 289
\reference{v1} Verner, D. A., Tytler, D., \& Barthel, P. D. 1994, 
ApJ, 430, 186
\reference{v2} Viegas, S. M. 1995, MNRAS, 276, 268
\reference{v3} Vogel, S., \& Reimers, D. 1995, A\&A, 294, 377
\reference{w1} Wampler, E. J. 1991, 368, 40
\reference{w2} Wampler, E. J., Bergeron, J., \& Petitjean, P. 1993, 
A\&A, 273, 15
\reference{w3} Weymann, R. J., Carswell, R. F., \& Smith, M. G. 
1981, ARA\&A, 19, 41
\reference{w4} Wheeler, J. C., Sneden, C., \& Truran, J. W. 1989, 
ARA\&A, 27, 279
\reference{w5} Womble, D. S. 1995 in ESO Workshop on QSO 
Absorption Lines, ed. G. Meylan, (Berlin: Springer), 157
\end{references}
\end{document}